\documentclass[11pt,a4paper]{article}
\pdfoutput=1

\usepackage{jcappub}
\usepackage[latin1]{inputenc}
\usepackage{graphicx}
\usepackage{epsfig}
\usepackage{subfigure}
\usepackage{color}
\usepackage{comment}
\usepackage{slashed}
\usepackage{soul}
\usepackage{tabularx}

\newcommand{\be}{\begin{equation}} 
\newcommand{\ee}{\end{equation}}
\newcommand{\bea}{\begin{equation}\begin{aligned}} 
\newcommand{\eea}{\end{aligned}\end{equation}}

\def\lsim{\mathrel{\raise.3ex\hbox{$<$\kern-.75em\lower1ex\hbox{$\sim$}}}}
\def\gsim{\mathrel{\raise.3ex\hbox{$>$\kern-.75em\lower1ex\hbox{$\sim$}}}}

\usepackage{array}
\newcolumntype{C}[1]{>{\centering\let\newline\\\arraybackslash\hspace{0pt}}m{#1}}

\newcommand{\tr}{{\rm tr}}
\newcommand{\yr}{{\rm yr}}
\newcommand{\au}{{\rm au}}

\newcommand{\pc}{{\rm pc}}

\newcommand{\Gpc}{{\rm Gpc}}

\newcommand{\td}{{\rm d}}
\newcommand{\Msun}{M_\odot}

\begin{document}

\mbox{} \hfill KCL-PH-TH/2018-70 \\
\mbox{} \hfill CERN-TH-2018-266

\title{Formation and Evolution \\of Primordial Black Hole Binaries \\in the Early Universe}

\author[a]{Martti Raidal,}
\author[a]{Christian Spethmann,}
\author[b]{Ville Vaskonen}
\author[a,c]{\\and Hardi Veerm\"ae}

\affiliation[a]{NICPB, R\"avala 10, 10143 Tallinn, Estonia}  
\affiliation[b]{Physics Department, King's College London, London WC2R 2LS, United Kingdom}
\affiliation[c]{Theoretical Physics Department, CERN, CH-1211 Geneva 23, Switzerland}
                            
\emailAdd{martti.raidal@cern.ch}
\emailAdd{christian.spethmann@kbfi.ee}
\emailAdd{ville.vaskonen@kcl.ac.uk}
\emailAdd{hardi.veermae@cern.ch}

\abstract{
The abundance of primordial black holes (PBHs) in the mass range $0.1 - 10^3 M_\odot$ can potentially be tested by gravitational wave observations due to the large merger rate of PBH binaries formed in the early universe. To put the estimates of the latter on a firmer footing, we first derive analytical PBH merger rate for  general PBH mass functions while imposing a minimal initial comoving distance between the binary and the PBH nearest to it, in order to pick only initial configurations where the binary would not get disrupted. We then study the formation and evolution of PBH binaries before recombination by performing N-body simulations. We find that the analytical estimate based on the tidally perturbed 2-body system strongly overestimates the present merger rate when PBHs comprise all dark matter, as most initial binaries are disrupted by the surrounding PBHs. This is mostly due to the formation of compact N-body systems at matter-radiation equality. However, if PBHs make up a small fraction of the dark matter, $f_{\rm PBH} \lesssim 10\%$, these estimates become more reliable. In that case, the merger rate observed by LIGO imposes the strongest constraint on the PBH abundance in the mass range $2 - 160 M_\odot$. Finally, we argue that, even if most initial PBH binaries are perturbed, the present BH-BH merger rate of binaries formed in the early universe is larger than $\mathcal{O}(10)\,{\rm Gpc}^{-3} {\rm yr}^{-1}\, f_{\rm PBH}^3$.
}

\maketitle

%-------------------------------------------------------------------------------
\section{Introduction}
%-------------------------------------------------------------------------------

The Advanced Laser Interferometer Gravitational-Wave Observatory (LIGO) observed ten black hole (BH) binary mergers during its first runs~\cite{Abbott:2016blz,Abbott:2016nmj,Abbott:2017vtc,Abbott:2017gyy,Abbott:2017oio,LIGOScientific:2018mvr}. The constituents of these binaries were relatively heavy, with masses in the range $7-50 \Msun$, and it remains unclear whether they have astrophysical or primordial origin~\cite{TheLIGOScientific:2016htt,Belczynski:2016obo,Bird:2016dcv}. Observables that may probe the origin of these BH binaries include the distribution of their eccentricities~\cite{Cholis:2016kqi}, masses~\cite{Kovetz:2016kpi,Kovetz:2017rvv,Raidal:2017mfl,Clesse:2017bsw,Bai:2018shq,LIGOScientific:2018jsj} and spins~\cite{Chiba:2017rvs,Postnov:2017nfw,LIGOScientific:2018jsj}, the redshift dependence of the BH merger rate~\cite{, LIGOScientific:2018jsj} and the correlation between the gravitational wave (GW) sources and galaxies~\cite{Raccanelli:2016cud,Scelfo:2018sny} or dark matter (DM) spikes~\cite{Nishikawa:2017chy}.

A population of primordial black holes (PBHs) may also contribute to the DM abundance, but the possibility that PBHs of any mass make up all of the DM is strongly constrained by a wide range of experimental observations (see e.g. Refs.~\cite{Carr:2016drx,Carr:2017jsz,Sasaki:2018dmp}). However, recent revisions of the femtolensing~\cite{Barnacka:2012bm} and the HSC/Subaru~\cite{Niikura:2017zjd} surveys have opened new windows for PBH DM for light PBHs with masses below $10^{-11} \Msun$~\cite{Katz:2018zrn,Inomata:2017vxo}. Heavier PBH DM, in the mass range $0.1 - 10^3 \Msun$, is constrained by microlensing~\cite{Tisserand:2006zx,Allsman:2000kg}, survival of stars in dwarf galaxies~\cite{Koushiappas:2017chw,Brandt:2016aco}, the distribution of wide binaries~\cite{Monroy-Rodriguez:2014ula} and by modification of the cosmic microwave spectrum due to accreting PBHs~\cite{Ricotti:2007au,Horowitz:2016lib,Ali-Haimoud:2016mbv,Poulin:2017bwe}. Also the recent 21cm observations by EDGES experiment~\cite{Bowman:2018yin} constrain the energy injection from accreting PBHs putting bounds on the abundance of PBHs in this mass window~\cite{Hektor:2018qqw}. 

The strongest potential bounds on the PBH abundance in a mass range around $10\Msun$ can be derived from LIGO observations. If PBHs made up a significant fraction of DM, they would produce a BH binary merger rate and a gravitational wave (GW) background much larger than what is observed by LIGO~\cite{Wang:2016ana,Raidal:2017mfl,Ali-Haimoud:2018dau}. These constraints are, however, subject to large theoretical uncertainties. A reliable estimate of the GW signatures from PBH binary mergers requires a good understanding of both the formation of the PBH binaries and of their subsequent interactions with the surrounding matter that may disrupt the binaries.

In the most common scenario, PBHs are formed when large curvature fluctuations in the early universe directly collapse gravitationally to form BHs~\cite{Hawking:1971ei,Carr:1974nx}. Independent of the formation mechanism, they are thereafter dynamically coupled to cosmic expansion and thus their peculiar velocities are negligible.  PBHs become gravitationally bound to each other roughly when their local density becomes equal to the density of surrounding radiation, which generically happens after matter-radiation equality. However, due to large Poisson fluctuations at small scales, some PBHs can decouple much earlier. When this happens, the closest PBH pairs start falling towards each other and their head-on collision may be prevented only by the torque caused by the gravitational field of the surrounding PBHs and other matter inhomogeneities. As a result a population of PBH binaries is formed.

This mechanism was first described about two decades ago in Ref.~\cite{Nakamura:1997sm}. It was assumed that all PBHs have the same mass, were uniformly distributed in space, all torque was provided by the PBH closest to the binary, and the early binaries were not disrupted between their formation and merger. This model has since been significantly improved~\cite{Ioka:1998nz,Nakama:2016gzw,Sasaki:2016jop,Raidal:2017mfl,Ali-Haimoud:2017rtz,Kavanagh:2018ggo,Chen:2018czv}. In this paper we provide a self contained extension of the merger rate derivation of Ref.~\cite{Ali-Haimoud:2017rtz}, that included the torques from surrounding PBH and matter inhomogeneities, to a general PBH mass function. We also account for the necessary separation between the initial binary and the surrounding PBHs.

The merger rate may be modified by interactions of the binaries with surrounding matter after their formation.  Estimates based on hierarchical 3-body systems show that excluding initial conditions where the pair becomes bound to the nearest PBH cuts the merger rate in half~\cite{Ioka:1998nz}. The required initial distance between the binary and the third PBH was found to be larger than the average distance between the PBHs. In that case, however, it is likely that the 3-body system is also coupled to other surrounding PBHs, so a full $N$-body analysis is needed to determine which binaries are disrupted by the nearest PBH. In Ref.~\cite{Ali-Haimoud:2017rtz} the disruption of PBH binaries after the formation of the first virialised haloes consisting of 10 or more PBHs was estimated, using simple analytic arguments, to have a negligible effect on the merger rate. However, the period between formation of the binary and formation of the first haloes was not considered. In Ref.~\cite{Kavanagh:2018ggo} the interaction with the surrounding CDM was shown to have only a mild effect on the merger rate, because of the high eccentricity of the PBH binaries merging today.

In this paper, we focus on the earliest stages of structure formation before recombination. To check if the binaries are significantly disrupted during this epoch, we perform $70$-body simulations that model the evolution of PBHs for the first 377\,kyr. A large number of surrounding PBHs makes it possible to numerically test the analytical predictions for the statistical distribution of orbital characteristics of the initial binaries. Because most binaries with a close third PBH are disrupted, we find that the distribution of eccentricities is better approximated by a Gaussian distribution than the broken power law found in Ref.~\cite{Ali-Haimoud:2017rtz}. If PBHs comprise most of the DM, then they (including any binaries) will rapidly form small $N$-body systems beginning matter-radiation equality. Due to this effect, the simulations show a much larger disruption rate than predicted in Ref.~\cite{Ali-Haimoud:2017rtz}.  On the other hand, if PBHs make up only a small fraction of DM, bound structures of PBHs form much later and the early disruption rate becomes negligible.

We re-evaluate the LIGO constraints on the PBH abundance from the observed BH-BH merger rate and from non-observation of the stochastic GW background by accounting for the possible suppression of the merger rate due to the interactions of the binary with the surrounding matter. The BH-BH mergers observed by LIGO require a relatively narrow mass function centred around $20 \Msun$ while the merger rate can be reproduced if about $0.2 \%$ of DM consists of PBHs. We stress, that these constraints are tentative, as they neglect the late disruption rate, which is likely significant for a large PBH fraction, and they also ignore the contribution from the population of perturbed initial PBH binaries. We argue, however, that, due to the hardness of initial binaries, the stochastic GW background and the merger rate should be observable even if the initial binary population is almost completely perturbed, which is the case if PBHs make up most of DM.

The paper is organised as follows. The analytical estimates of the PBH merger rate are derived in Sec.~\ref{sec:earlyPBBH}. In Sec.~\ref{sec:simulations} we compare the analytical results to numerical simulations and study the early interactions of PBH binaries with the surrounding PBHs. In Sec.~\ref{sec:phenomenology} we consider the gravitational wave phenomenology of PBH mergers and revise the constraints on the fraction of DM in PBHs. We summarise our main results in Sec.~\ref{sec:conclusions}. 

\vspace{2mm}
\noindent
Geometric units, $G = c = 1$, are used throughout this work.

%-------------------------------------------------------------------------------
\section{Merger rate from early PBH binary formation}
\label{sec:earlyPBBH}
%-------------------------------------------------------------------------------

In this section we derive the PBH merger rate under the assumption that the initially formed binary population is not disturbed. The initial condition consists of two PBHs with masses $m_1$ and $m_2$, proper separation $r$, and a sphere of comoving radius $y$ which contains no other PBHs. The distribution of the initial PBH binaries is set by the distribution of their initial conditions and subsequent emission of GWs. To exclude the possibility of early disruption by nearby PBHs, we will consider a range of possible initial conditions that gives a conservative estimate for the merger rate.

The binary is formed from a pair of close PBH. The surrounding matter contributes with its gravitational potential, which we expand around the centre of mass of the pair, and acts on the pair via a tidal force that generates the angular momentum of the binary. The derivation of the merger rate can be divided into two parts: 
\begin{itemize}
	\item The calculation of the orbital parameters of the binaries for a given initial separation and a known distribution of surrounding BHs and other matter inhomogeneities. From the orbital parameters we can estimate the coalescence time of the binary. This will be discussed in Sec.~\ref{sec:dynamics}.
	\item From the probability distribution of possible initial conditions that will lead to binary formation we can then derive the distribution of the orbital parameters and coalescence times. This will be the focus of Sec.~\ref{sec:distribution}.
\end{itemize}
We compare these analytic estimates for the formation of the initial binaries  with numerical results from $N$-body simulations in Sec.~\ref{sec:simulations}.

%-------------------------------------------------------------------------------
\subsection{Dynamics of early binary formation}
\label{sec:dynamics}
%-------------------------------------------------------------------------------

Consider first the dynamics of a PBH pair in an expanding background surrounded by other PBHs and matter with small inhomogeneities. At that time the Hubble parameter can be expressed as
\be\label{eq:H2}
	H^{2} = \frac{8\pi}{3} \left(\rho_{M} a^{-3} + \rho_{R} a^{-4}\right),
\ee
where $\rho_{\rm M}$ and $\rho_{\rm R}$ denote the comoving energy densities of matter and radiation, respectively, and the scale factor is chosen so that today $a = 1$. The scale factor at matter-radiation equality is $a_{\rm eq} \equiv \rho_{R}/\rho_{M} \approx 1/3400$. 

In the comoving Newtonian gauge and in the absence of anisotropic stress, the line element for spacetimes with small inhomogeneities is~\cite{Ioka:1998nz}
\be
	\td s^2 = - (1 + 2\phi({\bf x})) \td t^2 +  (1 - 2\phi({\bf x})) a(t)^2 \td x^2 \,,
\ee
where the potential $\phi$ is determined by
\be\label{eq:dphi}
	a^{-2}\Delta \phi = 4\pi \rho_{\rm PBH}(\bf x) + 4\pi\bar\rho_{\rm M}\delta_{\rm M}(\bf x) \,.
\ee
Here $\rho_{\rm PBH}$ and $\bar \rho_{\rm M}$ are the densities of PBHs and other types of matter respectively, and  $\delta_{\rm M}$ denotes the density fluctuations of the latter. Treating the PBHs as point masses, we obtain
\be\label{eq:phi}
	\phi({\bf x}) =
	- \sum_i \frac{m_i}{a |{\bf x}_i - {\bf x}|} 
	- 4\pi \rho_{\rm DM} \int \frac{\td^3 k}{(2\pi)^3} \frac{a^2}{k^2} e^{-i{\bf k} \cdot {\bf x} }\tilde{\delta}_{\rm DM}({\bf k}),
\ee
where $m_i$ and ${\bf x}_i$ denote the masses and comoving positions of the PBHs. The motion of test particles is encoded in the action $m\int \td s$. A system of $N$ non-relativistic PBHs, $a \td x/\td t \ll 1$, obeys the action 
\be\label{action_N}
	S^{(N)} 
	 = \int \td t \bigg[ \sum_{i} m_i \left( \frac{1}{2} \dot {\bf r}_i^2 +  \frac{1}{2} \frac{\ddot a}{a} {\bf r}_i^2 - \phi_{\rm ex}({\bf x})\right) + \sum_{i > j}\frac{m_im_j}{|{\bf r}_i - {\bf r}_{j}|} \bigg] \,,
\ee
where ${\bf r}_{i} \equiv a  {\bf x}_{i}$ is the proper distance. The last term accounts for the pairwise interaction of the PBHs and $\phi_{\rm ex}({\bf x})$ is an external potential describing the effect of the surrounding matter on the $N$-body system. The action of a two PBH system can then be approximated as
\bea\label{action_2}
	 S^{(2)}
	 \approx \int \td t \bigg[ &\frac{M}{2} \left(\dot{\bf r}_c^2 + \frac{\ddot a}{a} {\bf r}_c^2\right) - M \phi_{\rm ex}({\bf r}_c) + \frac{\mu}{2} \left(\dot{\bf r}^2 + \frac{\ddot a}{a} {\bf r}^2 + \frac{2M}{r} - {\bf r}\cdot {\bf T} \cdot  {\bf r} \right)
	\bigg] \,,
\eea
where $\bf r$ denotes the separation of the PBH, ${\bf r}_c$ is the centre of mass of the 2-body system, $M \equiv m_1 + m_2$ and $\mu \equiv m_1m_2/M$ are the total and reduced mass of the binary, and  ${\bf T}_{ij} \equiv \partial_i\partial_j\phi({\bf r}_c)$ results from the expansion of the external potential around the centre of mass of the system. %\footnote{This expansion reads: $m_1\phi({\bf r_c} + {\bf r} \mu/m_1) + m_2\phi({\bf r_c} - {\bf r} \mu/m_2) = M \phi({\bf r_c}) + {\bf r} \cdot{\bf T}\cdot{\bf r}/2 + \mathcal{O}(r^3).$
%\bea
%		m_1\phi({\bf r_c} + {\bf r} \mu/m_1) + m_2\phi({\bf r_c} - {\bf r} \mu/m_2) 
%	=	\phi_c + {\bf F} {\bf r_c} \\
%	+	\frac{m_1}{2} ({\bf r_c} + {\bf r} \mu/m_1){\bf T} ({\bf r_c} + {\bf r} \mu/m_1)
%	+	\frac{m_2}{2} ({\bf r_c} - {\bf r} \mu/m_2){\bf T} ({\bf r_c} - {\bf r} \mu/m_2)
%	=	\phi({\bf r_c}) + \frac{1}{2} {\bf r} \cdot ({\bf T}\cdot{\bf r}) + \mathcal{O}(r^3).
%\eea}
We assume that the centre of mass is stationary, so the time dependence of ${\bf T}$ can be estimated from the expansion only. The forces acting on the pair are summarised as follows
\be\label{eq:forces}
	{\bf F}/\mu
	= \underbrace{{\bf r} \ddot a/a}_{\mbox{Hubble flow}}
	-  \underbrace{M \hat {\bf r}/r^2}_{\mbox{self-gravity}}
	+ \underbrace{(\hat {\bf r} \cdot{\bf T}\cdot{\bf r}) \hat {\bf r}}_{\mbox{radial tidal forces}}
	+ \underbrace{({\bf r} \times ({\bf T}\cdot{\bf r}))}_{\mbox{tidal torque}} \times (\hat {\bf r}/r) \,,
\ee
where $\hat{\bf r} \equiv {\bf r}/r$ is the unit vector parallel to ${\bf r}$. The first three forces are radial (i.e. parallel to $\hat{\bf r}$), and the last term provides the torque that prevents the head on collision of the two PBHs. For the binary to form, the tidal forces are required to be much weaker than the gravitational attraction of the PBH pair, that is, the two bodies must be more tightly coupled to each other than to any other PBH. 

Decoupling from the expansion takes place when the second term in Eq.~\eqref{eq:forces} starts to dominate over the first one. We define
\be
	\delta_{\rm b} \equiv	\frac{M/2}{\rho_{\rm M} V(x_0)} \,,
\ee
where $V(x) \equiv (4 \pi/3)x^3$ denotes the comoving volume of comoving radius $x$. The quantity $\delta_{\rm b} - 1$ can be interpreted as the effective matter overdensity generated by a PBH pair of a total mass $M$ and a comoving separation $x_0$. We are interested in initial PBH pairs with $\delta_{\rm b} \gg 1$, as they produce tightly bound PBH binaries that may merge within the age of the universe. In the radiation dominated epoch such perturbations collapse roughly when $\rho_{\rm R}a^{-4} \approx \delta_{\rm b} \rho_{\rm M}a^{-3}$. So, we define the scale
\be\label{eq:a_dc}
	a_{\rm dc} \equiv a_{\rm eq}/\delta_{\rm b} \,,
\ee
as an approximate estimate of decoupling.

After decoupling, the separation of the PBH pair will stop growing and the tidal forces will be damped fast, ${\bf T} \propto a^{-3}$, as the universe expands. This is implied from Eq.~\eqref{eq:phi} because, in the first approximation, during radiation domination the surrounding PBHs are assumed to follow the Hubble flow and the matter density perturbations are constant. This means that the potential \eqref{eq:phi} scales roughly as $\phi \propto a^{-1}$, which translates to ${\bf T} \propto a^{-3}$, as claimed.

The angular momentum ${\bf L}$ of the two body system vanishes initially. It is generated by the tidal torque,
\be\label{eq:2body_L}
	{\bf L} %= \mu \int \td  t\, {\bf r} \times {\bf F}  
	= \mu \int \td t\,  {\bf r} \times ({\bf T}\cdot{\bf r}) \,.
\ee
It is more convenient to work with the dimensionless angular momentum, 
\be\label{eq:2body_j}
	{\bf j} \equiv \frac{{\bf L}/\mu}{\sqrt{r_a M}} \,,
\ee
where $r_a$ is the semi-major axis of the PBH binary. The orbital eccentricity $e$ of the binary is given by $e = \sqrt{1 - j^2}$, and the coalescence time for eccentric orbits, $j \ll 1$, by~\cite{Peters:1964zz}
\bea\label{eq:tau}
	\tau = \frac{3}{85} \frac{r_a^4}{\eta M^3} j^7 \,,
\eea
where $\eta \equiv \mu/M$ is the symmetric mass fraction. For non-eccentric orbits Eq.~\eqref{eq:tau} may overestimate the coalescence time by at most a factor of $1.85$~\cite{Peters:1964zz}.

Following Ref.~\cite{Ali-Haimoud:2017rtz} we estimate the effect of tidal torque perturbatively assuming that the decoupling from expansion takes place in the radiation dominated epoch and that the orbit of the binary remains eccentric, $j \ll 1$. The evolution of ${\bf j}$ can be evaluated perturbatively by first solving for purely radial motion using $\ddot{r} - r \ddot a/a + Mr^{-2} = 0$ and then plugging the solution into Eq.~\eqref{eq:2body_L}. As the relevant binaries are formed in the radiation dominated epoch, we may ignore the contribution of the matter density $\rho_{M}$ to the Hubble parameter \eqref{eq:H2} in the first approximation. The equation of motion for the comoving separation $x \equiv  r/a$ is then simply
\bea\label{eom:x_rad}
	x'' + \frac{a}{a_{\rm dc}} \frac{x_0^3}{x^2} = 0 \,,
\eea
where the prime denotes derivation with respect to $\ln(a)$ and the initial conditions are given by $x(a_0) = x_0$, $x'(a_0) = 0$. We must assume that $a_0 \ll a_{\rm dc}$. Eq.~\eqref{eom:x_rad} is solved by $x(a) \approx x_0 \chi(a/a_{\rm dc})$, where $\chi(y)$ is determined by $(y\partial_y)^2\chi+y\chi^{-2} = 0$ and the boundary conditions $\chi(y\to 0)=1$, $\chi'(y\to 0)=0$. The function $\chi(y)$ is positive and oscillates with an amplitude that decreases asymptotically as $0.2/y$. So, the semi-major axis of the fully decoupled binary is 
\be\label{eq:r_a}
	r_a = a x_a/2  
	\approx 0.1 a_{\rm dc} x_0 \,.
\ee
The first root of $\chi(y)$, indicating the first close encounter of the pair, lies at $y \approx 0.54$, which translates to $a \approx 0.54 a_{\rm dc}$, so the binary must decouple even earlier than our first estimate $a \approx a_{\rm dc}$ suggested.  Although Eq.~\eqref{eom:x_rad} ignores the contribution of matter to cosmic expansion, it yields a relatively good approximation even when $a_{\rm dc} \approx a_{\rm eq}$. Plugging the solution of Eq.~\eqref{eom:x_rad} into Eq.~\eqref{eq:2body_L}, and, again, neglecting the contribution of matter to the expansion, gives
\bea\label{eq:j}
	{\bf j} 
&	= \frac{1}{\sqrt{r_{a} M}}\int \frac{\td a}{a^{2}H}\, {\bf x} \times ({\bf T}a^{3} \cdot {\bf x})
%&	= 0.3 \frac{a_{\rm dc} x_{0}^{2}}{\sqrt{\frac{8\pi}{3} \rho_{R} r_a M}}\hat{\bf x}_{0} \times ({\bf T}a^{3}\cdot \hat{\bf x}_{0}) \,    \\	
	= \frac{0.95 \, x_0^3}{M} \, \hat {\bf r} \times ({\bf T}a^3\cdot \hat {\bf r}) \,,
\eea
where we used $\int^{\infty}_{0} \td y\,  \chi(y)^2 \approx 0.3$. Note that ${\bf T}a^3$ is constant under the assumptions made above.  Because ${\bf T}$ is rapidly damped by the expansion, the binary acquires about 90\% of its angular momentum during the first period.\footnote{The first period corresponds to the first root of  $\chi(y)$ at $y \approx 0.54$, during which the angular momentum generated is proportional to $\int^{0.54}_{0} \td y\,  \chi(y)^2 \approx 0.26$. The total angular momentum corresponds to $\int^{\infty}_{0} \td y\,  \chi(y)^2 \approx 0.3$.}

The coefficients in Eqs.~\eqref{eq:r_a} and~\eqref{eq:j} can be determined from the numerical simulations that we discuss in Sec.~\ref{sec:simulations}. So, to account for deviations from the simplified case, we will not fix them and define the coefficients $c_a$ and $c_j$ instead,
\be\label{eq:gen_rj}
	r_a 
%	= c_a a_{\rm dc} x_0\,,  \qquad
	=  c_a \frac{8 \pi \rho_R}{3}   \frac{\, x_0^4}{M} \,,  \qquad
	{\bf j} = c_j  \frac{\, x_0^3}{M} \, \hat {\bf r} \times ({\bf T}a^3\cdot \hat {\bf r}) \,.
\ee
For the above simplified scenario we have $c_a=0.1$ and $c_j=0.95$. 

The orbit of the binary is now explicitly determined by the PBH masses $m_1$ and $m_2$, the initial comoving separation $x_0$ and the tidal torque ${\bf T}$, and the coalescence time \eqref{eq:tau} of the initial binary can be written as
\be \label{eq:coalescence_time}
	\tau = \frac{4096 \pi^4 c_a^4 c_j^7 \rho_R^{4}}{2295} \, \frac{x_0^{37}}{\eta M^{14}} |\hat {\bf r} \times ({\bf T}a^3\cdot \hat {\bf r})|^7.
\ee
Of course, this prediction holds only when the orbit is not disturbed between the formation and merger of the binary. Note that the surrounding matter enters these expressions through the radiation density, which, by assumption, determines the Hubble parameter when the binary is formed, and the tidal torque, that implicitly accounts for the surrounding PBH and matter fluctuations. Both $x_0$ and ${\bf T}$ should be thought of as random variables that depend on the statistical properties of the initial PBH population. Next we will discuss how the distribution of masses and positions of the PBHs will determine the distribution of the orbital parameters derived above.

%-------------------------------------------------------------------------------
\subsection{Distribution of initial binaries}
\label{sec:distribution}
%-------------------------------------------------------------------------------

\begin{figure}
\begin{center}
\includegraphics[height=0.45\textwidth]{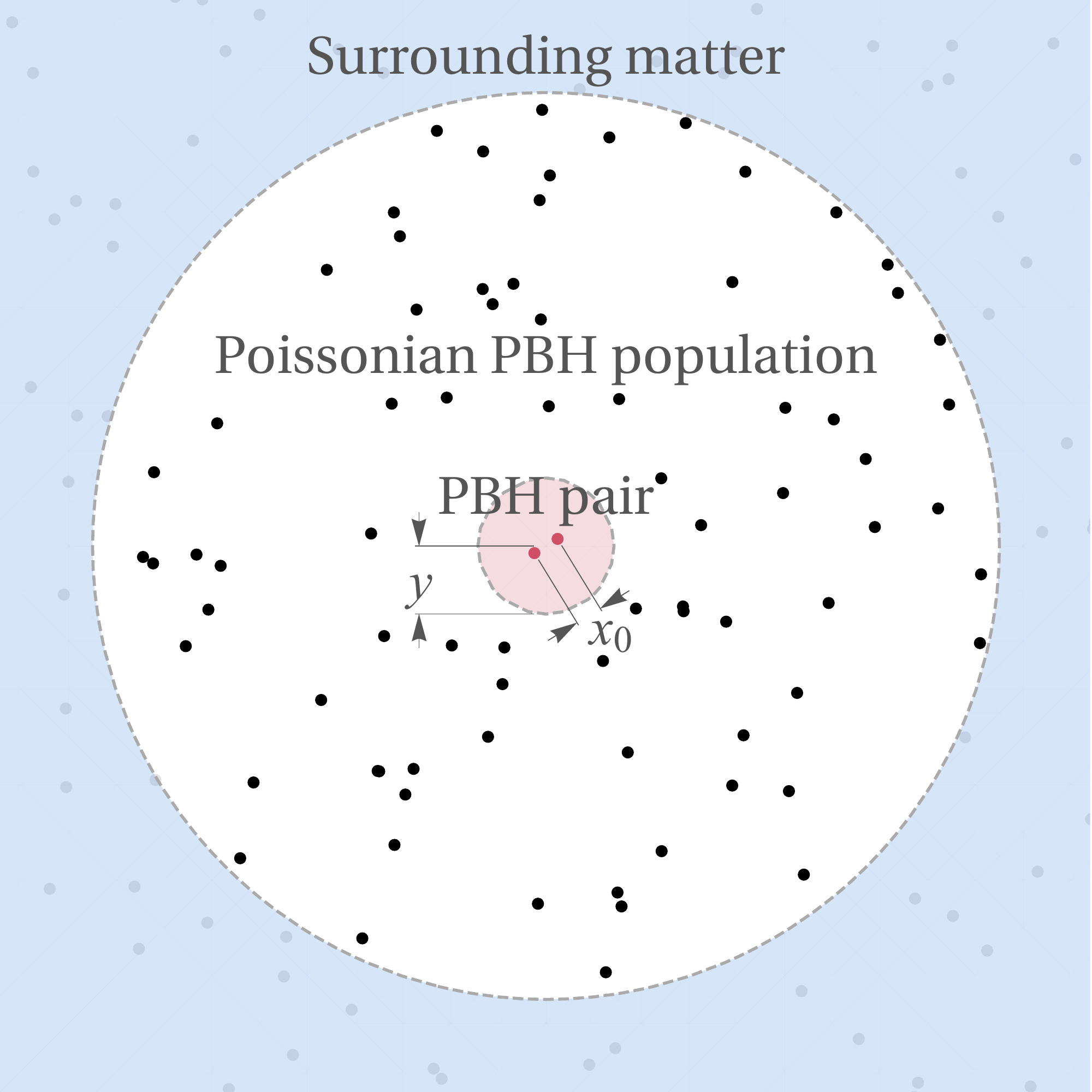}
\caption{Schematic description of the initial configuration for the simulation. The exterior region (blue) contains surrounding matter that has a uniform density and evolves only due to the expansion of the universe. A spherical region (white) contains a randomly distributed PBH population. The interior region (red) contains only the binary that is inserted so that, by using Eq.~\eqref{eq:coalescence_time}, its coalescence time matches the current age of the universe. A similar set-up applies for the analytic estimate for binary formation in Sec.~\ref{sec:dynamics}, but in that case the white region extends to infinity and all PBH in this region within the timescale of formation of the binary are assumed to evolve only due to cosmic expansion.}
\label{fig:simsetup}
\end{center}
\end{figure}

Having detailed the properties of the PBH binary given the initial condition with known separation $x_{0}$, masses $m_1$, $m_2$ and ${\bf T}$, we now turn to discuss the distribution of initial conditions which will eventually determine the distribution of $j$ and the merger rate of the PBH binaries.

Consider a PBH pair with masses $m_1$, $m_2$ at a comoving separation $x_{0}$ so that they are the only PBHs in spherical volume of comoving radius $y$. This set-up is shown in the interior region of Fig.~\ref{fig:simsetup}. The reason for forbidding surrounding PBHs closer than $y$ is to exclude initial configurations where the binary gets disrupted by  surrounding PBHs shortly after formation. In such cases the perturbative estimate of the coalescence time will inevitably fail. For the sake of generality, we will not fix the value of $y$ in the general discussion. The aim of the following is to estimate the density of viable initial conditions and, from it, the distribution of coalescence times and the merger rate. The spatial PBH distribution is assumed to be Poisson throughout the paper.\footnote{Based on general arguments, the spatial distribution at small scales has been shown to be well approximated by the Poisson distribution~\cite{Ali-Haimoud:2018dau}. It has been argued, however, that accounting for the two-point function of PBHs, $\xi_{\rm PBH}$, may affect the merger rate for wider mass functions~\cite{Ballesteros:2018swv,Desjacques:2018wuu,Raidal:2017mfl}. In Ref.~\cite{Desjacques:2018wuu} this effect was estimated to be irrelevant for PBHs in the LIGO mass range. In addition, a rough estimate yields that for $\xi_{\rm PBH} \leqslant 1$ the contribution of the PBH two-point function to the merger rate is generally subleading to the direct contribution of the width of the mass function~\cite{Raidal:2017mfl}. Only the latter will thus be considered in this paper. Formation of initially clustered PBH distributions, enabled by some more exotic PBH formation mechanisms, and the evolution of such clusters has been considered in Ref.~\cite{Belotsky:2018wph}. 
}
In this case the comoving number density of configurations producing a binary is
\bea\label{n_b}
	\td n_{\rm b}
&	=	\frac{1}{2} e^{-\bar{N}(y)} \td n(m_1) \td n(m_2) \td V(x_{0}) \,,
%&	= 	\frac{1}{2} \rho_{\rm DM}^2  e^{-\bar{N}(y)} 4\pi x^{2} \td x \, \, \psi(m_1) \frac{\td m_1}{m_1}  \, \psi(m_2) \frac{\td m_2}{m_2} 
\eea
where $\td n(m)$ is the comoving number density of PBH in the mass range $(m, m+ \td m)$, $\rho_{\rm DM}$ denotes the present DM energy density, $\bar{N}(y) \equiv n V(y)$ is the expected number of PBH in a spherical volume with radius $y$, and $n = \int \td n(m)$ is the PBH number density. The factor $1/2$ avoids overcounting.

The differential merger rate per unit time and comoving volume is then given by\footnote{In Ref.~\cite{Bringmann:2018mxj} it was argued, that the PBH binary merger rate is enhanced by a cascade of mergers in the early Universe. However, their application of the early binary formation mechanism beyond the first merger step is questionable as the PBHs may have peculiar velocities.}
\bea\label{dR_early}
	\td R
&	= \int \td n_{\rm b} \td j \frac{\td P}{\td j} \delta\left(\tau - \frac{3}{85} \frac{r_a^4}{\eta M^3} j^7\right)  \\
%&	= \int \td n_{\rm b} \td \tau \frac{\td j}{\td \tau} \frac{\td P (j(\tau))}{\td j} \\
&	= \frac{1}{14 \tau} \td n(m_1) \td n(m_2) \int \td V(x_{0}) e^{-\bar{N}(y)}  j \frac{\td P (j | x_0,y)}{\td j} \bigg|_{j = j(\tau)} \,,
\eea
where $j(\tau)$ is obtained from Eq.~\eqref{eq:tau} and $\td P/\td j$ is the distribution of the dimensionless angular momentum for a given $y$ and $x_0$. We derive $\td P/\td j$ in the following section. The choice of $\bar{N}(y)$ should guarantee that the pair is initially a 2-body system and thus not gravitationally coupled to a third PBH. Numerical simulations presented in Sec.~\ref{sec:simulations} indicate that initial configurations with a third PBH in the surrounding volume $V(y) \lesssim M/\rho_{\rm PBH}$ will produce binaries that are likely to be disrupted and will thus not contribute to the present merger rate.

%-------------------------------------------------------------------------------
\subsubsection{Angular momenta}
\label{sec:jdistribution}
%-------------------------------------------------------------------------------

The distribution of eccentricities or, equivalently, the dimensionless angular momenta $j$, gets contributions from surrounding PBH and matter fluctuations. The dimensionless angular momentum from the surrounding PBHs, for which the tidal tensor reads ${\bf T}_{\rm PBH}a^3 =  \sum_{i} m_i ({\bf 1} - 3\hat {\bf x}_i \otimes \hat {\bf x}_i )/x_i^3 \,$, is given by the sum, ${\bf  j}_{\rm PBH} \equiv \sum_i {\bf  j}_1({\bf x}_i, m_i)$, of contributions from individual PBHs (see Eq.~\eqref{eq:j})
\be\label{j1}
	{\bf  j}_1({\bf x}_i, m_i) 
%	=    j_0 \frac{3m_i}{\rho_{\rm PBH} V(x_i)}  \hat{\bf x}_i \times  \hat {\bf r}  \,(\hat {\bf x}_i \cdot \hat {\bf r}) \,,
	=    j_0 \frac{m_i}{\langle m \rangle} \frac{3}{\bar{N}(x_i)}  \hat{\bf x}_i \times  \hat {\bf r}  \,(\hat {\bf x}_i \cdot \hat {\bf r}) \,,
\ee
where $\langle m \rangle = \rho_{\rm PBH}/n$ is the average PBH mass\footnote{In general, the average over masses is defined as $\langle X \rangle \equiv n^{-1}\int X \td n$.} and we defined
\bea\label{def:j0}
	j_0 \equiv  c_j \, \bar{N}(x_{0}) \langle m \rangle/M \approx  0.4 f_{\rm PBH}/\delta_{b} \,,
\eea
where we used $\Omega_{\rm M}/\Omega_{\rm DM} = 1.2$. The quantity $j_0$ provides an order of magnitude estimate of the average dimensionless angular momentum, since, as most of the torque is likely generated by the PBH closest to the binary, for which $\bar{N}(x_{i}) \approx 1$ and $m_{i} \approx  \langle m \rangle$ on average,  so the other terms in Eq.~\eqref{j1} will on average contribute an order one factor. Binary formation requires $\delta_{b} \gg 1$, thus we must assume $j_0 \ll 1$. As we will see, this will be true for coalescence times less than the age of the universe.

We assume that the masses $m_i$ and positions ${\bf x}_i$ of the surrounding PBH and the matter density perturbations are statistically independent. The angular momentum, ${\bf  j} = {\bf  j}_{\rm PBH} + {\bf  j}_{\rm M}$, is then composed of a sum of independent variables, so its distribution is most conveniently estimated using the cumulant generating function $K({\bf  k}) \equiv \ln \left\langle e^{i {\bf k}\cdot {\bf  j}} \right\rangle$,
\bea\label{jdistr1}
	\frac{\td P }{\td^3 j}
	\equiv \langle \delta({\bf  j} - {\bf  j}_{\rm M} - {\bf j}_{\rm PBH} ) \rangle
	= \int  \frac{\td^3 k}{(2\pi)^3} e^{-i {\bf  k}\cdot{\bf  j} + K({\bf k})} \,,
%	= \lim_{V\to \infty}\int \delta({\bf  j} - \sum_i {\bf  j}_1 ({\bf x}_i, m_i) )  \prod^{nV}_{i=1} \frac{\td^3 x_i}{V} \frac{\td n(m_i)}{n}, \\	
\eea
because of the additive property of cumulants.

We start by considering the first two cumulants of ${\bf  j}$. Isotropy implies that 
\be
	\langle{\bf  j}\rangle = 0, \qquad
	\langle{\bf  j}\otimes{\bf  j}\rangle = \frac{1}{2} \sigma_j^2 ({\bf 1} - \hat {\bf  r} \otimes\hat {\bf r}),
\ee
where $\sigma_j^2 \equiv \langle {\bf  j}^2\rangle $ and the $\hat {\bf  r} \otimes\hat {\bf r}$ term arises because ${\bf  j} \perp \hat {\bf r}$ by construction \eqref{eq:gen_rj}. In general, the last feature implies that $K$ is a function of ${\bf  k}_{\perp} \equiv {\bf  k}\times \hat {\bf r}$, while isotropy further constrains it to be a function of $k_{\perp}$. Statistical independence implies that $\sigma_j^2 = \sigma_{j,\rm PBH}^2 + \sigma_{j,\rm M}^2$. The variance due to matter perturbations is obtained by applying Eq.~\eqref{eq:gen_rj}, averaging over orientations of $\hat {\bf r}$ and using that, by Eq.~\eqref{eq:dphi}, in Fourier space $a^3{\bf T} = \hat {\bf q} \otimes \hat {\bf q}  \, 4\pi \bar\rho_{\rm M} \delta_{\rm M} (q)$,\footnote{We assumed that $\bar{\rho}_M = \rho_{\rm M}$. This introduces a minor error as the contribution of $\sigma_{j,\rm M}$ becomes insignificant for $f_{\rm PBH} = 1$, while the distinction is negligible for $f_{\rm PBH} \ll 1$.}
\bea\label{sigma_j_M}
	\sigma_{j,\rm M}^2 
&	= \left\langle {\bf  j}_{\rm M}^{2}\right\rangle
	= \left(\frac{c_{j} x_0^3}{M}\right)^2 \frac{a^6}{5}\left\langle \tr({\bf T}\cdot {\bf T}) - \frac{1}{3}\tr({\bf T})^2\right\rangle \\
%	= \frac{6}{5} \left( c_{j}\frac{4\pi x_0^3}{3} \frac{\bar{\rho}_{\rm M}}{M}\right)^2 \left\langle \delta_{\rm M}^2 \right\rangle
&	= \frac{6}{5} j_0^2 \frac{\sigma_{\rm M}^2}{f_{\rm PBH}^2} ,
\eea
where $\sigma_{\rm M}^2 \equiv (\Omega_{\rm M}/\Omega_{\rm DM})^2 \left\langle \delta_{\rm M}^2 \right\rangle$ is the rescaled variance of matter density perturbations at the time the binary is formed and $f_{\rm PBH} \equiv \rho_{\rm PBH}/\rho_{\rm DM}$ is the fraction of PBHs. Following Ref.~\cite{Ali-Haimoud:2017rtz} we will use the value $\left\langle \delta_{\rm M}^2 \right\rangle = 0.005$ for numerical estimates. We stress, however, that differences from this estimate can arise when PBH formation is accompanied by enhanced density perturbations at small scales, as is often the case. 

A similar calculation yields, for the variance due to surrounding PBHs,
\bea\label{sigma_j_PBH}
	\sigma_{j,\rm PBH}^2 
&	= \lim_{\substack{V \to \infty \\ N/V = n}} N\left\langle {\bf  j}_{1}^{2}\right\rangle
	=  \frac{6}{5} j_0^2 \frac{\langle m^2 \rangle}{\langle m \rangle^2}\lim_{\substack{V \to \infty \\ N/V = n}} N \int^{V}_{V(y)} \frac{\td V(x)}{V} \frac{1}{\bar{N}(x)^2}   \\
&	= \frac{6}{5}  j_0^2  \frac{1 + \sigma_{m}^2/\langle m \rangle^2}{\bar{N}(y)} \,,
\eea
where $\sigma_{m} \equiv \sqrt{\langle m^2 \rangle - \langle m \rangle^2}$ is the width of the mass distribution. The new element in this computation is the average over positions of the surrounding PBHs. We first estimated it for a finite volume $V$ and then took the limit $V\to \infty$ by keeping the PBH number density, $n = N/V$, fixed. The contributions from each PBH are statistically independent and identical, so $\sigma_{j,{\rm PBH}}^2$ is just the contribution from a single PBH times the number of surrounding PBHs. The integral over positions has a lower bound because we exclude initial conditions where PBHs can be closer than $y$ to the centre of mass of the binary. 

In conclusion, the variance of ${\bf  j}$ is
\be\label{sigma_j}
	\sigma_{j}^2 
	= \sigma_{j,\rm M}^2  + \sigma_{j,\rm PBH}^2
	= \frac{6}{5}  j_0^2  \left(\frac{1 + \sigma_{m}^2/\langle m \rangle^2}{\bar{N}(y)} + \frac{\sigma_{\rm M}^2}{f_{\rm PBH}^2} \right).
\ee
The limit $\bar{N}(y)\to 0$, where the variance diverges, is never realised as, in order not to disrupt the binary, the distance to the closest PBH must be larger than the separation of the binary, that is $y > x_0$. Therefore, since $j_0 \approx \bar{N}(x_0) \ll 1$ we have $\sigma_{j}^2 \lesssim j_0$.
A comparison of Eqs.~\eqref{sigma_j_M} and \eqref{sigma_j_PBH} indicates that the variance due to the surrounding PBHs may be attributed to Gaussian matter perturbations with variance $(1 + \sigma_{m}^2/\langle m \rangle^2) f_{\rm PBH}^2 /\bar{N}(y)$. As $\bar{N}(y)$ is the expected number of PBHs in the empty volume around the binary, this result is not surprising for a Poisson distribution of PBHs. Moreover, when $\bar{N}(y) \gg 1$, it is expected that the distribution of ${\bf  j}$ is Gaussian with a width \eqref{sigma_j}. This is indeed so, as we will confirm shortly.

The cumulant generating function of ${\bf  j}$ decomposes as $K= K_{\rm PBH} + K_{\rm M}$. We will next calculate these separately. As matter density fluctuations are Gaussian only the first two cumulants contribute, thus $K_{\rm M}$ is simply
\be
	K_{\rm M}({\bf k}) 
	= -\frac{1}{2}\left\langle ({\bf k}\cdot {\bf j})^2 \right\rangle
	= - \frac{1}{4} \sigma_{j,\rm M}^2 k_{\perp}^2\,.
\ee
To compute $K_{\rm PBH}$ we proceed in a similar manner as in the case of $\sigma_{j,\rm PBH}^2$ -- starting with a finite volume and then taking the limit $V\to \infty$ by keeping $N/V$ fixed. This gives
\bea
	K_{\rm PBH}({\bf  k}) 
&	= \lim_{N\to \infty} N \ln \left\langle e^{i {\bf k}\cdot {\bf  j}_{1}({\bf x}, m)} \right\rangle
%	= \lim_{\substack{V \to \infty \\ N/V = n}} N \ln \left(\int \frac{\td^3 x}{V} \frac{\td n(m)}{n} e^{i {\bf k}\cdot {\bf  j}_{1}({\bf x}, m)}\right) 
	= \int \td^3 x \, \td n(m) \left(e^{i{\bf k}\cdot {\bf  j}_1({\bf x}, m)} - 1\right) \,.
\eea
After plugging in Eq.~\eqref{j1} and averaging over PBH positions in the region $|{\bf x}|>y$ we get\footnote{Defining $z \equiv k_{\perp} j_0 m/(\rho_{\rm PBH} V(y))$ and changing the integration variable to $u \equiv z V(y)/V(x)$, the spatial integration can be performed as follows 
\bea \nonumber
&	-z \int\frac{\td^2\Omega}{4\pi}\, \int^{z}_{0} \frac{\td u}{u^2}  \left[\exp\left(i 3 u (\hat{\bf  k}_{\perp} \cdot \hat{\bf x}) \,(\hat {\bf x} \cdot \hat {\bf r}) \right) - 1\right] \\
&	= -z \int^{z}_{0} \frac{\td u}{u^2}  \int^{\pi}_{0} \int^{2\pi}_{0} \frac{\td\cos(\theta) \, \td \phi }{4 \pi} \,   \left[\exp\left(i \frac{3u}{2} \sin^2(\theta)\sin(2\phi) \right) - 1\right] \\
%&	= -z \int^{z}_{0} \frac{\td u}{u^2}  \int^{1}_{0} \frac{\td v}{2\sqrt{1-v}}   \left(J_0 \left(\frac{3u v}{2}  \right) - 1\right) ,   \\
&	= -z \int^{z}_{0} \frac{\td u}{u^2}   \left[\frac{\pi}{2\sqrt{2}} J_{-\frac{1}{4}} \left(\frac{3u}{4}  \right) J_{\frac{1}{4}} \left(\frac{3u}{4}  \right) - 1\right]
	= {}_1F{}_2\left(-\frac{1}{2};\frac{3}{4},\frac{5}{4};-\frac{9z^2}{16}\right) - 1.
\eea
}
\bea \label{KPBH}
	K_{\rm PBH}({\bf  k}) 
%&	= \int \td n(m) \, \int_{|{\bf x}|>y} \td^3 x\left(\exp\left(i c \frac{m ({\bf  k}\cdot \hat {\bf r} \times \hat{\bf x}) \,(\hat {\bf x} \cdot \hat {\bf r})}{|{\bf x}|^3} \right) - 1\right)  \\
&	= \int \td n(m) \, \int_{|{\bf x}|>y}\frac{\td\Omega}{4\pi}\, \td V(x) \left[\exp\left(\frac{i3mj_0}{\rho_{\rm PBH} V(x)} ({\bf  k}_{\perp} \cdot \hat{\bf x}) \,(\hat {\bf x} \cdot \hat {\bf r}) \right) - 1\right] \\
%&	= -V(y)\int \td n(m) \,F\left( \frac{m}{\rho_{\rm PBH}V(y)} j_0 k_{\perp} \right).  \\	
&	= -\bar{N}(y)\int \frac{\td n(m)}{n} \,F\left( \frac{m}{\langle m \rangle} \frac{1}{\bar{N}(y)} j_0 k_{\perp} \right) \,,	
\eea
where $F(z) = {}_1F{}_2\left(-1/2;3/4,5/4;-9z^2/16 \right) - 1$ and ${}_1F{}_2$ is the generalised hypergeometric function. The function $F$ has the following properties
\bea\label{eq:Fasym}
	z - 1 \leq &F(z) \leq 3z^2/10, \quad F(z) \geq 0 \qquad &\mbox{when}& \quad z \geq 0, \\
	F(z) &\sim z - 1, \qquad &\mbox{when}& \quad z \to\infty, \\
	F(z) &\sim 3z^2/10, \qquad &\mbox{when}& \quad z \to 0, \\
\eea
which we list for later convenience. As a consistency check we find that the first two cumulants obtained from Eq.~\eqref{KPBH} match the earlier direct computation, i.e. $\left. -i\partial_{{\bf k}} K_{\rm PBH}\right|_{{\bf k} = 0} = 0$,  $\left.(-i\partial_{{\bf k}})^2 K_{ \rm PBH}\right|_{{\bf k} = 0} = \sigma_j^2$.

Both $K_{\rm PBH}$ and $K_{\rm M}$ depend only on $k_\perp$, so, as expected for an isotropic distribution of matter, it is sufficient to consider the distribution of $j$. From Eq.~\eqref{jdistr1} we then obtain
\bea \label{eq:Pj}
	j \frac{\td P}{\td j}  
%&	= 2\pi j \int \frac{k_{\perp}\td k_{\perp}\td \phi}{(2\pi)^2} e^{-i k_{\perp} j \cos(\phi) - K(k_\perp)}  \\
&	= \int^{\infty}_{0}  \td u \,u  J_{0}(u ) \exp\left[ {-\bar{N}(y) \int \frac{\td n(m)}{n} \,F\left(u \frac{m}{\langle m \rangle} \frac{1}{\bar{N}(y)} \frac{j_0}{j}\right) - u^{2} \frac{3}{10}\frac{\sigma_{\rm M}^{2}}{f_{\rm PBH}^{2}}\frac{j_{0}^{2}}{j^{2}}} \right]\,.
%&	= j \int^{\infty}_{0}  \td k_{\perp} \,k_{\perp}  J_{0}\left(j k_{\perp}\right) e^{-K_{{\bf j}}(k_{\perp})}\,.
\eea
This distribution peaks at $j \lesssim j_0$. Since for PBH binaries merging today $j_0 \ll1$, this result is consistent with the assumption of eccentric orbits. 

\begin{figure}
\begin{center}
\includegraphics[height=0.32\textwidth]{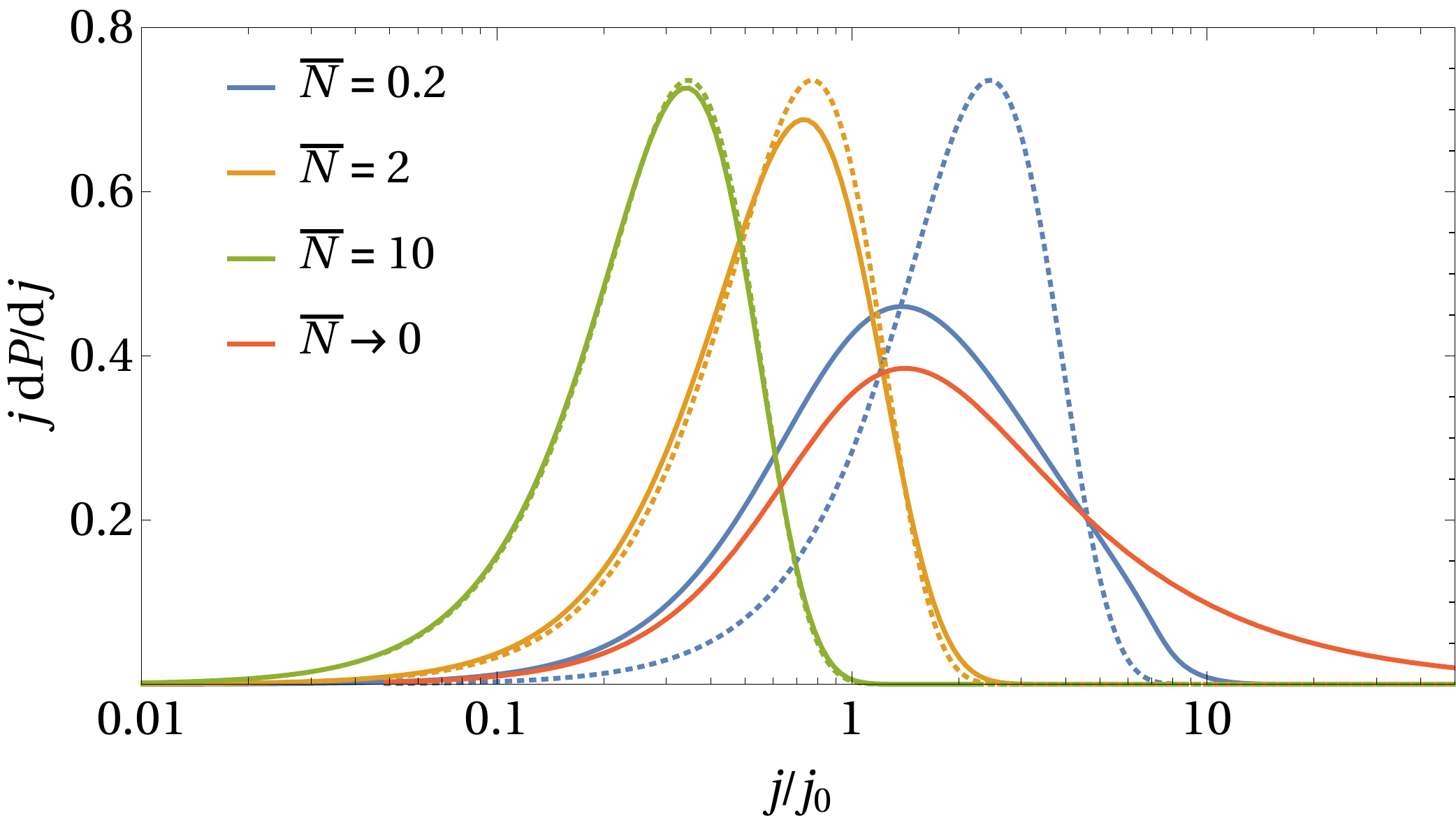}
\caption{The distribution of the logarithm of the dimensionless angular momentum, $j \td P/\td j$, for monochromatic mass functions and $f_{\rm PBH} = 1$. The exact distribution ~\eqref{eq:Pj} evaluated numerically for $\bar{N}(y) = 0.2, 2, 10$ is shown by the blue, yellow and green dotted line, respectively, while the red line corresponds to the $\bar{N}(y) \to 0$ limit~\eqref{eq:Pj_0}. The blue, yellow and green dotted line show the Gaussian approximation~\eqref{eq:Pj_inf} for $\bar{N}(y) = 0.2, 2, 10$, respectively.}
\label{fig:Pj}
\end{center}
\end{figure}

The logarithmic distribution $j\td P/\td j$ depends on $j$ only through the ratio $j/j_0$, so the characteristics of the binary, which affect only $j_0$, do not change the shape of the distribution but only shift it on the logarithmic scale. In particular, for monochromatic mass functions, $\td n(m) = n \delta(m - m_c) \td m$, the shape of this distribution does not explicitly depend on the PBH mass. %In this case the distribution~\eqref{eq:Pj} reduces to
%\bea \label{eq:Pj_mon}
%	j \frac{\td P}{\td j}  
%%&	= j^{-1} \int  \td u \,u J_{0}(u) \exp(-nV(y) F(u m_c c  y^{-3} j^{-1}) 
%&	= \int^{\infty}_{0} \td u \,u J_{0}(u) \exp\left(-\bar{N}(y) \, F\left(u\frac{1}{\bar{N}(y)} \frac{j_0}{j}\right) -u^{2} \frac{3\sigma_{\rm M}^{2}}{10f_{\rm PBH}^{2}}\frac{j_{0}^{2}}{j^{2}} \right) .
%\eea
The distribution \eqref{eq:Pj} for a monochromatic mass function is shown in Fig.~\ref{fig:Pj} for $f_{\rm PBH} = 1$ by the solid lines for different values of $\bar{N}(y)$.

In the limit $\bar{N}(y) \to 0$, $\sigma_{M} \ll f_{\rm PBH}$ we obtain a power law distribution with a break at $j_0$,
\bea\label{eq:Pj_0}
	j\frac{\td P}{\td j}
%&	=  j^{-1}  \int^{\infty}_{0}  \td u \,u  J_{0}(u ) \exp\left(-\,u \,j_0/j\right) 
	= \frac{ j^2/j_0^2}{(1+ j^2/j_0^2)^{3/2}} \,.
\eea
This limiting case matches the result of Refs.~\cite{Ali-Haimoud:2017rtz,Chen:2018czv}. Interestingly, this result does not depend on the PBH mass function as it drops out of the integral in Eq.~\eqref{eq:Pj}. Note, however, that the mass of the binary enters implicitly through $j_0$.  As shown in Fig.~\ref{fig:Pj}, Eq.~\eqref{eq:Pj_0} approximates the distribution~\eqref{eq:Pj} relatively well for $\bar{N}(y) \lesssim 0.2$ if $f_{\rm PBH} \gtrsim \sigma_{\rm M}$. However, if $f_{\rm PBH} \lesssim \sigma_{\rm M}$ matter fluctuations can dominate over the Poisson fluctuations of PBH. In conclusion, the approximation \eqref{eq:Pj_0} holds if the variance of the distribution \eqref{sigma_j} is dominated by PBHs, that is $\bar{N}(y) \ll f_{\rm PBH}^{2}/\sigma_{\rm M}^{2}$. 

In the limit $\bar{N}(y) \to \infty$ we obtain a Gaussian distribution,
\be\label{eq:Pj_inf}
	j\frac{\td P }{\td j} = \frac{2j^2}{\sigma_j^2} e^{-j^2/\sigma_j^2} \,,
\ee
where the width $\sigma_j$ is given by Eq.~\eqref{sigma_j}. The dotted lines in Fig.~\ref{fig:Pj} show this limiting case. We see that, for monochromatic mass functions, the Gaussian distribution is a decent approximation already for $\bar{N}(y)=2$. The latter corresponds to the case where the binary lies in an underdense region prompting matter to initially move away from the binary. It also matches the conclusion based on the analysis of 3-body systems which states that the distance to the closest PBH should be at least of the order of the average distance between PBHs~\cite{Ioka:1998nz}.

%-------------------------------------------------------------------------------
\subsubsection{Merger rate}
\label{sec:mergerrate}
%-------------------------------------------------------------------------------

Assuming negligible disruption between formation and merger, the merger rate of the PBH binaries can now be obtained from Eq.~\eqref{dR_early}. We take $\bar{N}(y)$ to be independent of $x_0$, that is, the binaries are expected to be disrupted if and only if initially there are surrounding PBH closer than $y$. This assumption is in good agreement with the numerical results of Sec.~\ref{sec:simulations}  in case $f_{\rm PBH} \ll1$. The integrals over $x_0$ and $u$ can then be factorised by replacing $x_0$ with $v = u j_0/j(\tau)$. The primordial merger rate reads
\bea\label{eq:R_tot}
	\td R
%&	= \frac{e^{-n V(y)}}{14 \tau} \int^{\infty}_{0} \td V(x_{0})  j \frac{\td P ( j(\tau) )}{\td j} 
%	\\
%&	= \frac{e^{-n V(y)}}{14 \tau} \int^{\infty}_{0} \td V(x_{0}) \int^{\infty}_{0}  \td u \,u  J_{0}(u ) \times \\
%&	 \times \exp\left(-nV(y) \int \frac{\td n(m)}{n} \,F\left(u \frac{m}{\langle m \rangle} \frac{1}{nV(y)} \frac{j_0}{j}\right) -u^{2} \frac{\sigma_{\rm M}^{2}}{2f_{\rm PBH}^{2}}\frac{j_{0}^{2}}{j^{2}}\right) 
%	\\
%&	= \frac{e^{-n V(y)}}{14 \tau} \int^{\infty}_{0} \td u \,u  J_{0}(u)     \times \\
%&	 \times \int^{\infty}_{0}  \td v \frac{\td V(x_{0})}{\td v} \exp\left(-nV(y) \int \frac{\td n(m)}{n} \,F\left(v \frac{m}{\langle m \rangle} \frac{1}{nV(y)} \right) -v^{2} \frac{\sigma_{\rm M}^{2}}{2f_{\rm PBH}^{2}}\right) 
%	\\
&	=	S \times \td R_0,
\eea
%where we made the change of variables $x_{0} \to v = u j_0/(f j(\tau))$, 
where
\bea\label{eq:R0}
	\td R_0	
&	=  \frac{0.65}{\tau} \left( \frac{\tau \eta  M^{14}}{f_{\rm PBH}^{7} c_j^7 c_a^4 \rho_{M}^{11}}\right)^{\frac{3}{37}} \td n(m_{1})\td n(m_{2}) \\
%&	\approx 2.1 \times 10^{6} \, \Gpc^{-3} \yr^{-1}  \times\, f_{\rm PBH}^{\frac{53}{37}} \eta^{\frac{3}{37}}  \left(\frac{M}{\Msun}\right)^{\frac{42}{37}} \left(\frac{\langle m \rangle}{\Msun}\right)^{-2} \left(\frac{\tau}{t_{0}}\right)^{-\frac{34}{37}} \,\frac{\td n(m_1)}{n} \frac{\td n(m_2)}{n} \\
&	\approx \frac{1.6 \times 10^{6}}{\Gpc^{3} \yr} \, f_{\rm PBH}^{\frac{53}{37}} \eta^{-\frac{34}{37}}  \left(\frac{M}{\Msun}\right)^{-\frac{32}{37}} \left(\frac{\tau}{t_{0}}\right)^{-\frac{34}{37}} \psi(m_1) \psi(m_2) \, \td m_1 \td m_2
\eea
is the rate in the limit $\bar{N}(y) \to 0$ and $\sigma_M/f_{\rm PBH} \to 0$. On the second line we have used the usual definition of a PBH mass function,
\be
	\psi(m) \equiv \frac{m}{\rho_{\rm PBH}} \frac{\td n}{\td m} \,,
\ee 
that is normalised to unity, $\int \psi(m) \td m = 1$, and the numerical values $c_a = 0.1$, $c_j = 1$, $t_{0} = 13.8 \times 10^{9} \, \yr$. The suppression factor
\bea\label{def:S}
	S
%&	= \frac{e^{-\bar{N}(y)}}{\Gamma(21/37)} \int \td v \, v^{-\frac{16}{37}} \exp\left( -\bar{N}(y) \int \frac{\td n(m)}{n} \,F\left(\frac{m}{\langle m \rangle} \frac{1}{\bar{N}(y)} v\right)  - \frac{3\sigma_{\rm M}^{2} v^{2}}{10 f_{\rm PBH}^{2}}  \right) \,
&	= \frac{e^{-\bar{N}(y)}}{\Gamma(21/37)} \int \td v \, v^{-\frac{16}{37}} \exp\left[ -\bar{N}(y) \langle m \rangle \int \frac{\td m}{m} \psi(m) F\left(\frac{m}{\langle m \rangle} \frac{v}{\bar{N}(y)}\right)  - \frac{3\sigma_{\rm M}^{2} v^{2}}{10 f_{\rm PBH}^{2}}  \right] \,
\eea
quantifies the contribution from matter density fluctuations and modifications due to the size of the empty region assumed around the pair. Note that, since $S$ does not depend on the coalescence time, the merger rate has an universal time dependence given by $\tau^{-\frac{34}{37}}$.

From the asymptotics of $F$ given in Eq.~\eqref{eq:Fasym} we obtain the following limiting cases: First, in the limit $\bar{N}(y) \to 0$ the suppression factor reads
\bea\label{limS:0}
	S_{\rm max} = \left(\frac{5f_{\rm PBH}^{2}}{6 \sigma_{\rm M}^{2}}\right)^{\frac{21}{74}} U\left(\frac{21}{74},\frac{1}{2},\frac{5f_{\rm PBH}^{2}}{6 \sigma_{\rm M}^{2}}\right),
\eea
where $U$ is the confluent hypergeometric function. Note that this is again independent of the shape of the mass function. The approximate merger rate reported in Ref.~\cite{Ali-Haimoud:2017rtz} corresponds to $S = \left(1 + \sigma_{\rm M}^{2}/f_{\rm PBH}^{2}\right)^{-21/74}$ and underestimates the suppression factor~\eqref{limS:0} by at most a factor of 1.43 when $f_{\rm PBH} \ll \sigma_{\rm M}$. 

Second, in the limit $\bar{N}(y) \to \infty$ we get
\bea\label{limS:inf}
	S_{\rm min}  = \frac{\sqrt{\pi}}{\Gamma(29/37)}  \left( \frac{\sigma_{j}}{j_0} \right)^{-\frac{21}{37}} e^{-\bar{N}(y)} \,.
\eea
For narrow mass functions, Eq.~\eqref{limS:inf} is a good approximation already when $\bar{N}(y) \gtrsim 1$. The inequalities in Eq.~\eqref{eq:Fasym} imply that the suppression factor is bounded between the above asymptotics,
\be
	S_{\rm min} \leq S \leq S_{\rm max} \leq 1,
\ee
where in the last inequality can be saturated in the limit, $\sigma_{\rm M}/f_{\rm PBH} \to 0$. In particular, $\td R_0$ constitutes the upper bound on the merger rate of initial PBH binaries.

\begin{figure}
\begin{center}
\includegraphics[width=0.48\textwidth]{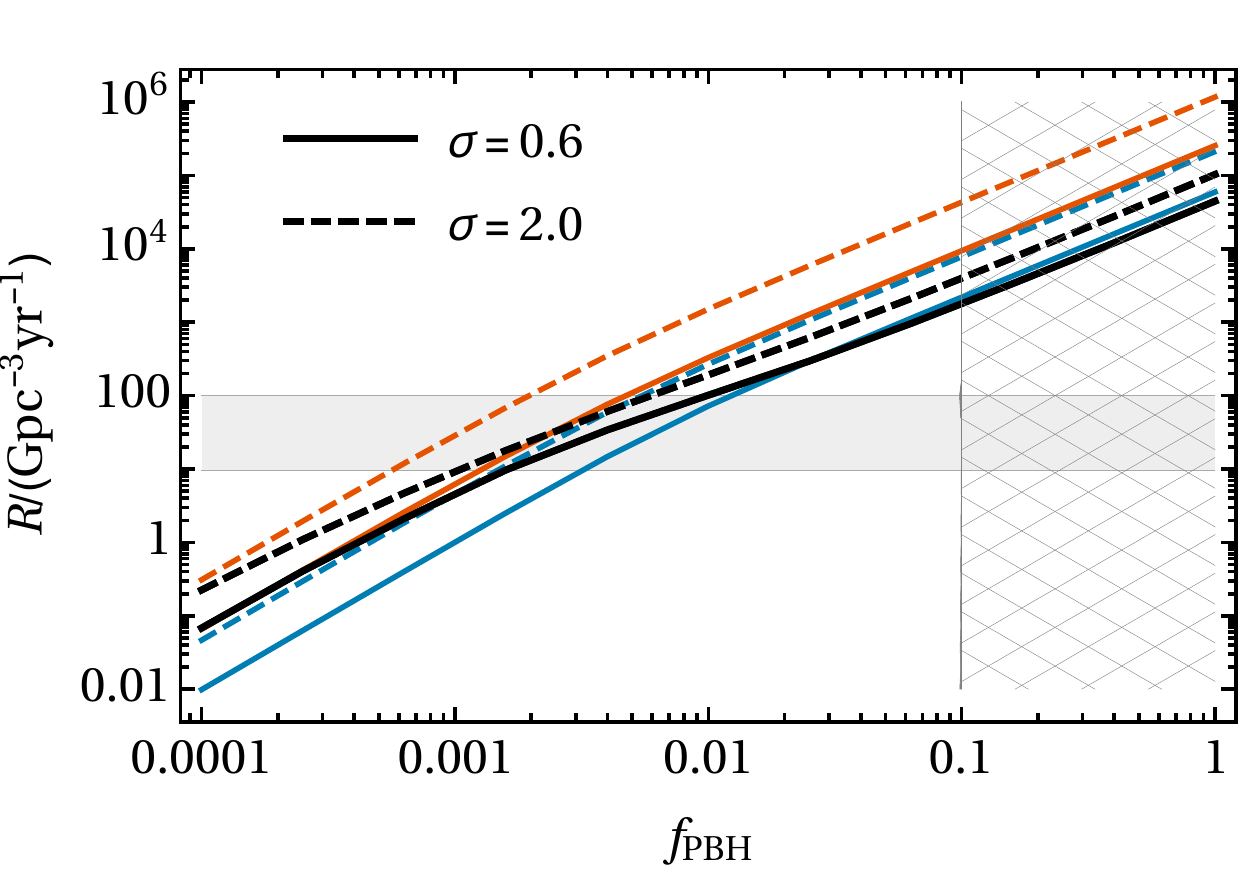}
\caption{The integrated merger rate $R$ as a function of the fraction of DM in PBHs for lognormal mass functions with $m_c = 20\Msun$. The dashed and solid lines correspond to different widths of the lognormal mass function. For the red lines $\bar{N}(y)\to 0$ and blue ones $\bar{N}(y)=2$, whereas for the black lines $\bar{N}(y)$ is given by Eq.~\eqref{eq:nVy_collapse}. The grey region shows roughly the rate at indicated by LIGO observations~\cite{LIGOScientific:2018mvr}. The hatched region at $f_{\rm PBH}>0.1$ indicates that the merger rate estimate~\eqref{eq:R_tot} is not reliable.}
\label{fig:R}
\end{center}
\end{figure}

The dependence of the integrated merger rate on the PBH fraction is shown in Fig.~\ref{fig:R}. We assumed a lognormal mass function for the PBHs,
\be\label{lognormalmf}
	\psi(m) = \frac{1}{\sqrt{2\pi}\sigma m} \exp\left(-\frac{\log^2(m/m_c)}{2\sigma^2}\right) \,,
\ee
where, following the notation of \cite{Carr:2017jsz},  $m_c$ denotes the peak mass of $m \psi(m)$, and $\sigma$ characterises the width of the mass spectrum. This class of mass functions has been found to provide a good representative of a large class of extended mass functions~\cite{Dolgov:1992pu,Green:2016xgy, Kannike:2017bxn,Horowitz:2016lib,Kuhnel:2017pwq}. We stress, however, that the log-normal mass function is not universal as several effects may cause deviations from it~\cite{Byrnes:2018clq,Germani:2018jgr,Yoo:2018kvb,Martin:2018lin,Biagetti:2018pjj,Franciolini:2018vbk}.

%-------------------------------------------------------------------------------
\section{Evolution of PBH binaries in the early universe}
\label{sec:simulations}
%-------------------------------------------------------------------------------

The PBH binaries are the first gravitationally bound structures to be formed in the early universe. Subsequently, interactions with nearby PBHs may disrupt the binaries. To estimate the effect of the surrounding PBHs on the binary population in the early universe and to determine which initial configurations produce undisrupted PBH binaries, we conducted $N$-body simulations of PBHs in an expanding background, using a custom-written C++ code. 

The simulations focus on the early evolution of PBH binaries, from their formation up to  $a = 3 a_{\rm eq}$, which corresponds roughly to the time of recombination. The initial conditions of the central PBH pair are chosen such that a binary with an expected coalescence time equal the age of the universe is formed according to Eq.~\eqref{eq:coalescence_time}. The aim is to study the interaction of this pair with the surrounding PBHs, and to test the validity of the formation mechanism described in Sec.~\ref{sec:dynamics} and the analytical predictions for the distribution of the PBH binary orbital parameters given in Sec.~\ref{sec:distribution}.

\subsection{Simulation set-up}

The simulations are performed in physical coordinates and solve the equations of motion resulting from the action~\eqref{action_N}. The initial state consists of a central PBH pair that will form a binary and $N-2$ randomly distributed PBHs in a sphere around the binary as shown in Fig.~\ref{fig:simsetup}. The individual particles in the simulation are subject to three different forces:
\begin{enumerate}
\item The gravitational attraction of the other $N-1$ particles is calculated in the Newtonian approximation without any regularisation at small distances. The individual contributions are summed.

\item The expansion of the universe is simulated by including a Hubble acceleration of the form $\ddot{\vec{r}} = \left( \dot{H} + H^2 \right) \, \vec{r}$.

\item The gravitational attraction of the PBHs in the rest of the universe is approximated by the gravitational potential of a spherical underdensity inside otherwise homogeneously distributed matter. Equivalently, this corresponds to a sphere with uniform negative mass density equal to the positive mass density of the simulated black holes. The \emph{comoving} radius and the total negative mass are kept constant.  
\end{enumerate}

We remark that the radius of the simulation is always much smaller than the Hubble radius, so the Newtonian approximation is justified. The simulation neglects the contributions of matter fluctuations as well as from the PBHs outside of the sphere.  These are not relevant for $f_{\rm PBH} \gg \sigma_{\rm M}$, as the dominant forces acting on the central binary arise in that case from the closest PBHs surrounding it. For $f_{\rm PBH} \lsim \sigma_{\rm M}$, in turn, these omissions underestimate the tidal forces. We also neglect the effect of emission of gravitational radiation. A binary that is loosing energy only by gravitational radiation satisfies $r_a \propto j^{-2}$ for $j \ll 1$~\cite{Peters:1964zz}. Eq.~\eqref{eq:tau} implies the time dependence $j(t) = j_0/(1 - t/\tau)$, where $\tau$ is the coalescence time at formation, so for binaries merging today the impact of gravitational radiation on the evolution of the orbit is of the order $10^{-4}$ within the first millions of years and can thus be safely neglected. We neglect baryons and, when $f_{\rm PBH} < 1$, treat the surrounding DM as non-dynamical, accounting it only in the expansion rate.

The simulation uses adaptive, individual time steps for the particles and implements the leapfrog integration algorithm in the drift-kick-drift formulation. The size of the individual time steps is chosen such that the simulation (when ignoring the expansion of the universe) approximately conserves the total energy of the system. To determine the required precision of the energy conservation, we consider a typical binding energy of $E_\mathrm{ref} = 10^{39}$ J (corresponding e.g.~to two lightly bound BHs with masses of 30 $\Msun$ circling each other with a distance of $\sim$4 mpc). The simulation is then required to conserve the total energy with a precision $\Delta E \le 3 \cdot 10^{-5} \, E_\mathrm{ref}$ in each global time step. We find that this level of precision allows for sufficiently accurate integration of binary orbits while keeping the total CPU time requirement manageable. 

To allow for exact reproduction of the black hole orbits, a large dynamical range is required. The simulation reduces the individual, adaptive time steps until the required precision or the maximal number of substeps is reached. If the minimal substep of $\sim0.63$\,seconds %simulation time 
cosmic time is not sufficient to reach the required precision, the code marks an exception in the output and continues with the minimal substep. 

To ensure that the simulations finish in finite time, it is furthermore necessary to define a minimal distance of the BHs, which is chosen at $d_{\mathrm{min}} = 3 \cdot 10^8$ m. If two BHs approach each other more closely, they are merged into a single object while conserving the mass and momentum of the previous two objects. In practice, we find that this minimal distance allows for the simulation of binaries with lifetimes corresponding to a small fraction of the age of the universe, with the exact number depending on the other orbital parameters. We also tested that changes in the results are insignificant if $d_{\mathrm{min}}$ is decreased.

The simulations are started with an initial scale factor of $a_{\mathrm{init}} = 10^{-3} \, a_{\mathrm{eq}}$. The initial set-up shown in Fig.~\ref{fig:simsetup} is achieved in the following way: First, the $N-2$ other particles in the simulation are randomly added to a spherical volume, such that the density of BHs in the spherical volume is equal to $f_\mathrm{PBH}$ times the total dark matter density. Then, the projection of the tidal force in a fixed direction from these $N-2$ particles is calculated for the centre of the sphere. Finally, the pair is added. The centre of mass of the pair is set to coincide with the centre of the spherical region and the vector joining the two PBH matches the fixed direction chosen before. The initial separation of the central BH binary is calculated by solving Eq.~\eqref{eq:coalescence_time} for $x_0$ and demanding that $\tau$ equals the current age of the universe. Since the simulation is performed in non-comoving coordinates, each object is introduced with an initial velocity given by the Hubble expansion, i.e.~$\vec{v}_0 = H_0 \, \vec{r}_0$, corresponding to a vanishing peculiar velocity. To guarantee that the initial conditions produce, on average, binaries that are expected to merge within the age of the universe, we used the values $c_a = 0.129$ and $c_j = 1.055$ found numerically from $f_{\rm PBH} = 1$ simulations. After the central PBH pair has been added, the simulation is run until a scale factor of $a_{\mathrm{final}} = 2.97 \, a_{\mathrm{eq}}$ is reached, corresponding to a total simulation time of 377\,kyr. The simulation therefore finishes after recombination. 

We ran 70-body simulations with a monochromatic mass function for $f_{\rm PBH} = 1$, $f_{\rm PBH} = 0.1$  and $f_{\rm PBH} = 0.01$ and with an extended mass function for $f_{\rm PBH} = 0.1$. Each set of simulations consisted of 3000 simulations. The data used in the following analysis only includes simulations that reached $a = 3 a_{\rm eq}$ within a runtime of 10 days for $f_{\rm PBH} = 1$ and 6 days for other simulations. We also excluded simulations where the central binary merged due to reaching the minimal distance set in the simulation. The latter makes up about 20\% of the finished simulations in the case of monochromatic simulations with $f_{\rm PBH} = 1$, 7\% for $f_{\rm PBH} = 0.1$ and 2\% for $f_{\rm PBH} = 0.01$, while the corresponding fraction is 2 -8\% for simulations with an extended mass function. We checked that increasing the minimal distance or decreasing the runtime of the simulation does not affect our conclusions about undisrupted binaries. However, we see in Figs.~\ref{fig:dist_nn} and ~\ref{fig:dist_nn_ext}, that the total number of used simulations (blue histogram) in the region where the central binary is expected to be disrupted remains below the expectation for Poisson distributed initial conditions (dashed line). This systematic effect can be attributed to the fact that $N$-body collisions are more likely to produce encounters where the distance briefly decreases below $d_{\mathrm{min}} = 3 \cdot 10^8$\,m while their accurate simulation requires more computational resources making the simulation less likely to finish.

\subsection{Properties of binaries merging within the age of the universe}

The binary may be disrupted shortly after formation by a nearby PBH or a small cluster of PBHs, which may appear due to the Poissonian nature of the spatial distribution. We find that a single nearby PBH is the dominant source for disruption at the earliest times, especially when $f_{\rm PBH} \ll 1$. When $f_{\rm PBH} \approx 1$, however, the formation of bound systems of several PBHs can be observed already at matter-radiation equality, and, if the initial binary finds itself close to such clusters, it is very likely to be disrupted. 

The initial binaries expected to merge today are highly eccentric, $j\ll1$. If they interact with other objects, the eccentricity decreases and, since $\tau \propto j^7$, the coalescence time will on average be increased by several orders of magnitude. So, in the first approximation, all disrupted binaries are removed form the population of binaries merging today.

Consider collisions between the binary and the PBH initially closest to it. Let $x_{\rm NN}$ denote the initial comoving distance of the nearest neighbour. We would like to choose the size of the empty region around to binary as the smallest $x_{\rm NN}$ that does not lead to the early disruption of the binary. A rough estimate can be obtained by noting that, since the volume $V(x_{\rm NN})$ surrounding the PBH pair does not contain any PBH by construction, the central pair corresponds to an effective matter density fluctuation $\delta_{\rm NN} \approx (M-\rho_{\rm PBH} V(x_{\rm NN}))/(\rho_{\rm M} V(x_{\rm NN})) $. As such configurations are expected to collapse at $a \approx a_{\rm eq}/\delta_{\rm NN}$ we estimate that, at a given $a$, the binary has collided with its neighbour if $x_{\rm NN} < y$, where $y$ is given by
\be\label{eq:nVy_collapse}
	\bar{N}(y) \approx \frac{M}{\langle m\rangle} \frac{f_{\rm PBH}}{f_{\rm PBH} + a_{\rm eq}/a} \,.
\ee
We stress that this estimate is intended to provide a maximal value for $x_{NN}$ below which most initial binaries will be disrupted by the PBHs initially surrounding it. It does not account for any sources of later disruption, e.g. interactions with PBH clusters. For monochromatic mass functions and $a \to \infty$ we obtain $\bar{N}(y) = 2$ consistent with the analysis of Ref.~\cite{Ioka:1998nz} which implied that $y$ should be of the order of the average distance between PBHs.

\begin{figure}
\begin{center}
\includegraphics[width=0.31\textwidth]{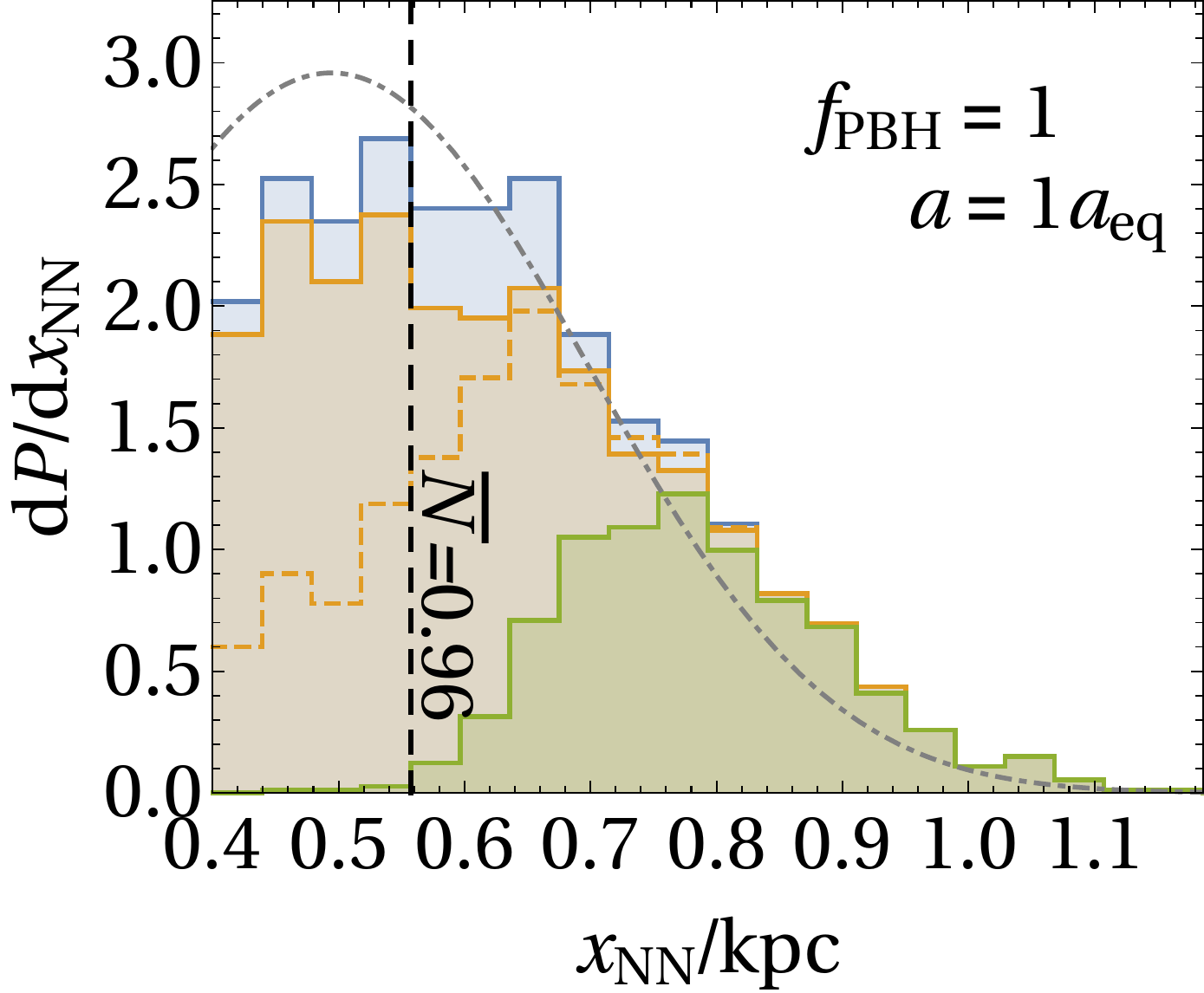} \hspace{1mm}
\includegraphics[width=0.31\textwidth]{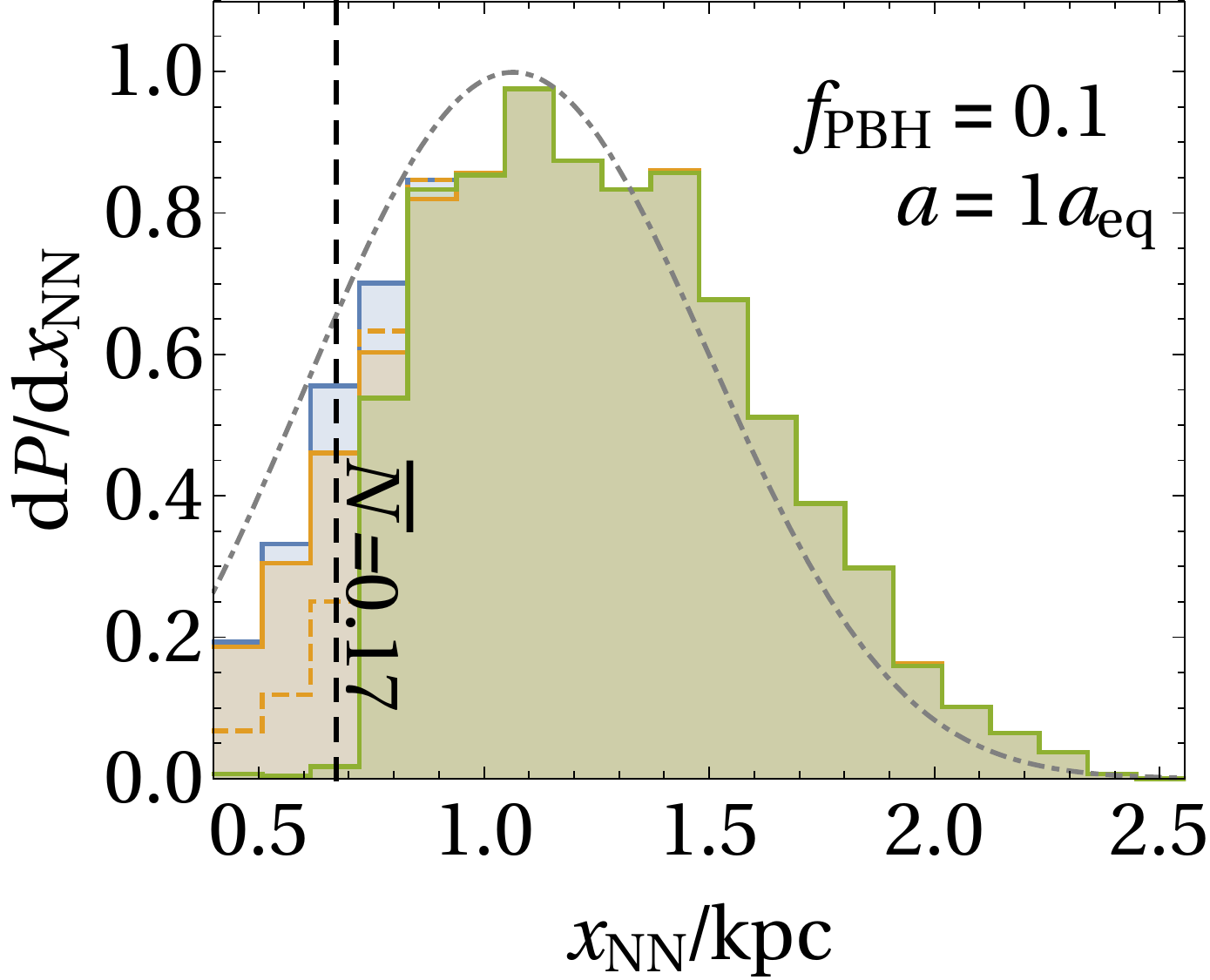} \hspace{1mm}
\includegraphics[width=0.31\textwidth]{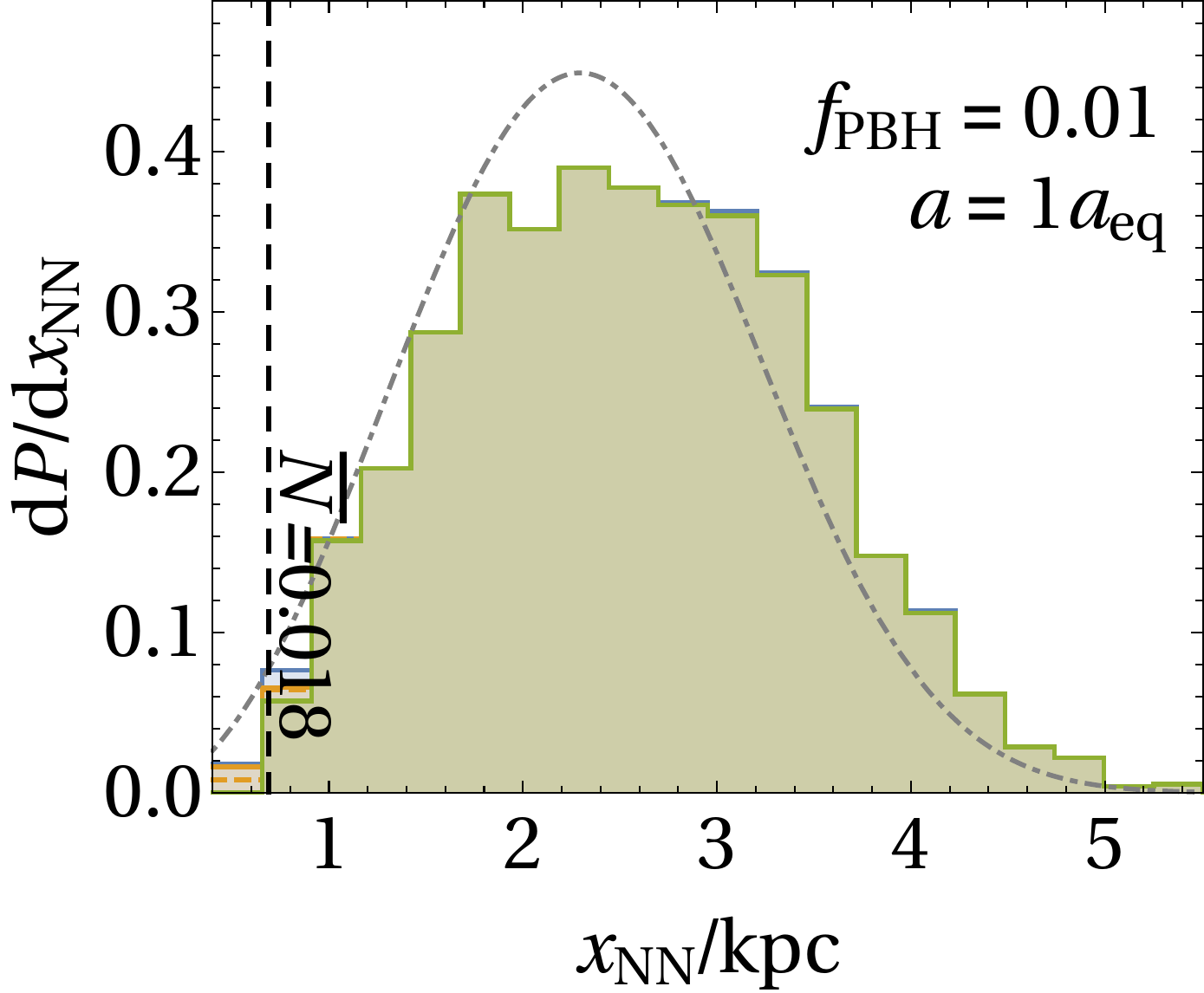}
\\
\includegraphics[width=0.31\textwidth]{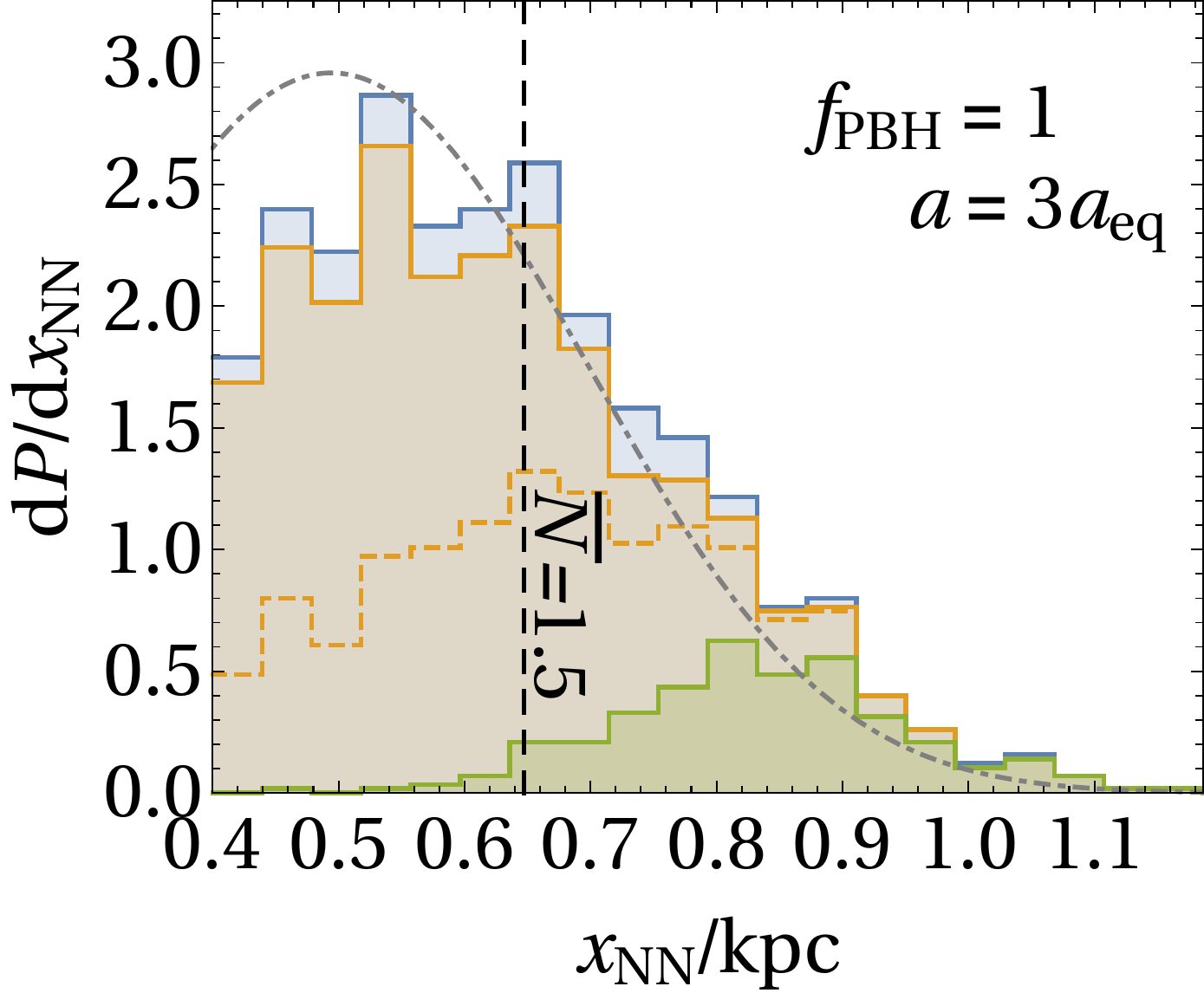} \hspace{1mm}
\includegraphics[width=0.31\textwidth]{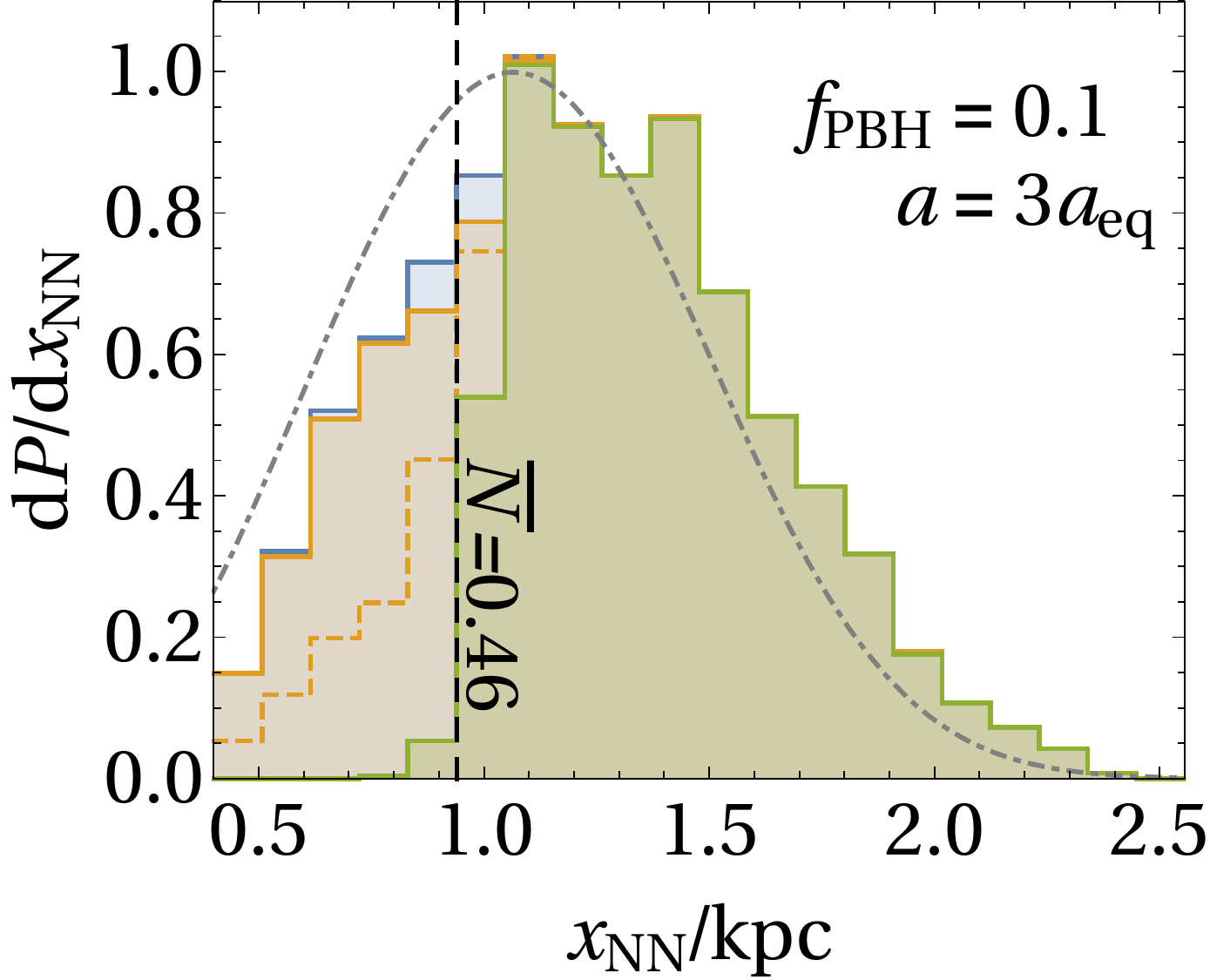} \hspace{1mm}
\includegraphics[width=0.31\textwidth]{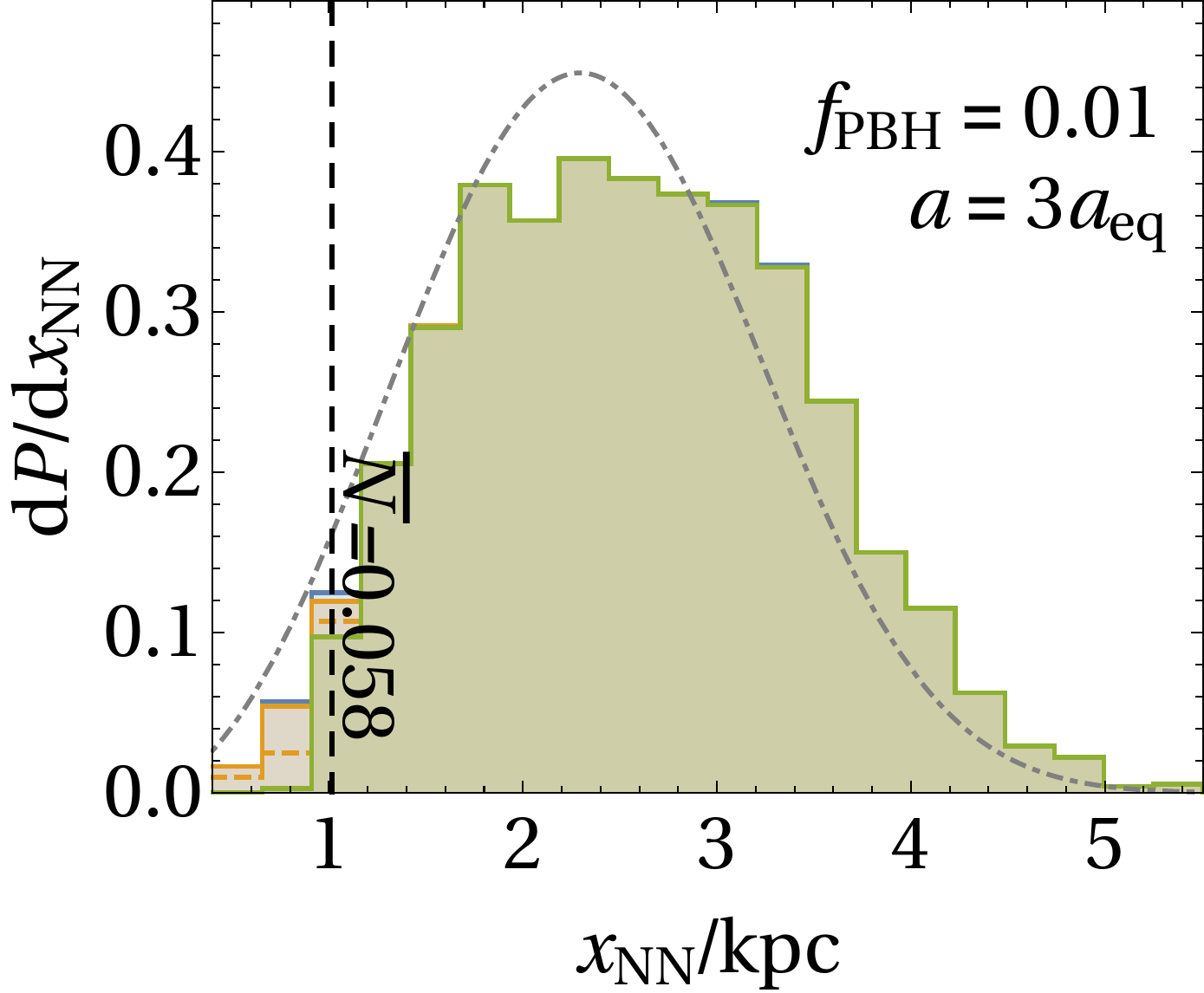}
\caption{The dependence of the state of the central pair at $a  = \, a_{\rm eq}$ (upper panels) and $a  = 3 \, a_{\rm eq}$ (lower panels) on the initial comoving distance of the PBH nearest to the binary for a monochromatic mass function at $m_c = 30 \Msun$ with different values of $f_{\rm PBH}$. The blue region shows all simulations, while simulations with bound and undisrupted central binaries at $a  = 3 \, a_{\rm eq}$ are shown in yellow and green, respectively. The yellow dashed line shows the pairs, where the total energy of the central pair is negative, while the solid yellow curve shows initial conditions where at least one of the central PBHs is in a binary system at the end of the simulation. The dashed vertical line corresponds to the estimate Eq.~\eqref{eq:nVy_collapse} for the minimal distance the nearest neighbour can have in order for not to disrupt the binary. The  dot-dashed line shows the expected distribution of the nearest neighbour distance. For comparison, the initial comoving separation $x_0$ of the binaries is of the order $200\,\pc$ for $f_{\rm PBH} = 1$.
}
\label{fig:dist_nn}
\end{center}
\end{figure}

\begin{figure}
\begin{center}
\includegraphics[width=0.31\textwidth]{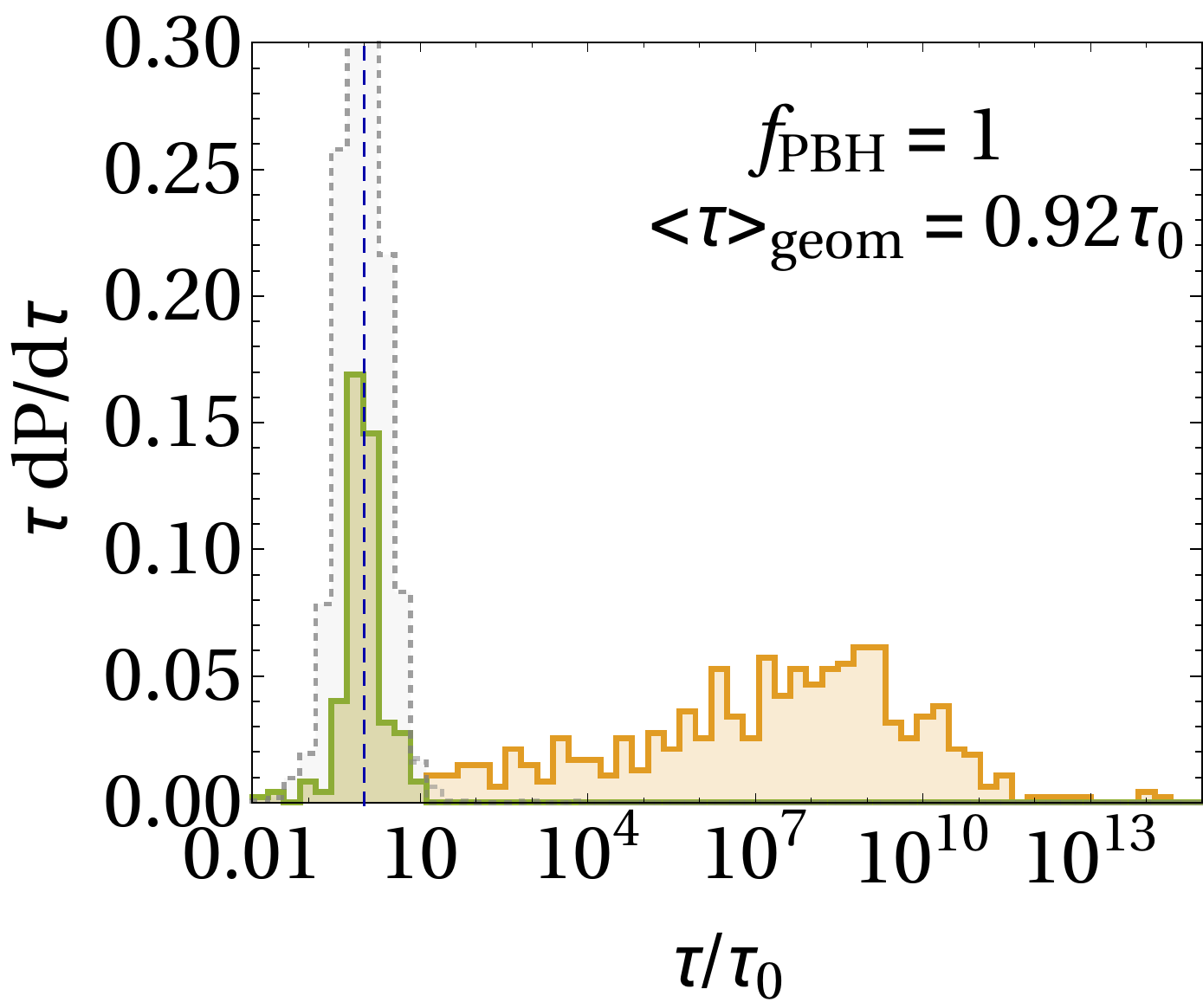} \hspace{1mm}
\includegraphics[width=0.31\textwidth]{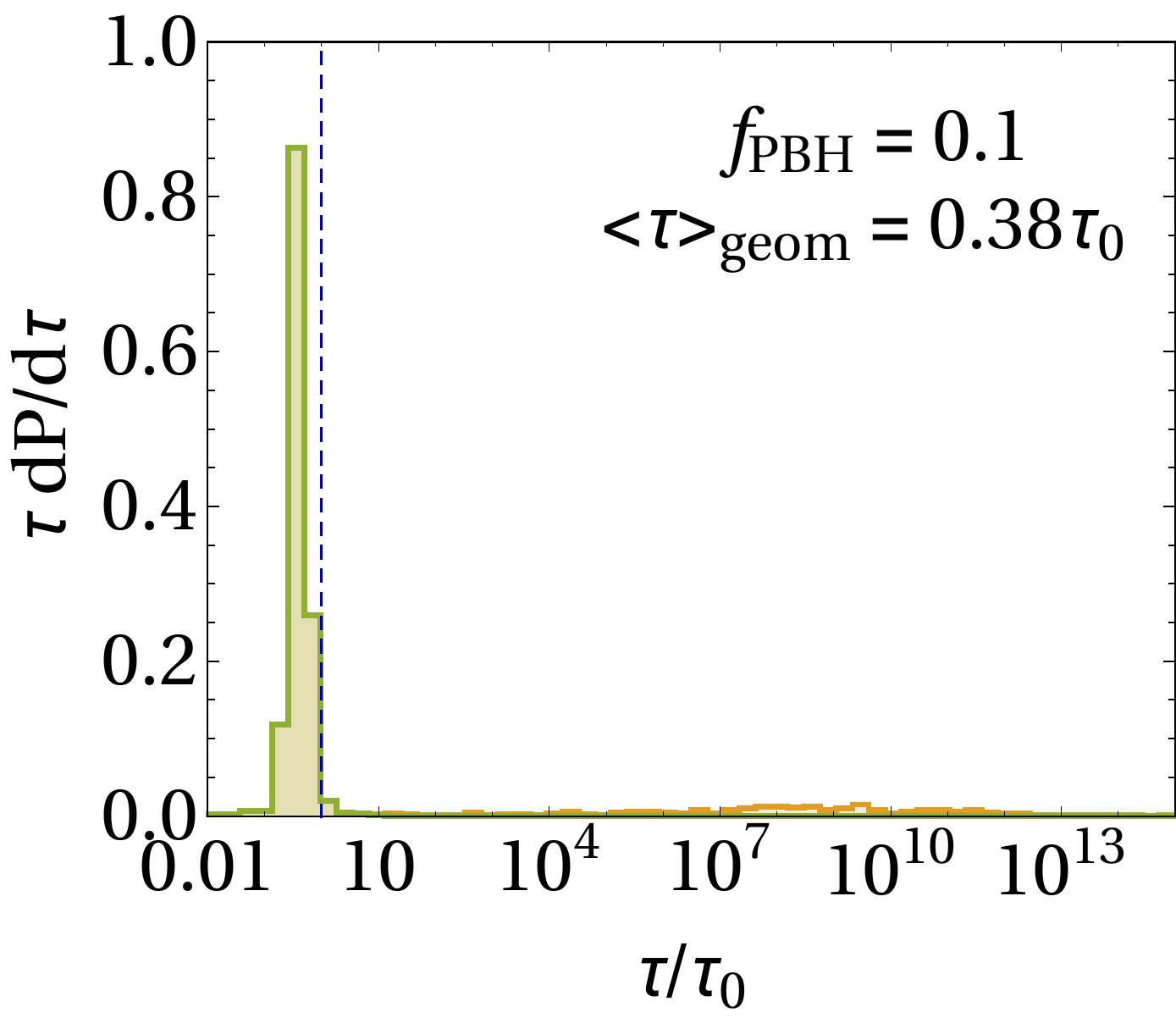} \hspace{1mm}
\includegraphics[width=0.31\textwidth]{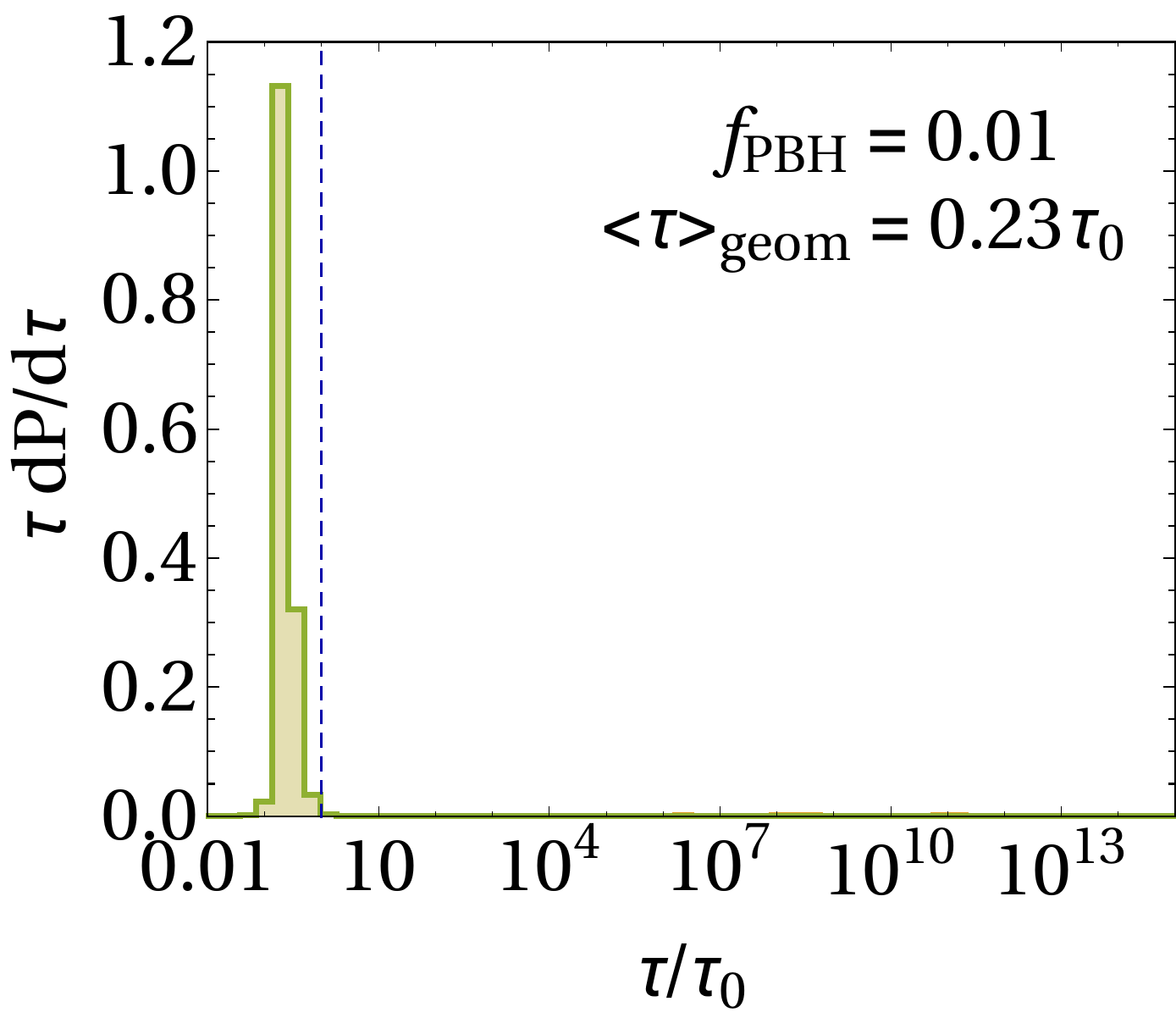}
\caption{Distribution of estimated coalescence times at $a  = 3 \, a_{\mathrm{eq}}$ for different PBH fractions and a monochromatic mass function with $m_c = 30 \Msun$. Bound and undisrupted central binaries are coloured in yellow and green respectively. The dashed vertical line indicates the age of the universe.  The geometric average of the expected coalescence times is evaluated only for unperturbed binaries. The dashed grey line in the $f_{\rm PBH} = 1$ plot shows the distribution of expected coalescence times at $a = 0.1 a_{\mathrm{eq}}$.
}
\label{fig:dist_tau}
\end{center}
\end{figure}

Using the simulations we can estimate if a binary with a nearest neighbour at distance $x_{\rm NN}$ gets disrupted. The results are summarised in Fig.~\ref{fig:dist_nn}. The numerical data is obtained from $2820$, $2423$, and $1382$ independent 70-body simulations with $f_{\rm PBH} = 0.01$, $f_{\rm PBH} = 0.1$ and $f_{\rm PBH} = 1$, respectively. In all simulations, the central pair initially forms a binary with a coalescence time of roughly the age of the universe, as can be seen from Fig.~\ref{fig:dist_tau}. The blue region in Fig.~\ref{fig:dist_nn} shows all initial configurations. As a consistency check we find that the initial distance of the nearest neighbour follows a Poisson distribution. Simulations in which at least one of the central BHs is in a binary system at $a=3a_{\rm eq}$ (upper panels) or $a=3a_{\rm eq}$  (lower panels), i.e. simulations where the initial binary may have swapped a BH, are shown by the yellow line, while the dashed yellow line shows simulations where the central BH remain bound to each other. More precisely, under the solid yellow line we included simulations where the energy of the system containing one of the central BHs and the BH closest to it was negative and we additionally imposed that the binding energy between the binary and its closest neighbour is less than 10\% of the binary binding energy. For the dashed yellow line we only require that the total energy of the initial binary is negative. The green regions show simulations with central binaries with a coalescence time $\tau < 10 t_0$, which we use as the working definition for undisrupted binaries. This definition is justified by the fact that encounters with surrounding PBHs dominantly increase the coalescence time by several orders of magnitude, as can be clearly seen from the fist panel in Fig.~\ref{fig:dist_tau}. 

Fig.~\ref{fig:dist_nn} shows that the numerical result is consistent with the estimate~\eqref{eq:nVy_collapse} for the minimal value of $x_{\rm NN}$ and for its dependence on the scale factor. For $f_{\rm PBH} \ll 1$ the sharp transition at $y$ in Fig.~\ref{fig:dist_nn} from almost all initial binaries being disrupted to almost all of them being non-perturbed confirms the underlying assumption in the computation of the rate~\eqref{eq:R_tot} that $y$ is independent of $x_0$. For $f_{\rm PBH} = 1$, however, half of the PBH with $x_{\rm NN} > y$ are disrupted already at $a = 3 a_{\rm eq}$. This implies an additional suppression factor on top of~\eqref{def:S}, because the latter accounts for disruption by the closest PBH only. Visual inspection of randomly selected simulations with $f_{\rm PBH} = 1$ and $x_{NN}>y$ where the binary was disrupted indicates that the disruption is mostly, but not always, due to the interaction of the binary and a nearby $N$-body cluster as opposed to a direct collision with the nearest PBH. As the binaries will continue to be disrupted in the early clusters when $a > 3 a_{\rm eq}$, i.e. after the end of the simulation, it is expected that nearly all initial binaries will be disrupted within the age of the universe in case $f_{\rm PBH} \approx 1$. It is, however, not possible to draw definite conclusions from our numerical results beyond stating the need for a careful revision of early binary formation when $f_{\rm PBH} \approx 1$. We will return to the merger rate in that case in Sec.~\ref{subsec:rate}.

The expected coalescence time $\tau$ distribution shown in Fig.~\ref{fig:dist_tau} for the $f_{\rm PBH} = 0.1$ and $f_{\rm PBH} = 0.01$ simulations is somewhat smaller than the age of the universe. This discrepancy is due to inaccuracies in the idealised analytic estimates \eqref{eq:gen_rj} and \eqref{eq:coalescence_time} used to determine the initial separation $x_0$ for the central PBH pair. Notably, since $\tau \propto x_0^{37}$ small deviations in the initial separation estimate can lead to large deviations in $\tau$ from $t_0$. Based on test simulations we fixed the numerical parameters $c_a=0.129$ and $c_j=1.055$ to obtain merger times close to $t_0$ in the case $f_{\rm PBH} =1$. However, Fig.~\ref{fig:dist_tau} shows that this choice overestimates $\tau$ for $f_{\rm PBH} =0.1$ and $f_{\rm PBH} =0.01$ by roughly a factor of 3 and 4, respectively. On the other hand, by Eq.~\eqref{eq:coalescence_time} $\tau \propto c_a^4 c_j^7$, so the obtained values $c_a=0.1$ and $c_j=0.95$ would have underestimated $\tau$ by a factor of 2. In all, in agreement with previous studies, we find that the analytic approach gives a reliable estimate of the coalescence time of the initial binary. Moreover, we stress that, because $\td R \propto c_j^{-21/37}c_a^{-12/37}$  by Eq.~\eqref{eq:R0},  the merger rate of initial PBH binaries is considerably less sensitive than $\tau$ to $\mathcal{O}(1)$ changes in $c_j$ and $c_a$.

\begin{figure}
\begin{center}
\includegraphics[width=0.31\textwidth]{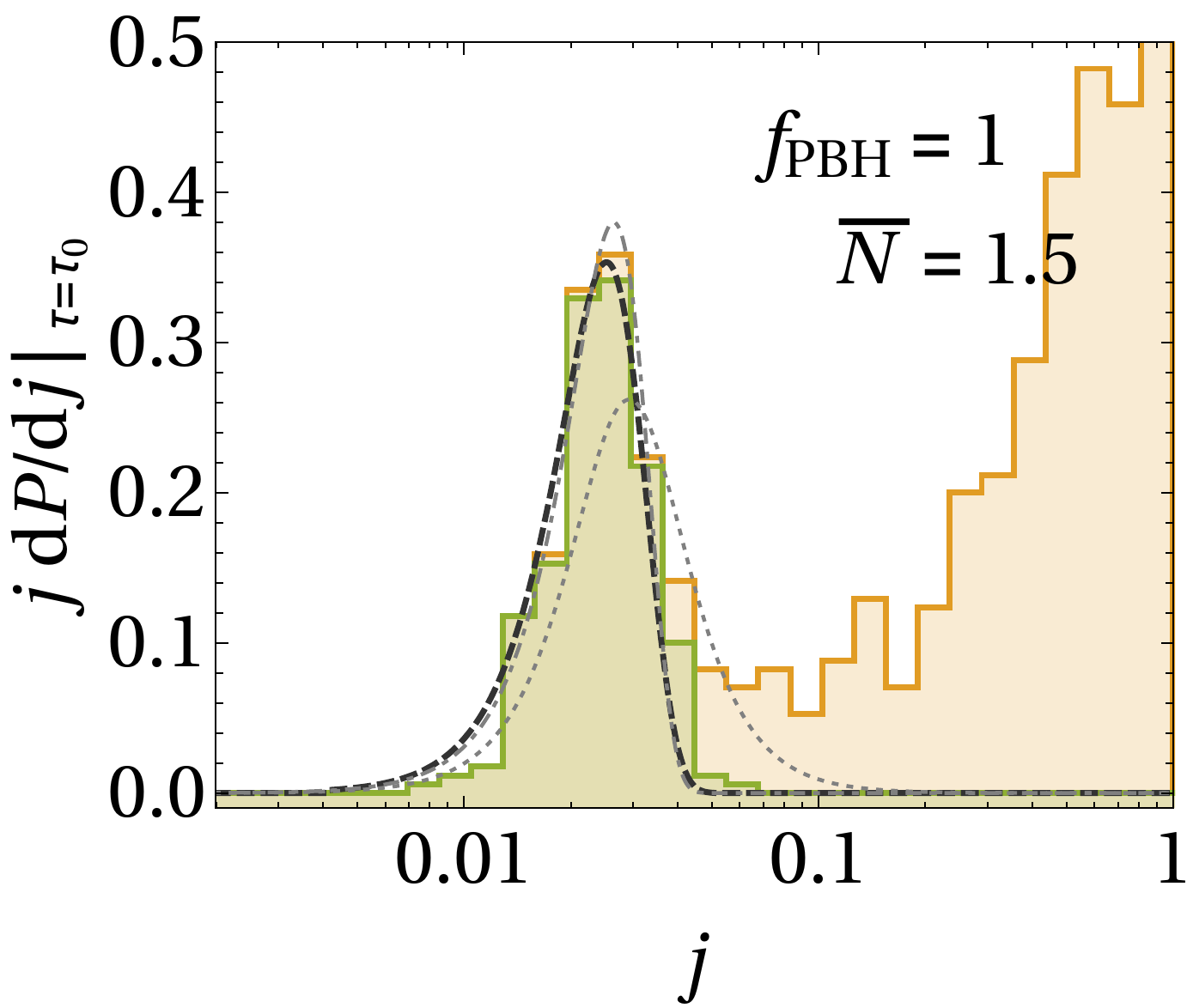} \hspace{1mm}
\includegraphics[width=0.31\textwidth]{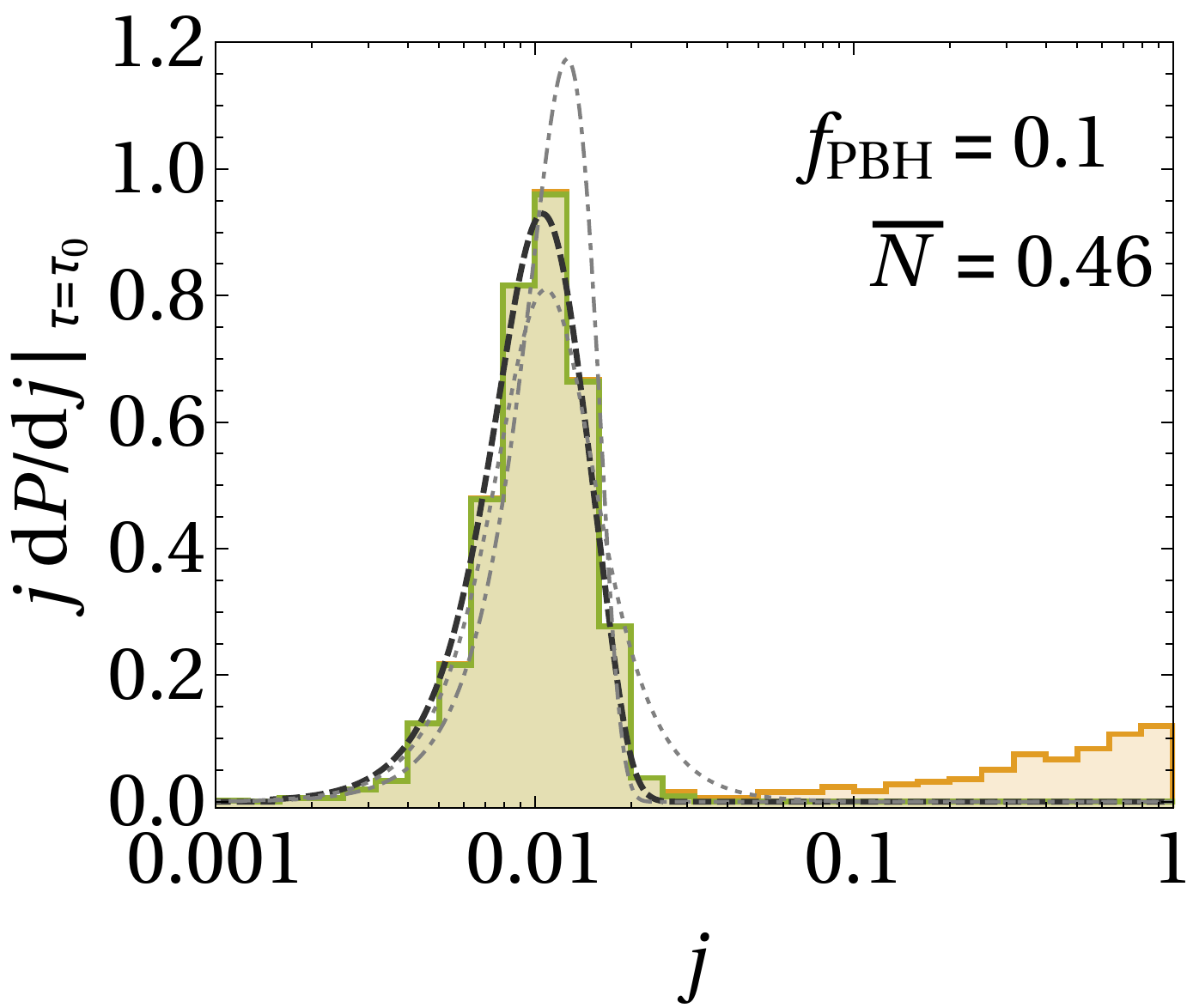} \hspace{1mm}
\includegraphics[width=0.31\textwidth]{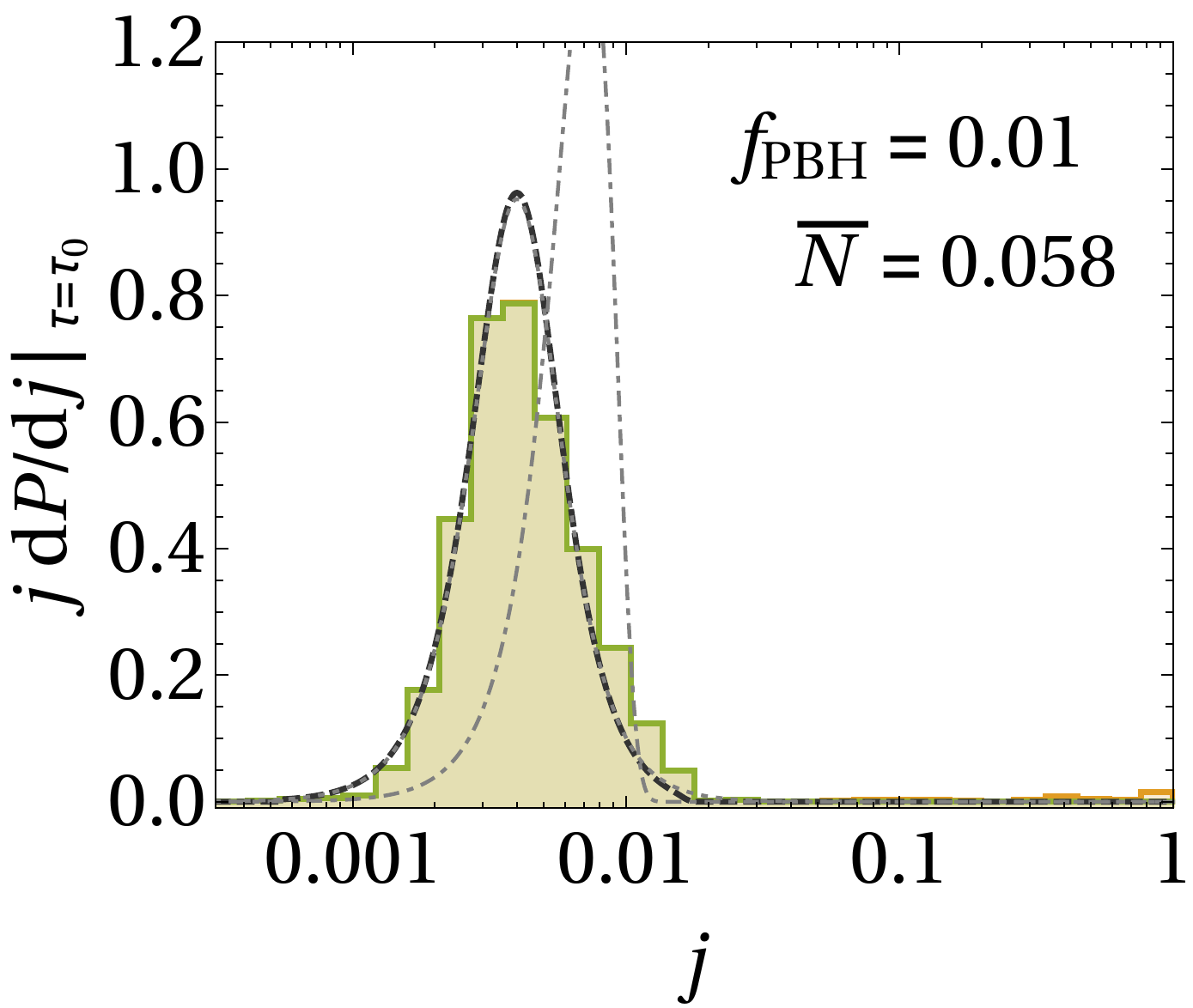}
\caption{Distribution of angular momenta at $a  = 3 \, a_{\mathrm{eq}}$ of PBH binaries expected to merge today for different PBH fractions and a monochromatic mass function with $m_c = 30 \Msun$. The bound and undisrupted central binaries are coloured in yellow and green respectively. The thick dashed line show Eq.~\eqref{eq:Pj_tau} evaluated from the exact distribution Eq.~\eqref{eq:Pj} using $\bar{N}(y)$ estimated from Eq.~\eqref{eq:nVy_collapse}, while the thin dot-dashed and dotted lines indicate the limiting cases $\bar{N}(y) \ll 1$ in Eq.~\eqref{eq:Pj_0} and $\bar{N}(y) \gg 1$ in Eq.~\eqref{eq:Pj_inf}, respectively. The distribution of bound binaries is normalised to unity.}
\label{fig:dist_j}
\end{center}
\end{figure}

Consider now the analytic estimate of the orbital parameters of binaries with a coalescence time $t_0$. The distribution \eqref{eq:Pj} gives the conditional probability for a fixed initial separation $x_0$ and the size of the empty region $y$. Since the initial separation is distributed simply as $n \,\td V(x_0)$, we obtain
\bea\label{eq:Pj_tau}
	\left. \frac{\td P}{\td j} \right|_{\tau = t_0} 
&	\propto \int \td V(x_0) \delta\left(t_0 - \tau(r_{a}(x_0),j)\right) \frac{\td P(j/j_0(x_0))}{\td j} \\
&	\propto \left.   j_0(x_0) \frac{\td P(j/j_0(x_0) )}{\td j}   \right|_{t_0 = \tau(r_{a}(x_0),j)} \,,
\eea
where we dropped overall factors independent of $j$ as they will be determined by properly normalising the distribution. In the last step we used $\td \tau/\td x_0 \propto \tau/x_0$ and $j_0 \propto x_0^3$. The initial separation $x_0$ is determined from the semimajor axis $r_a$ by Eq.~\eqref{eq:gen_rj} which in turn is fixed by the coalescence time \eqref{eq:tau} and the angular momentum $j$. This gives that $j_0 \propto j^{-21/16}$. The distribution of $\ln j$, that is $j\td P/\td j$, depends on $j$ through the combination of $j/j_0$, thus we can define a similar characteristic angular momentum $j_\tau$ for fixed $\tau$ so that $\left. j \td P/\td j \right|_{\tau = t_0}$ will be a function of $j/j_\tau$ only, i.e. $j/j_0|_{\tau = t_0} =  (j/j_\tau)^{37/16}$. This gives\footnote{We used the analytic predictions $c_a = 0.1$ and $c_j = 0.95$. Since $j_\tau \propto c_j^{16/37} c_a^{-12/37}$, it is relatively insensitive to $\mathcal{O}(1)$ changes in these parameters.}
\be
	j_\tau = 0.02 f_{\rm PBH}^{\frac{16}{37} }  (4\eta)^{\frac{3}{37}}  \left(\frac{M}{20 \Msun}\right)^{\frac{5}{37}}\left(\frac{t_0}{tau_0 }\right)^{\frac{3}{37}}
\ee
for binaries with total mass $M$ and mass asymmetry $\eta$. This quantity gives an order of magnitude estimate of the dimensionless angular momentum of initial binaries merging today. 

In Fig.~\ref{fig:dist_j} we compare the distribution of angular momenta at $a=3 a_{\rm eq}$ obtained numerically from the simulation and the corresponding analytic estimates based on Eqs.~\eqref{eq:Pj} and~\eqref{eq:Pj_tau} with $\bar{N}(y)$ fixed using Eq.~\eqref{eq:nVy_collapse}. The analytic prediction for unperturbed binaries is in good agreement with the numerical results (green). The $j$ distribution of perturbed binaries (yellow) is roughly uniform, i.e. $j\td P/\td j \propto j$. 

Interestingly, the analytic prediction for the distribution of $j$ of the unperturbed binaries (solid line in Fig.~\ref{fig:dist_j}) works well also when $f_{\rm PBH} = 1$ although most of the initial binaries with $x_{NN} > y$ are disrupted. This observation can be used to shed some light on the cause of disruption of the binaries with $x_{NN} > y$. The choice~\eqref{eq:nVy_collapse} for the size of the empty region around the binary provides a rough estimate of the minimal $x_{\rm NN}$ for which the binary is almost certainly disrupted by the infall of the nearest PBH, but it does not predict what happens to the binaries for which $x_{\rm NN} > y$. As the analytic prediction for the distribution of $j$, which assumes that all initial binaries with $x_{NN} < y$ are disrupted while binaries with $x_{NN} > y$ remain undisrupted, matches the numerical one, the process disrupting the binaries with $x_{NN} > y$ must be statistically nearly independent of $j$ and thus also from $x_{\rm NN}$. This is because the nearest PBHs give the dominant contribution on the tidal torque which determines $j$. In conclusion, the disruption in simulations with $x_{\rm NN} > y$ is less likely to take place immediately after the surrounding PBHs have decoupled from expansion, but later, as they interact with clusters of PBHs. We have also observed this by visually studying individual simulations with $x_{\rm NN} > y$ that produce a disrupted central binary.

The distribution of semimajor axis is easily obtained by noting that $r_a \propto j^{-7/4}$ when the coalescence time is fixed, so $\left. j \td P/\td j \right|_{\tau = t_0} = (7/4)\left.  r_a \td P/\td r_a \right|_{\tau = t_0}$. Analogously to the characteristic angular momentum $j_\tau$, we find the characteristic scale for the semimajor axis
\be
	r_{a,\tau} = 0.3 {\rm mpc} \times f_{\rm PBH}^{-\frac{28}{37} }  (4\eta)^{\frac{4}{37}}  \left(\frac{M}{20 \Msun}\right)^{\frac{19}{37}}\left(\frac{t_0}{tau_0 }\right)^{\frac{4}{37}}. 
\ee 
The semiminor axis $r_{b} = j r_a$ and the periapsis $r_{\rm per} \approx j^2 r_a  /2$ are therefore of the order of $\au$ and $10^{-2} \au$, respectively.
 
 \begin{figure}
\begin{center}
\includegraphics[height=0.32\textwidth]{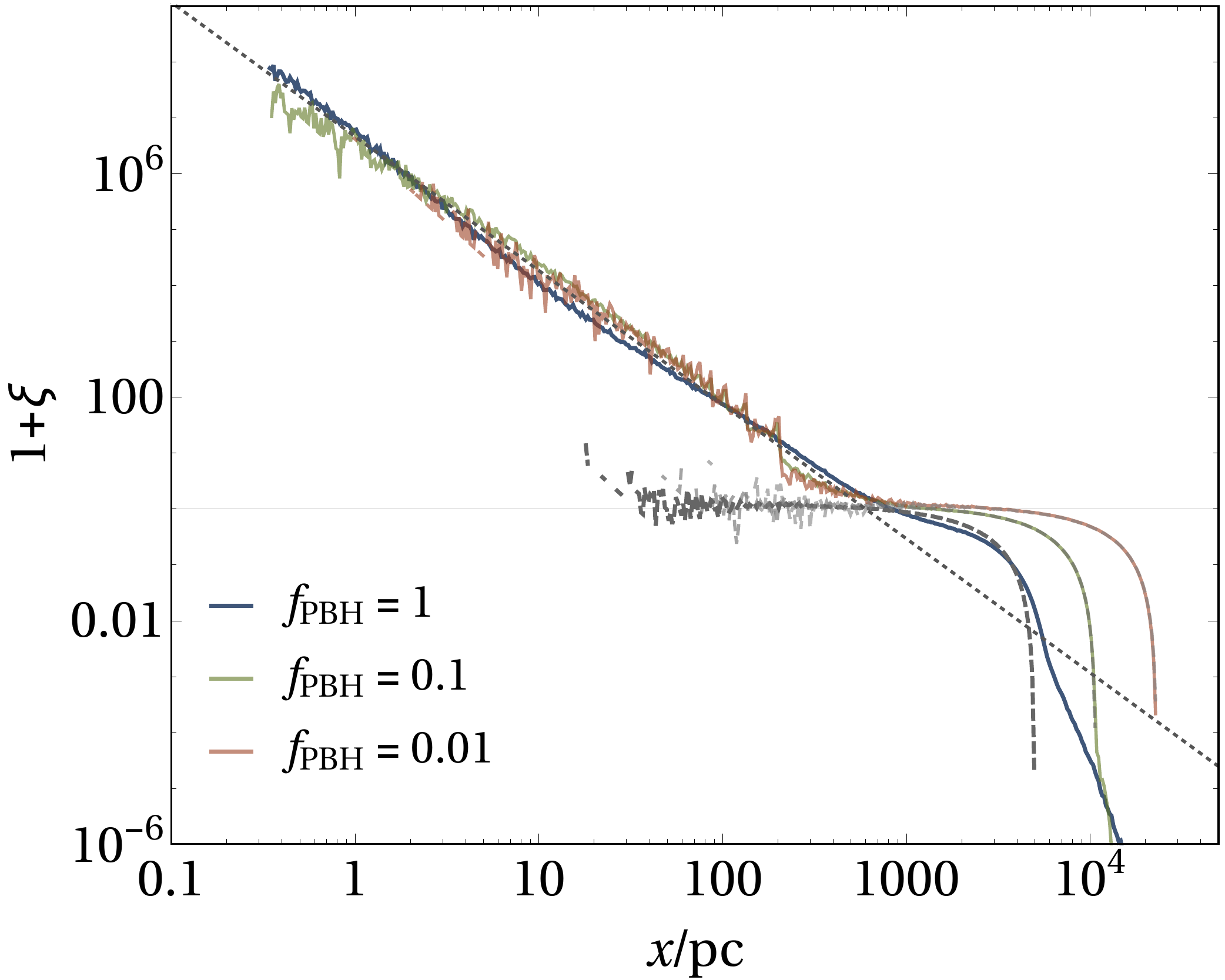} \hspace{4mm}
\includegraphics[height=0.32\textwidth]{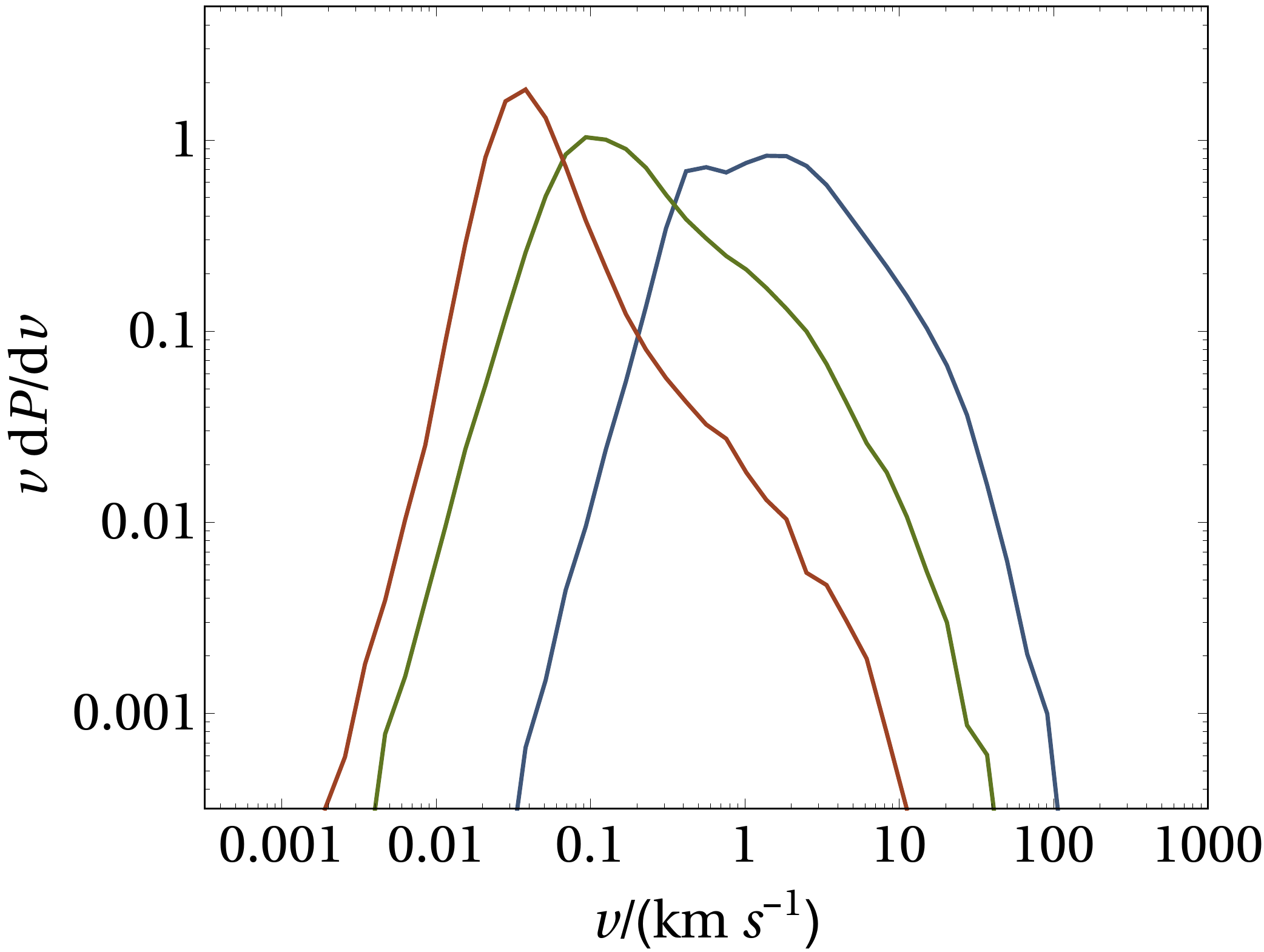} \\
\caption{
{\it Left panel:} The two point function at $a = 3 a_{\rm eq}$ obtained from simulations with $f_{\rm PBH} = 1$ ({\it blue line}), $f_{\rm PBH} = 0.1$ ({\it green line}), , $f_{\rm PBH} = 0.01$ ({\it red line}). The dashed grey lines show the two point function for the initial configuration. The dotted line shows the fit $\xi \propto x^{-2.38}$ at small scales.
{\it Right panel:} The velocity distribution at $a = 3 a_{\rm eq}$ for $f_{\rm PBH} = 1$ ({\it blue line}), $f_{\rm PBH} = 0.1$ ({\it green line}), , $f_{\rm PBH} = 0.01$ ({\it red line}).
}
\label{fig:xi_v}
\end{center}
\end{figure}

 \subsection{Properties of surrounding PBHs}
 \label{surrounding_BH}
 
 Let us briefly examine the properties of the PBHs surrounding the central binary. By combing the data, excluding the central pair, from all simulations with a given $f_{\rm PBH}$, we estimate the two point function of the PBH spatial distribution of surrounding PBH. The result is shown in the left panel of Fig.~\ref{fig:xi_v}. Initially, the two point function, shown by the grey dashed lines, is flat as expected for a Poisson distribution. The drop in $1 +\xi(x)$ for large $x$ appears because of the finite size of the simulation. At $a = 3 a_{\rm eq}$ we observe a power law behaviour at small scales, $\xi(x) \approx  \left( x/x_{c} \right)^{-\gamma}$, with $x_{c} =  604(8)\pc$ and $\gamma = 2.38(1)$, for the data from $f_{\rm PBH} = 1$ simulations. This shape does practically not depend on $f_{\rm PBH}$. Note that at $a = 3 a_{\rm eq}$ the flat region has almost disappeared for the $f_{\rm PBH} = 1$ simulations and $1+\xi(x)$ has developed an extended tail at large $x$, as PBHs can be found outside the initial comoving volume. Thus, larger simulations are needed to probe the small scale structure beyond $a = 3 a_{\rm eq}$.
 
The distribution of velocities of the surrounding PBH is shown in the right panel of Fig.~\ref{fig:xi_v} for different values of $f_{\rm PBH}$. It has a peak which decreases as a power law on both sides and is followed by a roughly log-normal tail. The velocity dispersion in all simulations is approximately constant already after $0.1a_{\rm eq}$ and it takes the values 6 km/s, 1.5 km/s and 0.4 km/s for $f_{\rm PBH} = 1$, $f_{\rm PBH} = 0.1$ and $f_{\rm PBH} = 0.01$, respectively.

\begin{figure}
\begin{center}
\includegraphics[width=0.31\textwidth]{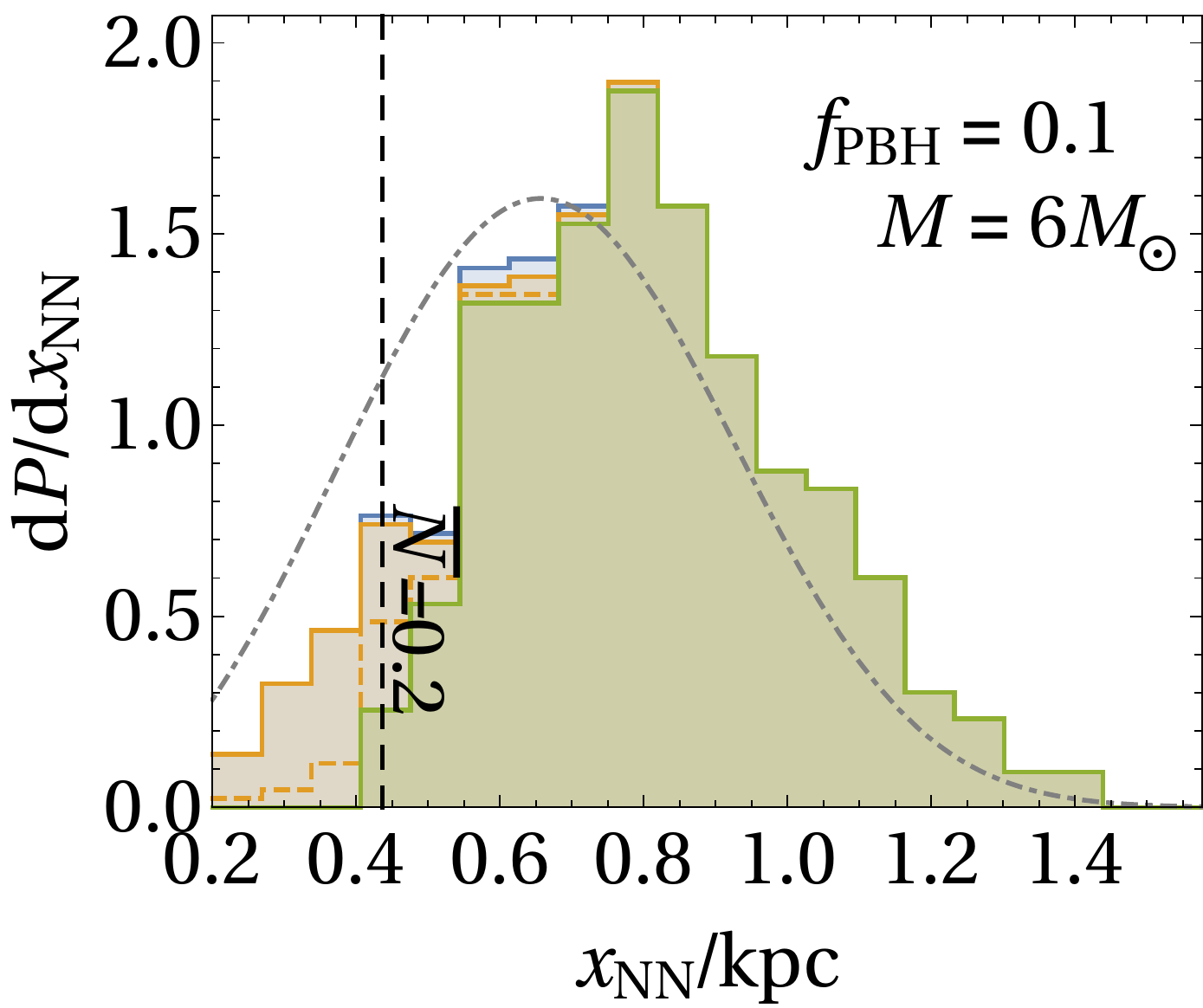} \hspace{16mm}
\includegraphics[width=0.31\textwidth]{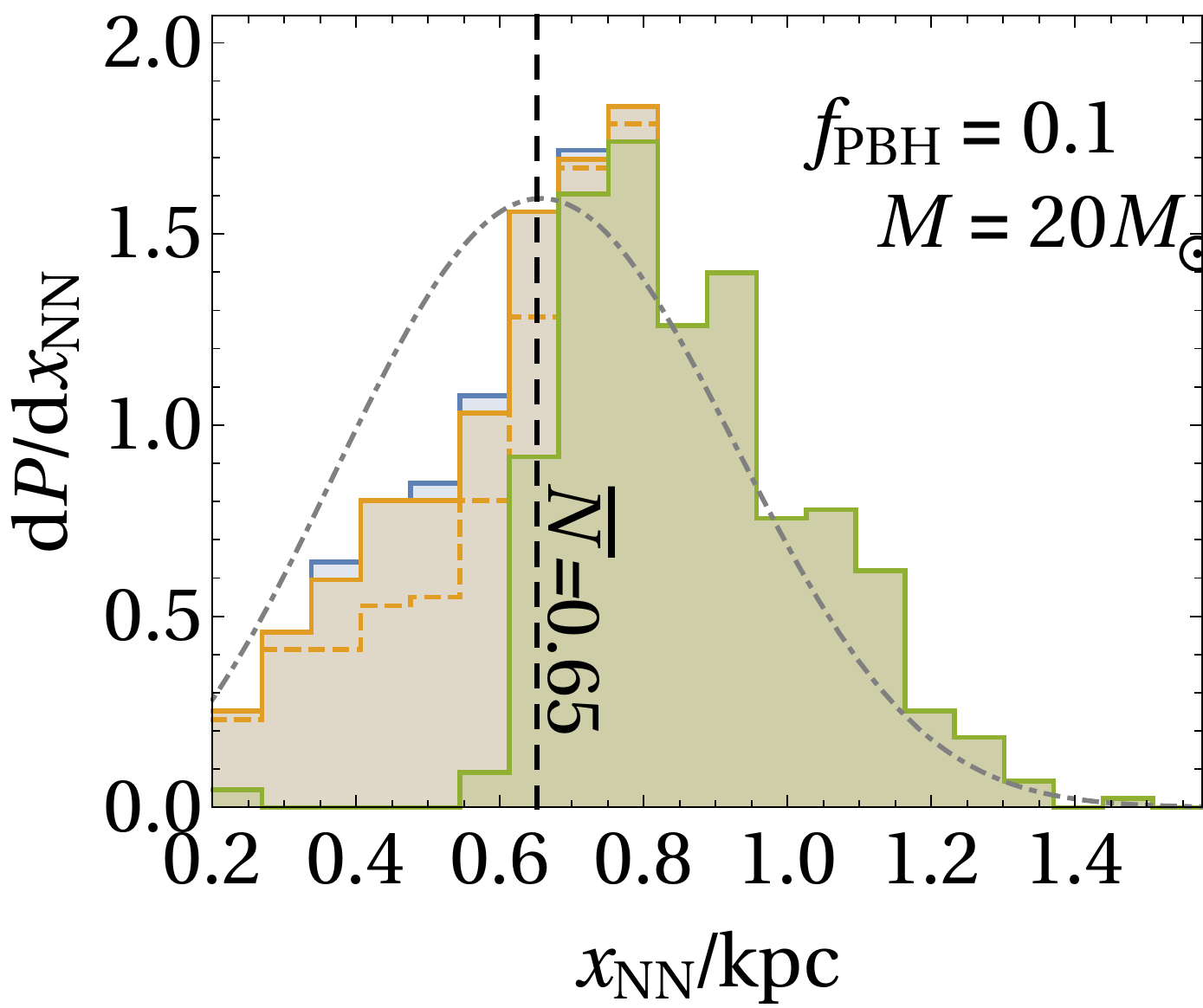} \\ \vspace{4mm}
\includegraphics[width=0.31\textwidth]{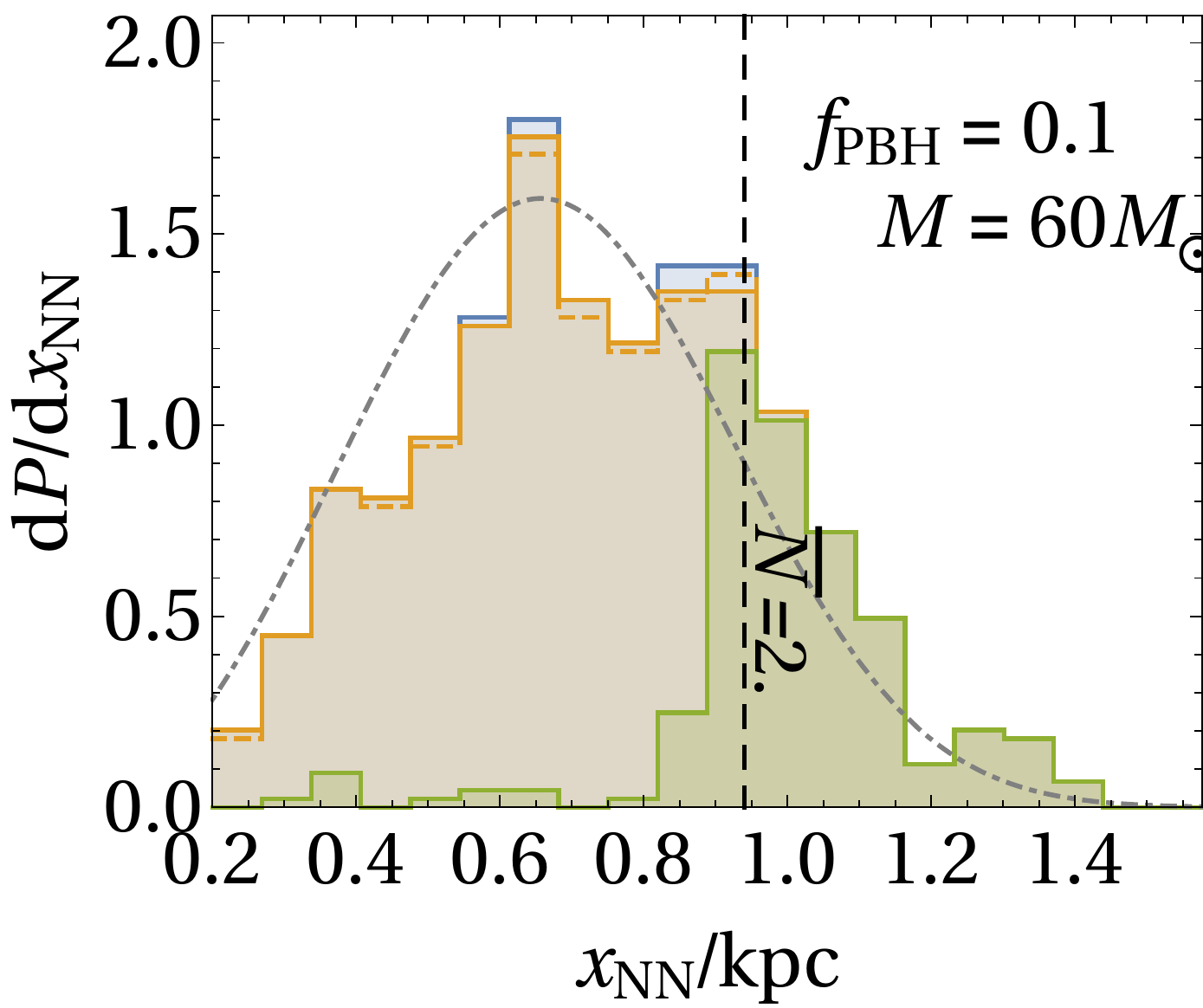}\hspace{16mm}
\includegraphics[width=0.31\textwidth]{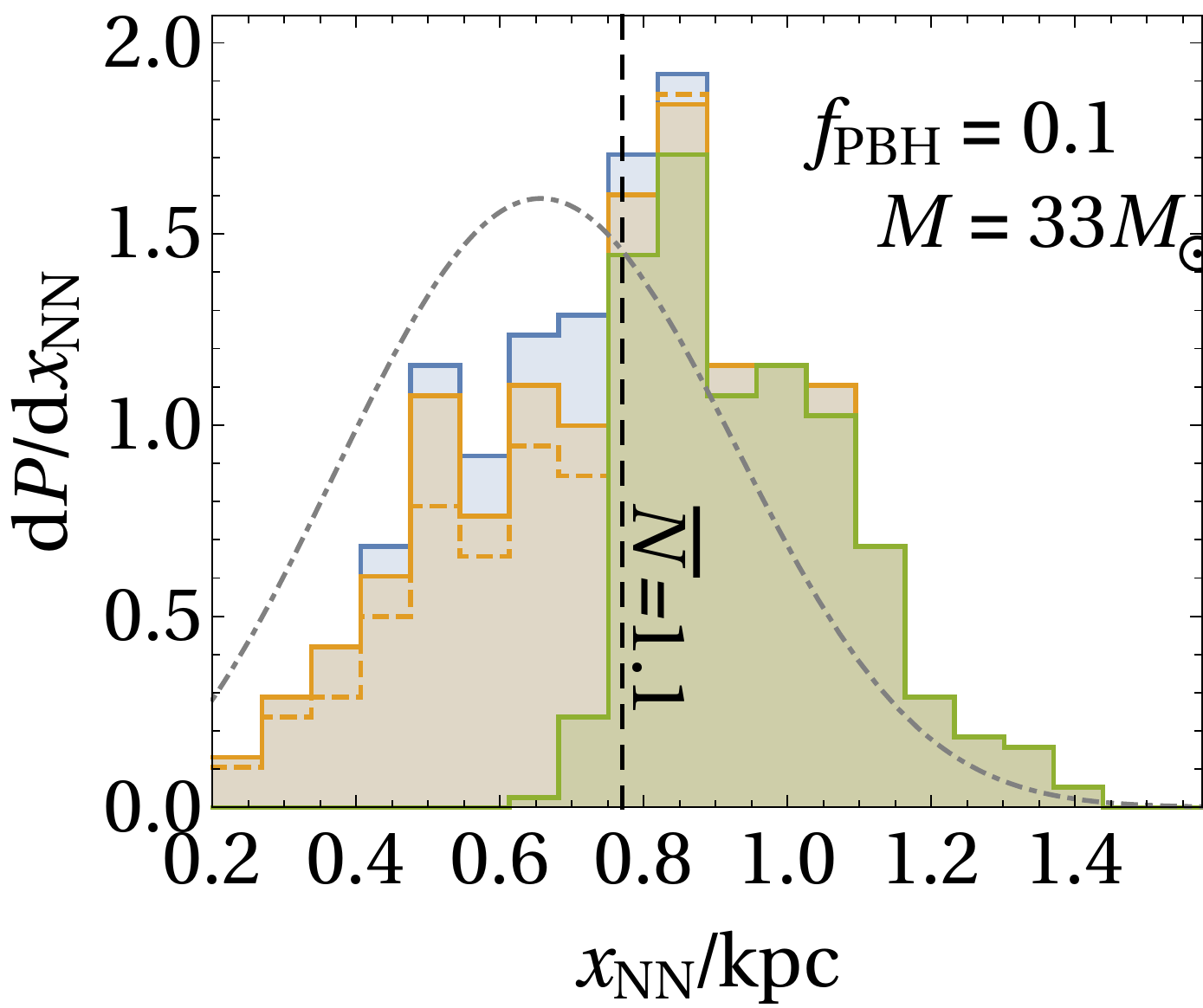}
\caption{The dependence of the state of the central pair at $a  = 3 \, a_{\rm eq}$ on the initial comoving distance of the PBH nearest to the binary in a PBH population with an extended mass function and $f_{\rm PBH} = 0.1$. The different panels show the fate of a central PBH pair with masses $(3+3) \Msun$ ({\it top left}),  $(10+10) \Msun$ ({\it top right}), $(3+30) \Msun$ ({\it bottom left}) and $(30+30) \Msun$ ({\it bottom right}). The blue region shows all simulations, while simulations with bound and undisrupted central binaries at $a  = 3 \, a_{\rm eq}$ are shown in yellow and green, respectively.  The yellow dashed line shows the pairs where the total energy of the central pair is negative, while the solid yellow curve shows initial conditions where at least one of the central PBH forms a binary at the end of the simulation.  The dashed vertical line is the estimate~\eqref{eq:nVy_collapse} for the minimal distance the nearest neighbour can have in order for not to disrupt the binary. The  dot-dashed line corresponds to the expected distribution of the nearest neighbour distance.
}
\label{fig:dist_nn_ext}
\end{center}
\end{figure}

\subsection{Extended mass functions}
\label{ext_mf}

To study the effect an extended mass function has on the survival of the binary, we simulated a simplified scenario where $f_{\rm PBH} = 0.1$ and the PBH population consists of $3\Msun$, $10\Msun$ and $30\Msun$ BH with mass density fractions $25\%$, $50\%$ and $25\%$, respectively. In detail, the simulation included 41, 25 and 4 surrounding PBHs of mass $3\Msun$, $10\Msun$ and $30\Msun$ and a central pair with different combinations of these masses. The state of the central pair at $a = 3 a_{\rm eq}$ with a mass $(3+3) \Msun$,  $(10+10) \Msun$, $(3+30) \Msun$ and $(30+30) \Msun$ is shown in Fig.~\ref{fig:dist_nn_ext}. The histograms are based on 629, 636, 555 and 646 simulations, respectively.  The definitions of the blue, yellow and green regions corresponding to all, bound and non-separated binaries are the same as for Fig.~\ref{fig:dist_nn}. Note that the dashed yellow line can exceed the solid one because the binding energy of the binary with its nearest neighbour is constrained only in the second case. In addition, the most massive, in this case $M = 60\Msun$, binaries can be bound to several light PBHs which provide only a small fraction of the binding energy of the system. The lighter PBHs will, in general, be eventually ejected from the system.

Fig.~\ref{fig:dist_nn_ext} shows that the estimate \eqref{eq:nVy_collapse} for the disruption by the nearest neighbours works relatively well also for non-monochromatic mass functions with $f_{\rm PBH} \ll 1$. We checked that the angular momentum distribution follows the analytic prediction \eqref{eq:Pj}. The coalescence times of the binaries in these simulations were approximately $0.4 t_0$, which is consistent with the monochromatic simulations (see the second panel of Fig.~\ref{fig:dist_tau}).

An important feature specific to extended mass functions is that, although heavy binaries are more easily disrupted, almost all of them will remain bound to each other. As a result, a population of less eccentric heavy binaries with a wide coalescence time distribution, peaked around $10^{10} t_0$, appears. On the other hand, the disruption rate is slightly increased when compared to the monochromatic case, as can also be seen from Fig.~\ref{fig:binary_evol}, because initial binaries containing light PBH are more easily disrupted.

\subsection{Implications for the present merger rate}
\label{subsec:rate}

Let us now turn to the merger rate of initial PBH binaries in the late universe. To estimate which initial conditions will produce undisrupted binaries merging today we will rely on Eq.~\eqref{eq:nVy_collapse} for the choice of an appropriate value for $\bar{N}(y)$. The simulations indicate that this approach works when $f_{\rm PBH} \ll 1$ as for  $f_{\rm PBH} \approx 1$ the disruption of initial binaries is not determined by the closest PBH, but their interaction with (or within) the compact $N$-body systems that form in the early universe. Extrapolating Eq.~\eqref{eq:nVy_collapse} to $a \gg 3 a_{\rm eq}$ and considering only initial conditions that produce binaries expected to survive until the collapse of the first DM structures, i.e. until $a/a_{\rm eq} = 1/\sigma_M$, we obtain
\be\label{eq:nVy}
	\bar{N}(y) \approx \frac{M}{\langle m \rangle}\frac{f_{\rm PBH}}{f_{\rm PBH} + \sigma_{\rm M}}.
\ee
By Eq.\eqref{def:S}, the contribution of binaries much heavier than the average PBH mass will therefore be exponentially suppressed. 

We stress that this is a rough approximation as it assumes that all initial PBH binaries with $x_{NN} \gtrsim y$ survive while all binaries with $x_{NN} \lesssim y$ are disrupted and will not merge within the age of the universe. This is certainly not the case when $f_{\rm PBH} \approx 1$. In the case of a monochromatic mass function, where $\bar{N}(y) \approx 2$ according to Eq.~\eqref{eq:nVy}, we see from the left panel of Fig.~\ref{fig:dist_nn} that about half of the binaries with $x_{NN} \gtrsim y$ are disrupted already after 0.4\,Myr.  The future of the remaining binaries will depend on their interaction with the surrounding PBH clusters and, given a half-life of the order of only 0.4\,Myr, nearly all of these binaries are expected to be disrupted within the age of the universe. The fraction of unperturbed initial binaries will, of course, depend on the subsequent evolution of the PBH clusters, which is not accessible by our simulation. The disruption rate may decrease at later times when, for example, the dense clumps containing a few PBHs are dissolved within larger structures.  By using Eq.~\eqref{eq:R0} as a rough estimate, we note that when $f_{\rm PBH} \approx 1$ the initial binaries can produce a large enough merger rate to be consistent with LIGO even if only $0.1\%$ of the initial binaries remain unperturbed.

\begin{figure}
\begin{center}
\includegraphics[width=0.31\textwidth]{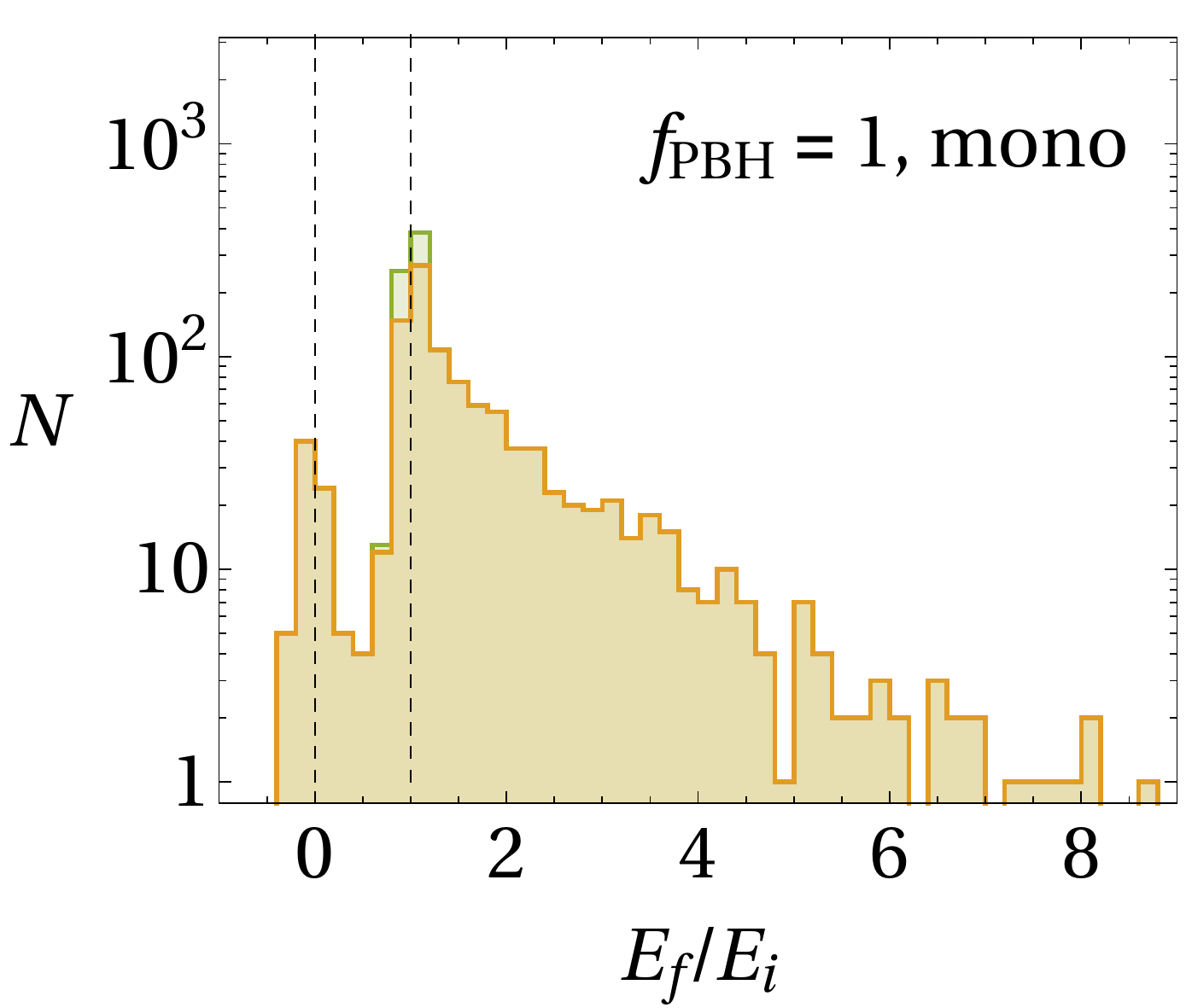} \hspace{1mm}
\includegraphics[width=0.31\textwidth]{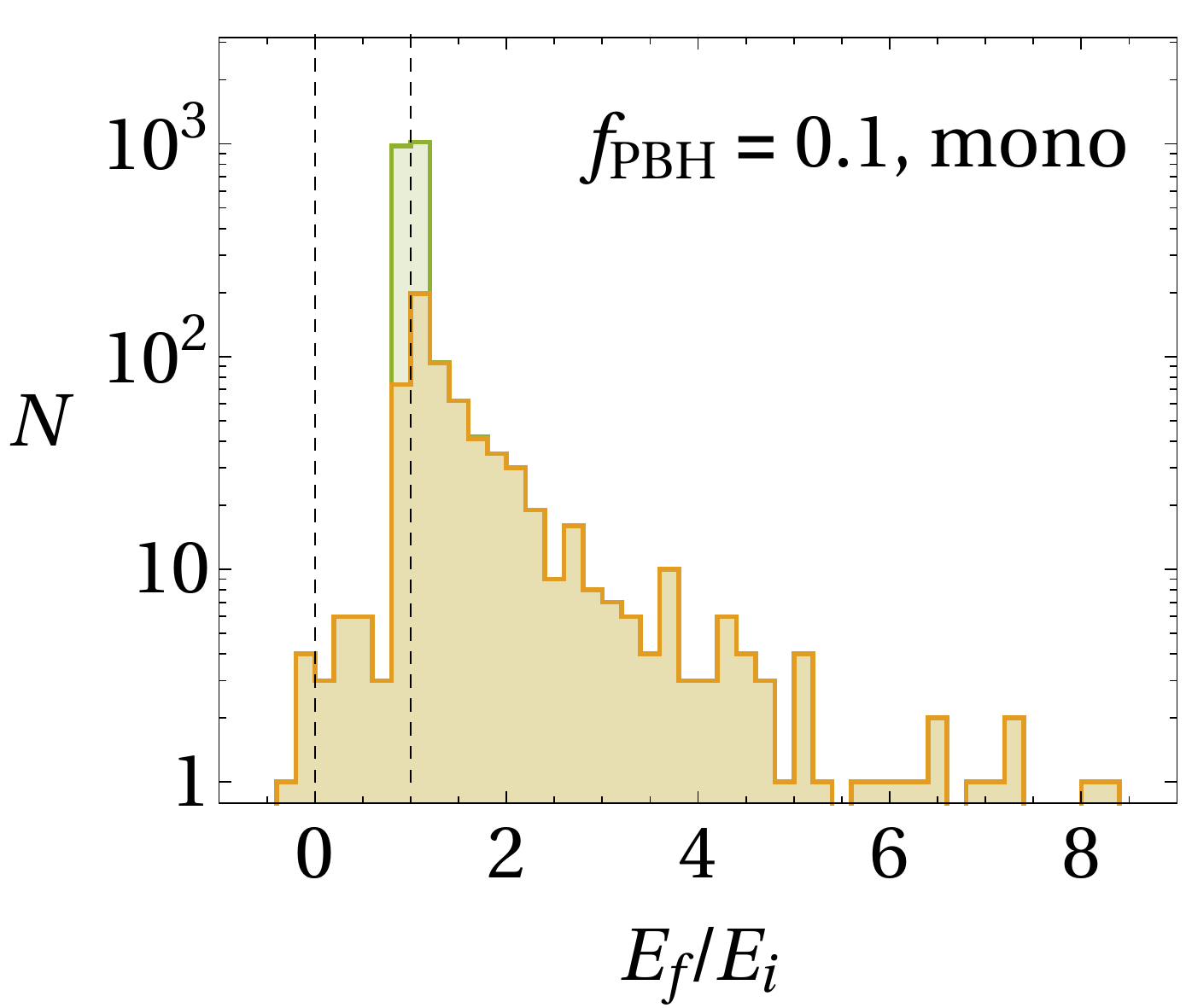} \hspace{1mm}
\includegraphics[width=0.31\textwidth]{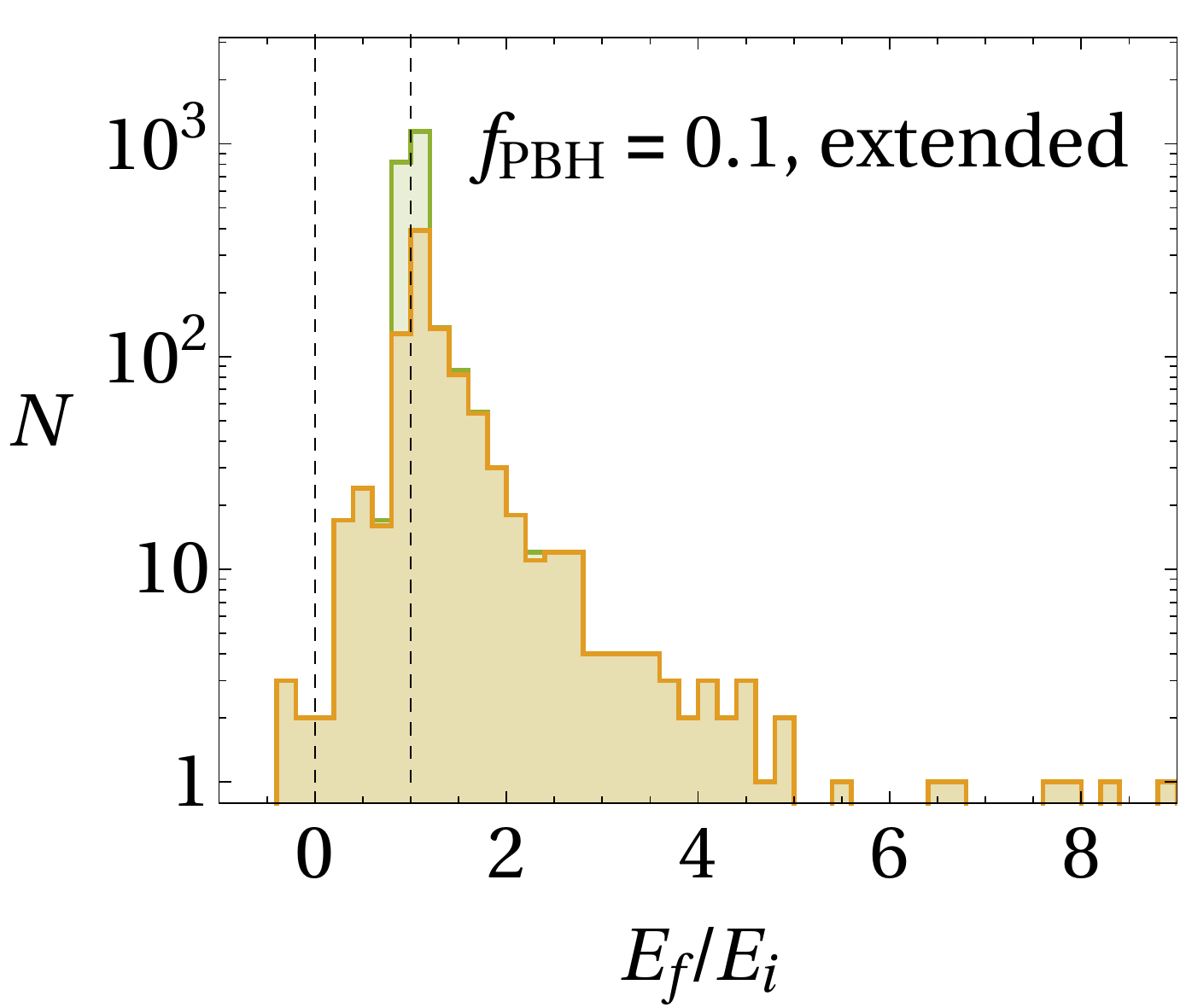}
\caption{The distribution of the relative change of the energy of the central binaries. The initial energy $E_{i}$ corresponds to the energy of the central binary, if it is bound, evaluated at $a=0.3 a_{\rm eq}$, while $E_{f}$ is the total energy of the two body system formed by at least one of the initial PBH and the PBH closest to it, evaluated at $a=3 a_{\rm eq}$. Negative ratios indicate that the binary was separated, $0< E_{f}/E_{i} < 1$ corresponds to softened binaries and $E_{f}/E_{i} > 1$ to hardened ones. The green histogram shows all binaries (disrupted and undisrupted) while the yellow histogram shows only the disrupted binaries containing at least one central PBH. \emph{Left panel:} Simulations with $f_{\rm PBH} = 1$ and $30\Msun$ PBHs. \emph{Middle panel:} Number of simulations with $f_{\rm PBH} = 0.1$ and $30\Msun$ PBHs.  \emph{Right panel:}  Simulations with $f_{\rm PBH} = 0.1$ combining all simulations with extended mass functions described in Sec.~\ref{ext_mf}.}
\label{fig:binary_evol}
\end{center}
\end{figure}

Most of the discussion focuses on a small fraction of all initial binaries -- the ones that are expected to merge within the age of the universe. Even if nearly all binaries are disrupted, a large fraction of PBHs will still form binaries, as can be seen in Figs.~\ref{fig:dist_nn} and \ref{fig:dist_nn_ext}. These binaries may contribute to the present merger rate.

The velocity dispersion estimates in Sec.~\ref{surrounding_BH} indicate that initial binaries merging within the age of the universe are hard  -- a hard binary is defined by having a binding energy, $E = m_1 m_2/(2r_a)$, that is larger than the average kinetic energy of the surrounding bodies. A well known result from studies of globular clusters is the Heggie-Hills law: hard binaries tend to get harder, while soft binaries get softer~\cite{1975MNRAS.173..729H,1975AJ.....80..809H}. The binding energy in binary-single body encounters is, on average, increased by an $\mathcal{O}(1)$ factor, which implies that the semimajor axis will decrease, but not by much. Binary-binary encounters, however, will generally lead to the ionisation of one of the binaries~\cite{1538-3873-104-681-981}. For wide mass spectra, the binary emerging from binary-single body collisions will generally comprise the heaviest PBH. Since the energy is roughly preserved, the final binary will expand so that $r_{a,\rm out}/r_{a,\rm in} \approx m_{\rm heaviest}/m_{\rm lightest}$. 

The distribution of the relative change in the binding energy in different sets of simulations is shown Fig.~\ref{fig:binary_evol}. The data consists of simulations where the central binary was bound at $a = 0.3 a_{\rm eq}$. Hardened binaries are $\mathcal{O}(10)$ times more abundant than softened ones. In detail, the ratio of perturbed binaries (yellow) whose energy increased by at least $5\%$ to the number of perturbed binaries where the energy decreased by more than $5\%$ was 720/114, 493/40 and 622/92 in the panels of Fig.~\ref{fig:binary_evol} counting from left to right, respectively. The frequency of hardened binaries in Fig.~\ref{fig:binary_evol}  drops exponentially fast as $E_{f}/E_{i}$ increases confirming that the binding energy will grow by  a $\mathcal{O}(1)$ factor on average. In both sets of $f_{\rm PBH} =0.1$ simulations, the set of perturbed binaries (yellow) consist mainly of binaries that are expected to be perturbed at $a=3 a_{\rm eq}$ by the estimate \eqref{eq:nVy_collapse} as can be seen in Figs.~\ref{fig:dist_nn} and~\ref{fig:dist_nn_ext}.

Let us now attempt to derive a conservative estimate for the merger rate from perturbed initial binaries in the case $f_{\rm PBH} \approx 1$.  For simplicity we will consider a monochromatic population of PBHs with mass $m_{\rm c}$. In that case, even if most of the initial binaries are disrupted, they will not be ionised. They will have roughly the same energy and semimajor axis as they had initially, but their eccentricity will be considerably decreased. Thus, in order to obtain a rough estimate for the merger rate of such binaries we may consider the following idealised scenario: 
\begin{enumerate}
	\item Most of the initial binaries remain bound. They may have exchanged a PBH.
	
	\item The energy of the perturbed binaries is of the order of the initial binary. This means that we may estimate the semimajor axis from the initial conditions Eq.~\eqref{eq:r_a}. To obtain the largest increase in coalescence time and thus the smallest rate we will assume that $r_a$ remains the same.
	
	\item	The eccentricity of the perturbed binaries is significantly decreased. Again, to obtain the smallest rate, we use $j =1$ as the final value. For final binaries with a coalescence time $\tau$, the initial comoving separation is then fixed by Eq.~\eqref{eq:tau} and the previous assumption, $x_0 \, \approx 10 {\rm pc}\, (m_{\rm c}/\Msun)^{7/16}(\tau/t_0)^{1/16}$. The horizon scale at PBH formation, $0.3 \pc\, (m/\Msun)^{1/2} $, is about an order of magnitude smaller, thus such initial separations are viable.
	
	\item The initial binary should be able to interact with the surrounding PBHs before it merges. Thus we assume that the coalescence time of the initial binary is larger than some time period $\tau_c$. When $r_a$ is fixed, the condition $\tau_{\rm init}  > \tau_c$ implies that we need to consider initial binaries with $j_{\rm init} \geq \left(\tau_c/\tau\right)^{\frac{1}{7}}$. We do not constrain the distance of the closest PBH, thus we can use the initial $j$ distribution in the limit $y\to 0$, given by Eq.~\eqref{eq:Pj_0}. Together with Eq.~\eqref{def:j0} we find that the probability of finding such angular momenta is
\be
	P(\tau_{\rm init} > \tau_c| x_0) 
	= P\left(j_{\rm init}>\left(\tau_c/\tau\right)^{\frac{1}{7}}| x_0, y\to0\right)
%	= j_0 \left(\tau_c/t_0\right)^{-\frac{1}{7}}
	\approx 0.5 \bar{N}(x_0)  \left(\tau/\tau_c\right)^{-\frac{1}{7}}
\ee
for $\bar{N}(x_0) \ll 1$. By the assumptions above $\bar{N}(x_0) \approx 2\times10^{-4} f_{\rm PBH} (m_{\rm c}/\Msun)^{5/16} $.
\end{enumerate}

In all, with a final distribution of angular momenta peaked at $j=1$ and $\tau_c \approx 1\, \rm Myr$ we obtain from Eq.~\eqref{dR_early} that the present merger rate,\footnote{Formally, the evolution of the orbital parameters of initial binaries can expressed as
\be \nonumber
	P(j,r_a) = \int \td j_{\rm init} \td r_{a,\rm init} \, P(j,r_a|j_{\rm init},r_{a,\rm init})P(j_{\rm init},r_{a,\rm init})\,,
\ee
where according to the simplifying assumptions
\be \nonumber
	P(j,r_a|j_{\rm init},r_{a,\rm init}) =  \delta(j-1) \delta(r_{a}-r_{a,\rm init}) \theta\left(j_{\rm init} -  \left(\tau_c/\tau\right)^{\frac{1}{7}}\right).
\ee
Note that the transition probability does not have to be normalised to unity since the number of binaries is not conserved. The rate can then be evaluated by replacing the distribution $P(j_{\rm init},r_{a,\rm init})$ with $P(j,r_a)$ in Eq.~\eqref{dR_early}.}
\bea\label{eq:dR_pert}
	R 
&	\gtrsim \frac{3n}{32 \tau}  \bar{N}(x_0) P(\tau_{\rm init} > \tau_c|x_0)  \\
&	\approx 6 \, \Gpc^{-3} \yr^{-1} 
	 \, f_{\rm PBH}^{3}\left(\frac{m_{\rm c}}{10 \Msun}\right)^{-\frac{3}{8}} \left(\frac{\tau}{t_{0}}\right)^{-\frac{27}{56}},
\eea
can be only slightly below the merger rate reported by the LIGO and Virgo collaborations, $9.7-101\,\Gpc^{-3} \yr^{-1}$~\cite{LIGOScientific:2018mvr}. We stress that this is a very rough lower bound as more realistic scenarios will, on average, have final values of $j$ or $r_a$ which will increase this rate. Also the interactions with other surrounding matter will lead to an increase of $j$ and of the binding energy~\cite{Kavanagh:2018ggo}. With $\tau_c = 1 \,\rm Myr$ we have that $j_{\rm initial} \geqslant 0.26$.  The corresponding relatively high initial angular momentum will generally require the nearest neighbour to be initially much closer than the average separation between PBHs. Thus, it is expected that it couples to the binary early and, moreover, a more accurate description of the binary emerging from this configuration likely reduces to solving the full 3-body problem on an expanding background. In any case, the perturbative approach outlined in Sec.~\ref{sec:dynamics} has limited value in such situations. Another indicator that such binaries originate from early 3-body systems is the $f_{\rm PBH}^{3}$ dependence of the rate. To obtain an observable merger rate only a very small fraction of initial conditions have to produce binaries -- the probability of the required initial 3-body configurations appear is roughly $\bar{N}(x_0)^2 \approx \mathcal{O}(10^{-8})$. The merger rate can decrease mainly by ionising these binaries. According to Fig.~\ref{fig:binary_evol}, this will take place with a small probability, and note that the binaries in our simulations are much softer than the ones contributing to the rate \eqref{eq:dR_pert}, i.e. they are easier to disrupt.

In conclusion, the perturbative estimate \eqref{eq:R_tot} becomes more accurate when $f_{\rm PBH} \ll 1$. For $f_{\rm PBH} \approx 1$ most of the initial binaries are disrupted\footnote{In contrast to Ref.~\cite{Bringmann:2018mxj}, our results indicate that initial conditions for which the local PBH density is comparable or larger than the average density of DM likely lead to a decrease of the merger rate instead of a cascade of mergers. This is because such initial conditions tend to form compact $N$-body systems in which the initial binaries are rapidly disrupted.}, yet the disrupted binaries should still produce a merger rate consistent with LIGO. In both cases a more careful study of the formation and evolution of small scale structures and their interaction with the binaries is needed.

%-------------------------------------------------------------------------------
\section{Phenomenology of PBH mergers} 
\label{sec:phenomenology}
%-------------------------------------------------------------------------------

Now we turn to discuss the implications of the PBH mergers to the phenomenology of the PBH mergers. Using the results of the previous sections, we first study the temporal behaviour of the PBH merger rate, and by comparing it to the case of astrophysical BHs we confirm the earlier statements (see e.g. Ref.~\cite{Sasaki:2018dmp}) that the merger rates of primordial and astrophysical BHs are very different especially at large redshifts. Then, we perform a likelihood fit of the PBH mass function on the observed merger events, and finally derive potential constraints on the fraction of DM in PBHs from the rate of observed events and from the non-observation of the stochastic GW background.

%-------------------------------------------------------------------------------
\subsection{Temporal behaviour of the merger rate} 
\label{sec:rates}
%-------------------------------------------------------------------------------

\begin{figure}
\begin{center}
\includegraphics[width=0.5\textwidth]{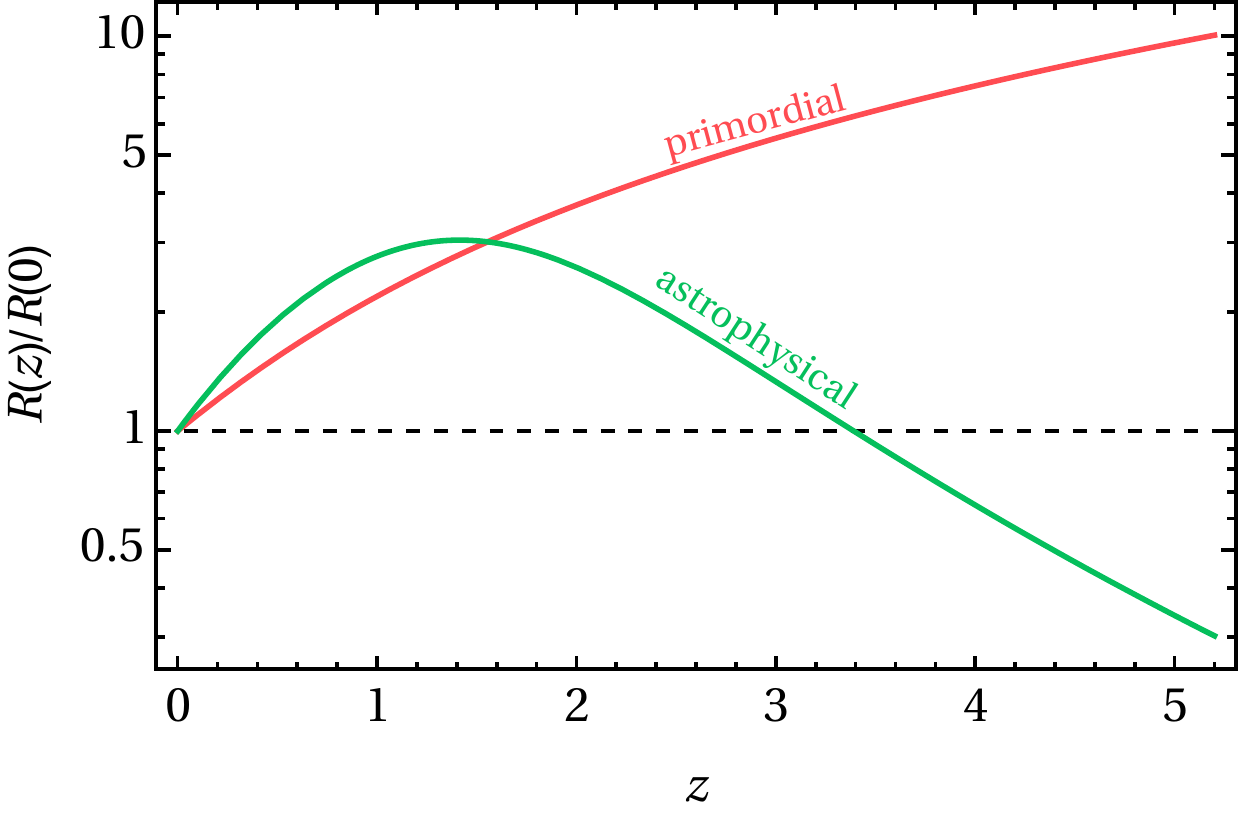}
\caption{The lines show the $z$ dependence of the BH merger rates normalised to one at $z=0$ for early PBH binary formation (red) and for the astrophysical BH binaries (green).}
\label{fig:rates}
\end{center}
\end{figure}

The $z$ dependence of the PBH merger rate~\eqref{dR_early} is shown by the red line in Fig.~\ref{fig:rates}. For comparison, we show by the green line the $z$ dependence of the merger rate of astrophysical BHs. The properties of astrophysical BH binaries are assumed to inherit the properties of stellar binaries from which they are formed. In that respect, the $z$ dependence follows from the star formation rate and the delay from the formation of a stellar binary to merging of the BHs. The star formation rate is well approximated by the empirical formula~\cite{Madau:2014bja,Belczynski:2016obo}
\be
	{\rm SFR}(z) \propto \frac{(1 + z)^{2.7}}{1 + ((1 + z)/2.9)^{5.6}} \equiv P_b(z)\,,
\ee
and, as in Refs.~\cite{Belczynski:2016obo,TheLIGOScientific:2016wyq}, we assume that the delay time distribution is $P_d(t) \propto t^{-1}$ for $t > 50\,\rm{Myr}$ and zero otherwise. Then, the differential merger rate of BH binaries is given by
\be\label{RA}
	\td R_A \propto \left(\int \td t_d \td z_b P_b(z_b) P_d(t_d) \delta(t(z)-t(z_b)-t_d) \right) M^\alpha \eta^\beta \psi(m_1)\psi(m_2) \td m_1 \td m_2 \,,
\ee
where $\alpha$ and $\beta$ parametrise the mass dependence of the astrophysical binary formation. These are not relevant for the $z$ dependence that is given by the part in brackets.

The temporal behaviour of the merger rates for the primordial and astrophysical BHs are very different especially at large redshifts. Whereas the PBH merger rate monotonically increases as a function of $z$ due to the universal time dependence $R \propto \tau^{34/37}$ predicted by Eq.~\eqref{eq:R0}, the astrophysical BH merger rate starts to drop above $z\simeq1$. LIGO is able to observe BH mergers only at low redshifts, $z\lsim 0.4$, but future GW observatories such as LISA or DECIGO can instead probe much higher redshifts, $z\simeq10$. As pointed out e.g. in Ref.~\cite{Sasaki:2018dmp}, the observations of high redshift BH mergers can reveal if a significant fraction of DM is in PBHs.

%-------------------------------------------------------------------------------
\subsection{Likelihood fit to LIGO observations}
\label{sec:fit}
%-------------------------------------------------------------------------------

Next we study the LIGO observations in the light of the derived merger rate. We start by finding a fit for the lognormal PBH mass function. We perform a maximum likelihood analysis considering the ten observed BH-BH merger events listed in Table III of Ref.~\cite{LIGOScientific:2018mvr}. The log-likelihood function reads 
\be
\ell = \sum_j \ln \frac{ \int \td P(m_1,m_2,z) p_j(m_{j,1} | m_1) p_j(m_{j,2} | m_2) p_j(z_j | z) \theta(\rho(m_1,m_2,z)-\rho_c)}{\int \td P(m_1,m_2,z) \theta(\rho(m_1,m_2,z)-\rho_c)} \,,
\ee
where $\td P(m_1,m_2,z) \propto \td R(m_1,m_2,z) \td V_c(z)$ is the differential probability of having a BH binary consisting of individual masses $m_1$ and $m_2$ merging at redshift $z$, and $V_c(z)$ is the comoving volume. The experimental uncertainties are accounted by $p_j(m_j | m)$ denoting the probability to observe a BH mass $m_j$ given that the BH has mass $m$, and $p_j(z_j | z)$, that is the probability to observe a BH-BH merger at redshift $z_j$ when it happens at redshift $z$. We take both $p_j$ to be Gaussian. The unit step function implements a detectability threshold based on the signal-to-noise ratio $\rho$ of the GW events.

Our estimate of the detectability of BH-BH mergers by LIGO is based on the signal-to-noise ratio of the GW events as in Refs.~\cite{TheLIGOScientific:2016pea,Abbott:2017vtc}. We characterise the sensitivity of LIGO by the total strain noise, $S_n(f)$, of the detectors using the fit given in Ref.~\cite{Bai:2018shq}. The actual strain sensitivity of the detectors fluctuates and has increased over time, whereas the total strain noise that we use is constant in time, and its amplitude is slightly above the best performance of the LIGO detectors. We therefore believe that our analysis may underestimate the detectability of the events, so a more careful treatment may produce slightly stronger constraints.

The signal-to-noise ratio is given by~\cite{TheLIGOScientific:2016pea},
\be
	\rho^2 \equiv \int_0^\infty \frac{4|\tilde{h}(\nu)|^2}{S_n(\nu)} \td \nu \,,
\ee
where $\tilde{h}(\nu)$ is the Fourier transform of the signal~\cite{Cutler:1993vq,Chernoff:1993th}, that we compute using the results of Ref.~\cite{Ajith:2007kx}. For the detectability threshold we take $\rho > \rho_c = 8$~\cite{TheLIGOScientific:2016pea,Abbott:2017vtc}, implying roughly that mergers taking place at $z>0.4$ are not detectable.

\begin{figure}
\begin{center}
\includegraphics[height=0.46\textwidth]{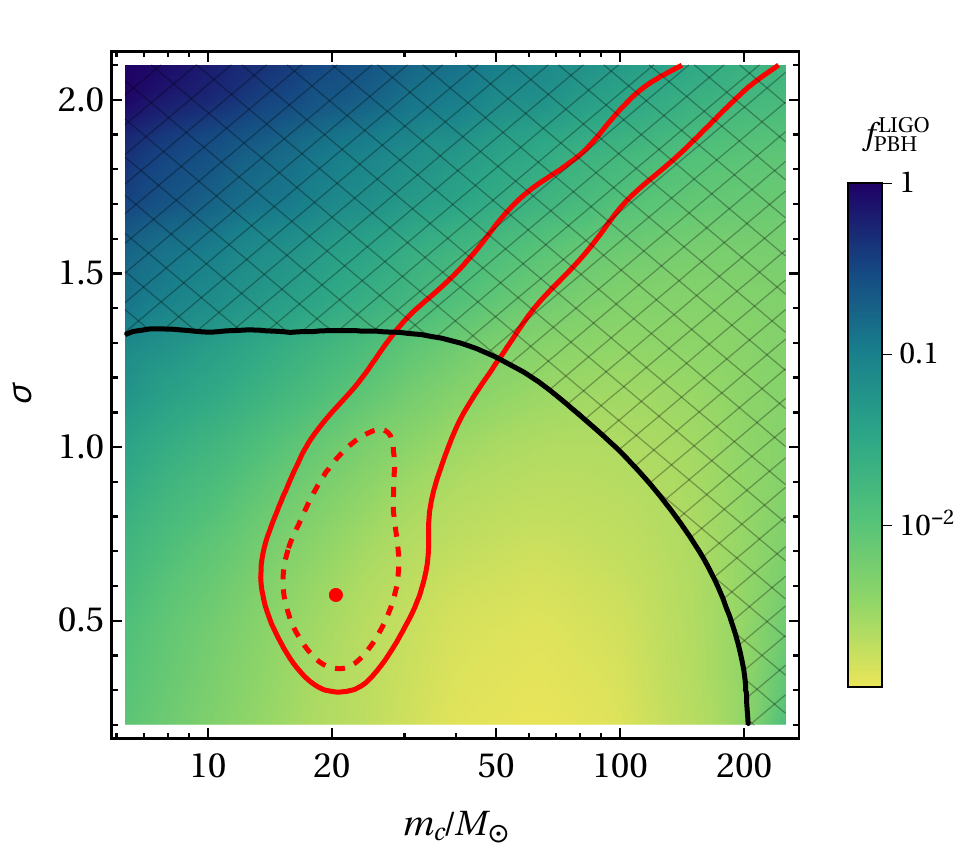} %\hspace{4mm}
\caption{%\emph{Left panel:} 
The red dot shows the best fit, $m_c = 20\Msun$, $\sigma=0.6$, of the maximum likelihood fit to the LIGO BH-BH merger events for the lognormal mass function, and the red dashed and solid lines correspond respectively, the relative likelihoods $\ell-\ell_{\rm max}=-3.09,\,-5.91$, which we refer as $2\sigma$ and $3\sigma$ confidence levels. The colour coding indicates the upper bound on the fraction of DM in PBHs from the observed BH-BH merger event rate. In the hatched region above the black solid line the PBH explanation of the observed GW events is excluded by existing constraints on the fraction of DM in PBHs. %\emph{Right panel:} The solid and dashed lines show, respectively, the relative likelihoods $\ell-\ell_{\rm max}=-3.09,\,-5.91$ for the fit of the mass function parameters $\gamma$ and $m_{\rm max}$ for astrophysical BH mergers.
}
\label{fig:fit}
\end{center}
\end{figure}

Assuming that all LIGO BH-BH merger events are created by the early PBH binary formation mechanism, we can constrain the PBH mass function. The result of a maximum likelihood fit for a lognormal mass function~\eqref{lognormalmf} is shown in Fig.~\ref{fig:fit}, where the red dashed and solid lines correspond to the $2\sigma$ and $3\sigma$ confidence levels, respectively, and the best fit, $m_c = 20\Msun$, $\sigma = 0.6$, is indicated by the red dot. The best fit points to a relatively narrow mass functions, but the $3\sigma$ region extends to large widths. However, the constraints on the fraction of DM in PBHs disfavour very wide mass functions, as we will show in the next subsection.

%-------------------------------------------------------------------------------
\subsection{Constraints on the PBH abundance} 
\label{sec:constraints}
%-------------------------------------------------------------------------------

\begin{figure}
\begin{center}
\includegraphics[height=0.5\textwidth]{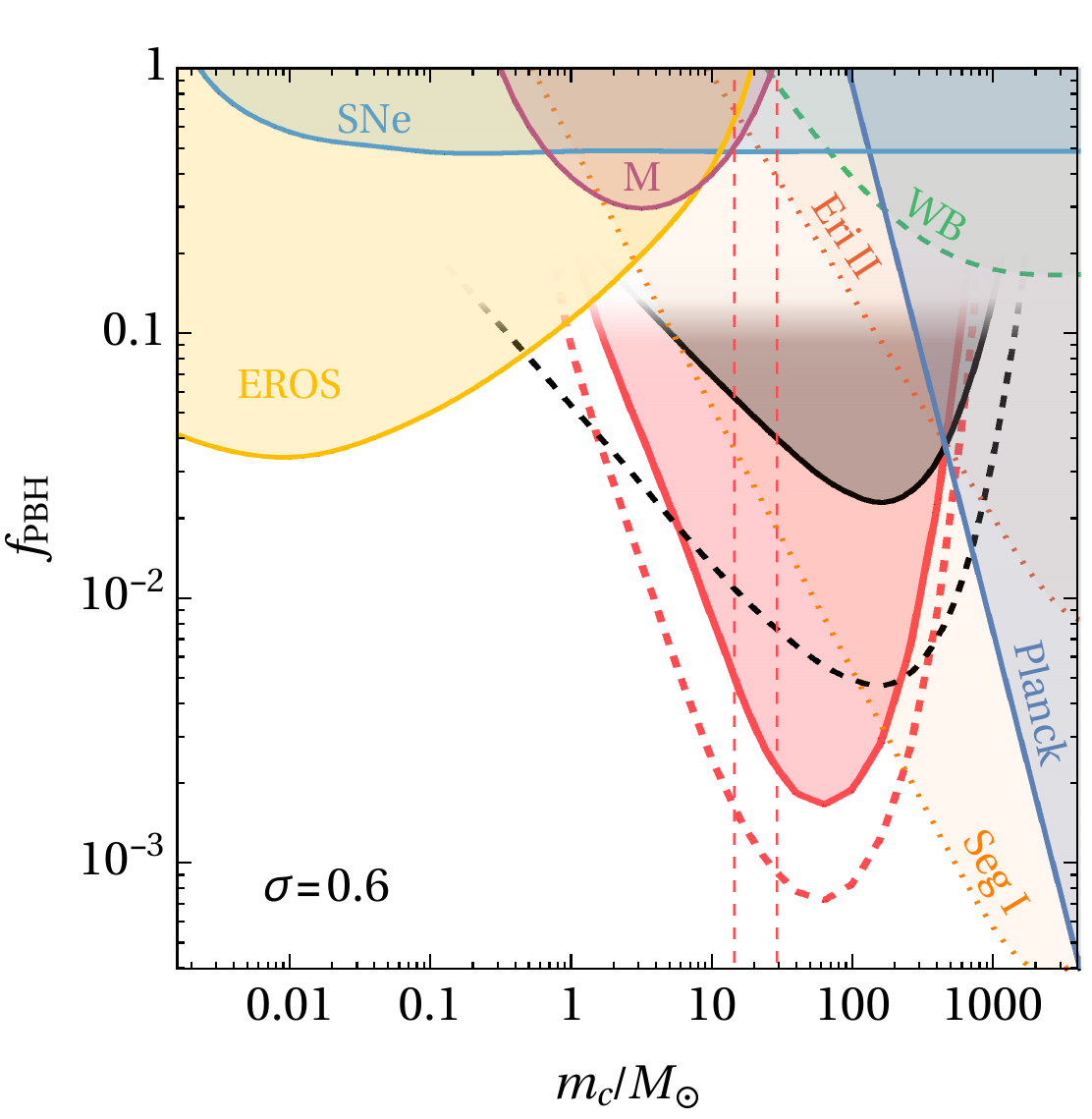}
\caption{Constraints for the lognormal mass function with $\sigma=0.6$. The red solid line shows the $2\sigma$ upper bound on the fraction of DM in PBHs, $f_{\rm PBH}$, from the observed merger rate by LIGO, and the red dashed line shows the $2\sigma$ lower limit on $f_{\rm PBH}$ assuming that all observed BHs are primordial. The vertical dashed lines show the $2\sigma$ confidence levels from the fit shown in Fig.~\ref{fig:fit}. The black solid curve shows the constraint on $f_{\rm PBH}$ from non-observation of the stochastic GW background by the second LIGO observation run, and the black dashed line the projected sensitivity of the final phase of LIGO. Since the analytic merger rate estimates are not reliable and the numerical simulations indicate that the merger rate is strongly suppressed for $f\gsim 0.1$, we do not show the above constraints in that region. The yellow, purple and light blue regions are excluded by the microlensing results from EROS~\cite{Tisserand:2006zx} and MACHO (M)~\cite{Allsman:2000kg}, and by lack of lensing signatures in type Ia supernovae (SNe)~\cite{Zumalacarregui:2017qqd}, respectively. The dark blue, orange, red and green regions are ruled out by Planck data~\cite{Ali-Haimoud:2016mbv}, survival of stars in Segue I (Seg I)~\cite{Koushiappas:2017chw} and Eridanus II (Eri II)~\cite{Brandt:2016aco}, and the distribution of wide binaries (WB)~\cite{Monroy-Rodriguez:2014ula}, respectively. 
}
\label{fig:constraints}
\end{center}
\end{figure}

The prediction for the number of BH-BH merger events that LIGO should see within the period $\Delta t$ is
\be
	N = \Delta t \int \td R(m_1,m_2,z) \td V_c(z) \theta(\rho(m_1,m_2,z)-\rho_c) \,.
\ee 
During $\Delta t \simeq 165$\,days, LIGO has observed $N_{\rm obs}=10$ BH-BH merger events~\cite{LIGOScientific:2018mvr}. We estimate the error of $N_{\rm obs}$ from Poisson statistics by taking the log-likelihood region $\ell - \ell_{\rm max} < -2$, which for Gaussian distribution would correspond to the $2\sigma$ confidence level. This gives $N_{\rm obs}=10\substack{+7.7 \\ -5.1}$, so by $2\sigma$ confidence level the number of observed events is smaller than $17.7$. This can be converted to an upper bound on the PBH abundance $f_{\rm PBH}$ for a given mass function using the merger rate estimate~\eqref{eq:R_tot}. In the case of lognormal mass function the resulting maximal value of $f_{\rm PBH}$ is shown by the colour coding in Fig.~\ref{fig:fit} in the $(m_c,\sigma)$ -plane, and for $\sigma = 0.6$ by the red solid line in Fig.~\ref{fig:constraints}. Moreover, the region between the red solid and red dashed lines in Fig.~\ref{fig:constraints} corresponds $4.9<N<17.7$, that is, it corresponds to values of $f_{\rm PBH}$ for which the LIGO BH-BH merger event rate can be obtained at a $2\sigma$ confidence level. We remark that the PBH abundance inside the $2\sigma$ region in Fig.~\ref{fig:fit} is $f_{\rm PBH} < 0.01$, confirming the validity of the perturbative merger rate estimate~\eqref{eq:R_tot}.

The non-observation of the stochastic GW background can further constrain the PBH abundance. The spectrum of the stochastic GW background from binary BH coalescences is~\cite{Phinney:2001di,Zhu:2011bd}
\be \label{eq:GWB}
	\Omega_{\rm GW}(\nu) = \frac{\nu}{ \rho_c} \int \frac{\td R(m_1,m_2,z)\,\td z}{(1+z) H(z)}  \frac{\td E_{\rm GW}(\nu_r)}{\td \nu} \, \theta(\rho_c-\rho(m_1,m_2,z))\,,
\ee
where $\rho_c$ denotes the critical density, $\nu_r=(1+z)\nu$ is the redshifted source frequency, $H(z)$ is the Hubble parameter, and $\td E_{\rm GW}$ is the total GW energy in the frequency range $(\nu,\nu+{\rm d}\nu)$ emitted in the coalescence of BHs~\cite{Cutler:1993vq,Chernoff:1993th,Ajith:2007kx}. The factor $\theta(\rho_c-\rho(m_1,m_2,z))$ subtracts the contribution of events which can be observed individually. We find, however, that this factor can safely be neglected.

We calculate the stochastic GW background arising from coalescences of the early formed PBH binaries. By comparing its strength to the sensitivity of LIGO, we then constrain the fraction of DM in PBHs. The resulting upper bound arising from the latest LIGO observation run is shown by the black solid line in Fig.~\ref{fig:constraints}. The black dashed line shows instead the projected final sensitivity of LIGO. We see that in the range where LIGO is most sensitive the constraint from the observed merger rate is significantly stronger than the one from non-observation of the stochastic GW background. We remark that the constraint from the GW background is also weaker than the previous estimate given in Ref.~\cite{Raidal:2017mfl}. This is due to the exclusion of the initial conditions where the initial binary is perturbed by its nearest neighbour. The constraint from the stochastic GW background is affected more strongly by the mass dependence of~\eqref{eq:nVy} and the resulting exponential suppression of the merger rate of heavy binaries. Note, however, that via the stochastic GW background it is possible to probe narrow mass functions with $m_{\rm c} < 1 \Msun$ for which the individual mergers are not observable by LIGO. The stochastic GW background from early PBH binary mergers was also studied in~\cite{Chen:2018rzo}.

The other constraints shown in Fig.~\ref{fig:constraints} arising from lensing (EROS, Macho and SNe), accretion (Planck), observations of compact stellar systems (Eridanus II and Segue I) and wide binaries are evaluated for the lognormal mass function using the method introduced in Ref.~\cite{Carr:2017jsz}. The strongest of these (EROS and Segue I) exclude the hatched region above the black solid line in Fig.~\ref{fig:fit} for the fraction of DM in PBHs which would give $N=4.9$. Thus, the PBH origin of the LIGO merger events is excluded in that region as the required minimal PBH abundance is larger than allowed by EROS or Segue I.

We emphasise that whereas the early merger rate estimate for $f\simeq 1$ is subject to serious uncertainties as discussed in the previous sections, the merger rate~\eqref{eq:R_tot} with the choice \eqref{eq:nVy} works well for $f_{\rm PBH} \ll 1$. As the $f_{\rm PBH} \lesssim 0.1$ is already excluded by EROS or Segue I, we conclude that the observed merger rate significantly strengthens the constraint in the mass range $2 - 160\Msun$. If the BHs observed by LIGO have a primordial origin, the fraction of DM in PBH and the peak mass of the distribution are contained within the region in Fig.~\ref{fig:constraints} bound by the dashed red vertical lines and the red solid and dashed curves. These follow from the $2\sigma$ fit of the mass function and the merger rate, respectively.

%-------------------------------------------------------------------------------
\section{Conclusions} 
\label{sec:conclusions}
%-------------------------------------------------------------------------------

We studied the formation and evolution of PBH binaries in the early universe, and revised the constraints from LIGO on the PBH abundance. Formed already before matter-radiation equality, the PBH binaries are the first gravitationally bound structures in the universe. We evaluated the characteristics of these binaries analytically allowing for extended PBH mass functions. The angular momentum was estimated perturbatively from the tidal force created by all surrounding matter, requiring that the PBH closest to the binary was sufficiently far. We then studied this scenario numerically using $N$-body simulations finding that the analytically obtained distributions of orbital characteristics of the binaries are in good agreement with the numerical result. We found that the angular momentum distribution is approximately Gaussian for the undisrupted PBH binaries.

The initially formed binary may be disrupted shortly after formation by a nearby PBH or a small cluster of PBHs. The latter may appear due to the Poissonian fluctuations, and we find that when $f_{\rm PBH} = 1$ a large fraction of the initial binaries expected to merge within the age of the universe were absorbed into these clusters already before recombination. If $f_{\rm PBH} \ll 1$ the $N>2$ body systems form later, and in this case we found that single nearby PBHs are the dominant source of disruption. We estimated analytically which initial configurations avoid this disruption, and derived the suppression factor for the merger rate caused by this. We verified these estimates for the binary disruption by comparing them with the results of the $N$-body simulations. 

We concluded that our analytical merger rate gives a good approximation for the merger rate of the initially formed PBH binaries in the case $f_{\rm PBH} \ll 1$, assuming that the disruption rate at later stages is not significant. For $f_{\rm PBH} = 1$  the present merger rate estimate is greatly suppressed already at recombination, and our analytical results for the merger rate fail. Nevertheless, as most initially close PBH pairs will form binaries that are unlikely to be ionised, we argue that the population of perturbed binaries can still generate a merger rate consistent with LIGO. Moreover, since our results depend on the local PBH density, we conclude that, in case the PBH are initially clustered, the early binaries are more likely to be disrupted and the merger rate is therefore suppressed.

We used our merger rate results to re-evaluate the constraints on the PBH abundance from LIGO observations. We calculated the constraints from both the observed merger rate and the non-observation of the stochastic GW background. We found that the former gives the strongest constraint on the PBH abundance in the mass range $2 - 160\Msun$. We also performed a likelihood fit for the lognormal PBH mass function on the 10 observed BH-BH merger events finding a best fit at $m_c = 20\Msun$ and $\sigma=0.6$. In that case the fraction of DM in PBHs needed to obtain the observed merger rate is $f_{\rm PBH} \simeq 0.002$.

%-------------------------------------------------------------------------------
\acknowledgments
%-------------------------------------------------------------------------------
The authors thank Yacine Ali-Ha\"{i}moud  for useful discussions. This work was supported by the grants IUT23-6, PUT799, by EU through the ERDF CoE program grant TK133,  and by the Estonian Research Council via the Mobilitas Plus grant MOBTT5. VV was supported by the United Kingdom STFC Grant ST/L000326/1.

\bibliography{PBBH}

\end{document}